\documentclass[]{aa}
\pdfoutput=1

\usepackage{graphicx}
\usepackage{txfonts}
\usepackage{natbib}
\usepackage{epsf}
\usepackage{graphics}
\usepackage{verbatim}
\usepackage{longtable}

\usepackage{deluxetable}
\usepackage{lscape}
\usepackage{pdflscape}

\usepackage[flushleft]{threeparttable}

\def \MJ{M$_{\mathrm{Jup}}$}

\def \msol{M$\mathrm{_\odot}$}

\def \1s{$1\,\sigma$}

\def \t0{T$_0$}

\def \teff{T$_{\mathrm{eff}}$}

\makeatletter

\newcommand{\Rmnum}[1]{\expandafter\@slowromancap\romannumeral #1@}
\makeatother

\begin{document}

\title{The SOPHIE search for northern extrasolar planets\thanks{Based on observations collected with the {\it SOPHIE}  spectrograph on the 1.93-m telescope at Observatoire de Haute-Provence (CNRS), France, by the {\it SOPHIE}  Consortium.}}
            
\subtitle{IX. Populating the brown dwarf desert}

\author{
Wilson,~P.~A.\inst{1}
\and H\'ebrard,~G.\inst{1,2}
\and Santos,~N.C.\inst{3,4}
\and Sahlmann,~J.\inst{5}
\and Montagnier,~G.\inst{1,2}
\and Astudillo-Defru,~N.\inst{6,7}
\and Boisse,~I.\inst{8}
\and Bouchy,~F.\inst{8}
\and Rey,~J.\inst{9}%
\and Arnold, L. \inst{2}
\and Bonfils,~X.\inst{6,7}
\and Bourrier,~V.\inst{9}%
\and Courcol~B.\inst{8}%
\and Deleuil,~M.\inst{8}%
\and Delfosse,~X.\inst{6,7}
\and D\'iaz, R. F.\inst{9}
\and Ehrenreich,~D.\inst{9}
\and Forveille,~T.\inst{6,7}
\and Moutou,~C.\inst{8,10}
\and Pepe,~F.\inst{9}
\and Santerne, ~A.\inst{3}
\and S\'egransan,~D.\inst{9}
\and Udry,~S.\inst{9}
}
\offprints{pwilson@iap.fr}

\institute{
Institut d'Astrophysique de Paris, UMR7095 CNRS, Universit\'e Pierre \& Marie Curie, 98 bis boulevard Arago, 75014 Paris, France
\email{pwilson@iap.fr}
\and Observatoire de Haute-Provence, CNRS, Universit\'e d'Aix-Marseille, 04870, Saint-Michel-l'Observatoire, France
\and Instituto de Astrof\'isica e Ci\^{e}ncias do Espa\c co, Universidade do Porto, CAUP, Rua das Estrelas, P-4150-762 Porto, Portugal
\and Departamento de F\'isica e Astronomia, Faculdade de Ci\^encias, Universidade do Porto,Rua do Campo Alegre, 4169-007 Porto, Portugal 
\and European Space Agency, European Space Astronomy Centre, P.O. Box 78, Villanueva de la Canada, 28691 Madrid, Spain 
\and Univ. Grenoble Alpes, IPAG, F-38000 Grenoble, France 
\and CNRS, IPAG, F-38000 Grenoble, France 
\and Aix Marseille Universit\'e, CNRS, LAM (Laboratoire d'Astrophysique de Marseille) UMR 7326, 13388 Marseille, France 
\and Observatoire de Gen\`eve, Universit\'e de Gen\`eve, 51 Chemin des Maillettes, 1290 Sauverny, Switzerland 
\and Canada-France-Hawaii Telescope Corporation, 65-1238 Mamalahoa Hwy, Kamuela, HI 96743, USA 
}

\date{Received 16 October 2015; Accepted 4 February 2016}
      
\abstract{Radial velocity planet search surveys of nearby Solar-type stars have shown a strong deficit of brown dwarf companions within $\sim5$~AU. There is presently no comprehensive explanation of this lack of brown dwarf companions, therefore, increasing the sample of such objects is crucial to understand their formation and evolution. Based on precise radial velocities obtained using the SOPHIE spectrograph at Observatoire de Haute-Provence we characterise the orbital parameters of 15 companions to solar-type stars and constrain their true mass using astrometric data from the Hipparcos space mission. The nine companions not shown to be stellar in nature have minimum masses ranging from $\sim$13 to 70~\MJ, and are well distributed across the planet/brown dwarf mass regime, making them an important contribution to the known population of massive companions around solar-type stars. We characterise six companions as stellar in nature with masses ranging from a minimum mass of $76\pm4~\textrm{M}_\textrm{Jup}$ to a mass of $0.35\pm0.03~M_\sun$. The orbital parameters of two previously known sub-stellar candidates are improved.}

\authorrunning{P.~A.~Wilson et al.}
\titlerunning{Massive companions}

\keywords{planetary systems -- techniques: radial velocities -- stars: brown dwarfs -- stars: individual: HD\,6664, HD\,283668, BD+730275, HD\,39392, HD\,50761, HD\,51813, HD\,56709, HD\,77065, BD+261888, HD\,122562, HD\,132032, HD\,134113, HD\,134169, HD\,160508, BD+244697.}
\maketitle

\section{Introduction \label{sect:intro}}

The upper mass boundary separating brown dwarfs from stars is well defined at the hydrogen burning limit, whereby an object with a central temperature insufficient for sustained hydrogen fusion in its core is considered a brown dwarf \citep{chabrier97,burrows01}. Without hydrogen fusion in the core, thermal equilibrium is never reached and the brown dwarf steadily cools with time. The boundary between brown dwarfs and planets is currently unclear and is still a subject of much debate. Numerous studies have used the deuterium burning limit as a lower brown dwarf limit, whereby an object is no longer able to fuse deuterium in its core. For a solar metallicity object, the lower limit for deuterium burning is $\sim13$~\MJ~\citep{burrows01}. The deuterium burning limit however, is dependant on a number of different factors such as the helium abundance, the initial deuterium abundance as well as the metallicity of the object \citep{spiegel11}. More importantly there is no robust physical justification for this limit. It has been suggested that the overlapping brown dwarf and planet regimes should be distinguished by their formation mechanisms \citep{chabrier14}, with planets forming via core accretion in protostellar discs and brown dwarfs akin to stars which form in molecular clouds.

Differentiating between the two formation scenarios observationally remains a challenge due to the lack of clear observational diagnostics. There are however ways of differentiating between the two formation mechanisms such as mean density measurements and the composition of the object \citep{chabrier14}. A measured radius which is significantly smaller than what is expected for a near solar metallicity object can indicate an enhanced abundance of heavy elements in favour of the core accretion scenario. The opposite is not necessarily true as a larger than expected radii does not necessarily mean there is an under abundance of heavy materials \citep{leconte09}, but could instead be due to a radius inflation which is observed in many hot-Jupiter atmospheres \citep{charbonneau00,bodenheimer03,burrows07}. A larger than expected radius could also be the result of a high-metallicity, high-cloud-thickness atmosphere  \citep{burrows11}. If the object has formed through disk instabilities we expect to have a chemical composition similar to the original star forming material, whereas if it formed via core accretion it may be enhanced or depleted in some elements. The limiting factor for spectroscopically differentiating between the two scenarios are spectral resolution and signal-to-noise which are ultimately governed by the brightness of the object. The high resolution observations are further limited to young self luminous objects with wide enough separations from their host star to allow for spectroscopic observations.

The brown dwarf desert is a term used to describe the absence of brown dwarf companions on tight orbits ($\lesssim 5$~AU, $P\leq 11$~yr) around solar-mass stars compared to both planetary and stellar mass companions \citep{marcy00,marcy05,grether06}. The lack of such close-in brown dwarf companions is well established as these objects are readily detected by planet search RV surveys which are most sensitive to massive objects on close-in orbits. Within the mass domain that is dominated by planets, the occurrence rate increases with decreasing mass \citep[e.g.][]{howard10, mayor11}, whereas for low mass stellar companions, on the other side of the desert, the occurrence rate increases with increasing mass \citep{grether06}. The sparse number of companions in the brown dwarf mass regime compared to planetary mass companions, free floating brown dwarfs and the initial mass function of individual stars, indicates a separate formation mechanism between the planetary and stellar companions \citep{chabrier14}. To test this hypothesis a larger number of objects in the brown dwarf desert are needed.

In this paper we present the detection of 15 companions to solar type stars straddling the brown dwarf desert. The host star magnitudes range from 7.56 to 9.76 in V-band and they are located at a distance of 31.46 to 94.16~pc away from Earth. The observations are part of the SOPHIE search for northern extrasolar planets sub-programme 2 \citep{bouchy09} that covers a volume-limited sample of $\sim2000$ stars orbiting main sequence stars of F, G and K spectral type. The work presented in this paper is a continuation of work by \cite{diaz12} who characterised seven new objects with minimum masses between $\sim10$~\MJ\ and $\sim90$~\MJ.

In section \S~\ref{sect:RV} we discuss the observations and the data reduction. In \S~\ref{sect:analysis} we present the analysis techniques used with the results being presented in \S~\ref{sect:results}. We summarise the results in \S~\ref{sect:conclusions}.

\section{SOPHIE radial velocities \label{sect:RV}}
\subsection{Observations \label{subsec:observations}}
The observations were conducted using the 1.93~m telescope at the Haute-Provence Observatory (OHP) together with the SOPHIE cross-dispersed {\'e}chelle spectrograph. The spectrograph, which is kept in a thermally controlled chamber with the dispersive elements kept at a constant pressure in a nitrogen filled chamber, covers the 3872-6943~\AA\ wavelength range, spanning 39 orders \citep{perruchot08}. Observations were performed in a high resolution mode with $\lambda/\Delta \lambda \sim 75000$ at 550~nm. ThAr calibrations were obtained approximately every two hours during the night and were used to calculate the wavelength calibration drift. Two fibers were used during the observations, a target fiber and a sky fiber. The sky fiber observations were used to remove the sky background contribution for those situations where the observations are contaminated by moonlight being scattered in Earth's atmosphere. The sky background was removed when deemed significant by a detectable cross correlation function (CCF) peak in the sky background fiber data \citep[e.g.][]{bonomo10}. A list of the timespan and number of the observations are listed in Table\,\ref{tab:observations}.

\begin{table}
\centering
\small{
\caption{Observations.}
\label{tab:observations}
\begin{center} 
\begin{tabular}{l c c} 	
\hline\hline
Name & Timespan of observations [days] & \# of observations\\  
\hline 
HD\,6664	&	1577	&	 25 \\ 
HD\,283668	&	2751	&	 16 \\ 
BD+73~0275 	&	3282	&	 25 \\ 
HD\,39392	&	804	&	 15 \\ 
HD\,50761	&	1472	&	 16 \\ 
HD\,51813	&	2301	&	 14 \\ 
HD\,56709	&	2971	&	 25 \\ 
HD\,77065	&	10388	&	 106 \\ 
BD+26~1888 	&	1476	&	 19 \\ 
HD\,122562	&	2962	&	 29 \\ 
HD\,132032	&	735	&	 18 \\ 
HD\,134113	&	10127	&	 103 \\ 
HD\,134169	&	382	&	 12 \\ 
HD\,160508	&	1236	&	 26 \\ 
BD+24~4697 	&	497	&	 17 \\ 
\hline
\end{tabular}
\end{center}
}
\end{table}

\subsection{Data reduction \label{subsec:datareduc}}
The data were reduced using the SOPHIE pipeline \citep{bouchy09} which processes the raw images into wavelength-calibrated 2D-spectra. It includes the localisation of the 39 orders followed by an optimal order extraction, cosmic-ray rejection, wavelength correction and spectral flat field corrections. The spectra were subsequently crosscorrelated with numerical cross correlation masks which included the F0, G2 and K5 spectral types. Orders 5 to 38 were used as they produced the smallest scatter on average. The radial velocities, contrast relative to the baseline, the full width at half maximum (FWHM) and the bisector velocity span of the CCF were all calculated using a Gaussian fit to the CCF. The bisector velocity span was calculated using the method described by \cite{queloz01}. The radial velocity observations obtained after the 17$^{th}$ of June 2011 instrument upgrade \citep{perruchot11}, are known as the SOPHIE+ measurements \citep{bouchy13}. \cite{courcol15} reported on a long-term systematic instrumental drift of the order of a few m/s. This drift was not removed from the data as it has a negligible impact on the RV variations reported in this paper. The radial velocity models presented here take into account any offset between the two observational epochs (before and after the instrument upgrade) by letting the radial velocity offset between each epoch of observations vary independently. This is to account for a systematic offsets which may have been introduce during the instrument upgrade. The average signal-to-noise for all objects is 49 with observations spanning 399 to 2936 days.

\begin{table*}
\centering
\small{
\caption{Stellar properties.}
\label{tab:stellar_properties}
\begin{center}
\begin{tabular}{lccccccccc}
\hline \hline
Name   &  $\pi$  & V  & temperature  & $\log \mathrm{g}$    & $\nu$micro & $\log R^\prime_{\mathrm{HK}}$ & [Fe/H]   & $v\sin i$ &mass    \\
   & [mas] &   & [K]    & [cgs]   &  [$\mathrm{km~s}^{-1}$]    &  [dex]  & [dex] & [$\mathrm{km~s}^{-1}$]  & M$_\odot$   \\
\hline
HD\,6664  & $27.29^{\,u}$ & 7.78 & $5854 \pm 22$ & $4.53 \pm 0.04$ & $0.96 \pm 0.03$ 		& 	$-4.50\pm0.10$	&	$-0.20 \pm 0.02$   	& $2.5\pm1.0$ &  	$0.92 \pm 0.07$ \\
HD\,283668 & 23.66 & 9.44 & $4845 \pm 66$ & $4.35 \pm 0.12$ & $0.02 \pm 0.30$ 		&	$-5.01\pm0.10$	&	$-0.75 \pm 0.12$ 	& $1.3\pm1.0$ & 	$0.68 \pm 0.06$ \\
BD+73~0275  & $26.18^{\,u}$ & 8.98 & $5260 \pm 25$ & $4.43 \pm 0.04$ & -- 				& 	$-4.82\pm0.10$	&	$-0.42 \pm 0.02$   	& $2.8\pm1.0$ & 	$0.77 \pm 0.05$ \\
HD\,39392  & 10.62 & 8.38 & $5951 \pm 42$ & $4.08 \pm 0.03$ & $1.28 \pm 0.06$ 		&	$-5.11\pm0.16$	&	$-0.32 \pm 0.03$	& $1.9\pm1.0$ &	$1.08 \pm 0.08$ \\
HD\,50761  & 11.52 & 8.13 & $6450 \pm 70$ & $4.47 \pm 0.17$ & $1.93 \pm 0.12$ 		& 	$-5.06\pm0.18$	&	$-0.02 \pm 0.05$   	& $8.2\pm1.0$ &  	$1.19 \pm 0.10$ \\
HD\,51813  & $15.13^{\,u}$ & 8.67 & $6012 \pm 32$ & $4.53 \pm 0.05$ & $1.24 \pm 0.04$ 		& 	$-4.60\pm0.11$	&	$0.13 \pm 0.02$   	& $4.9\pm1.0$ &  	$1.06 \pm 0.07$ \\
HD\,56709  & 14.99 & 7.56 & $6926 \pm 73$ & $4.63 \pm 0.09$ & $1.80 \pm 0.11$ 		&	--			&	$0.03 \pm 0.04$ 	& $5.1\pm1.0$ & 	$1.33 \pm 0.09$ \\
HD\,77065  & 31.52 & 8.82 & $4990 \pm 100$ & $4.28 \pm 0.20$ & --	 			& 	$-5.01\pm0.10$	&	$-0.51 \pm 0.10$   	& $2.8\pm1.0$ &  	$0.75 \pm 0.06$ \\
BD+26~1888  & 26.30 & 9.76 & $4748 \pm 87$ & $4.3 \pm 0.2$ & $0.74 \pm 0.24$ 		& 	--			&	$0.02 \pm 0.04$   	& $2.8\pm1.0$ &  	$0.76 \pm 0.08$ \\
HD\,122562 & 18.60 & 7.69 & $4958 \pm 67$ & $3.74 \pm 0.14$ & $1.05 \pm 0.09$ 		&	$-5.27\pm0.16$	&	$0.31 \pm 0.04$ 	& $3.4\pm1.0$ & 	$1.13 \pm 0.13$ \\
HD\,132032 & 17.86 & 8.11 & $6035 \pm 33$ & $4.45 \pm 0.05$ & $1.2 \pm 0.04$ 		&	$-5.13\pm0.22$	&	$0.22 \pm 0.03$ 	& $3.2\pm1.0$ &  	$1.12 \pm 0.08$ \\
HD\,134113$^\dagger$  & 13.91 & 8.26 & $5782 \pm 22$ & $4.25 \pm 0.03$ & $1.27 \pm 0.04$& 	$-5.09\pm0.23$	 & $-0.74 \pm 0.02$  & $1.5\pm1.0$ &	$0.81 \pm 0.01$ \\
HD\,134169  & 17.27 & 7.70 & $5940 \pm 40$ & $4.33 \pm 0.03$ & -- 				& 	$-5.09\pm0.14$	&	$-0.51 \pm 0.10$   	& $1.0\pm1.0$ &  	$0.91 \pm 0.05$ \\
HD\,160508 & 10.83 & 8.11 & $6212 \pm 30$ & $4.16 \pm 0.03$ & $1.5 \pm 0.04$ 		&	$-5.25\pm0.13$	&	$0.01 \pm 0.02$ 	& $3.7\pm1.0$ & 	$1.25 \pm 0.08$ \\
BD+24~4697 & 20.51 & 9.74 & $5077 \pm 32$ & $4.67 \pm 0.07$ & $1.04 \pm 0.05$ 	&	$-4.69\pm0.10$	&	$-0.11 \pm 0.02$ 	& $1.5\pm1.0$ & 	$0.75 \pm 0.05$ \\
\hline
\end{tabular}
    \begin{tablenotes}
      \small
      \item $^{\,u}$ Parallaxes have been updated following the astrometric analysis presented in this paper.
      \item $^\dagger$ Values are adopted from \citep{sausa11}. Alternate values computed in this study: \teff$ = 5940 \pm 40$~K, $\log g = 4.32 \pm 0.05$, $\mathrm{[Fe/H]}=-0.70 \pm 0.03$ and $m = \ 0.89\pm0.05$~M$_\odot$.
    \end{tablenotes}
\end{center}
}
\end{table*}

\section{Analysis \label{sect:analysis}}
\subsection{Stellar parameters \label{subsec:stellar}}

The stellar parameters, temperature, gravity, microturbulence and metallicity where all obtained using the spectroscopic analysis methods described in \citet{santos13}, and references therein. The method makes use of the equivalent widths of a list of $\mathrm{Fe~ \Rmnum{1}}$ and $\mathrm{Fe~ \Rmnum{2}}$ lines, as measured in the SOPHIE spectra. All the analysis is done in LTE using the 2010 version of the MOOG software \citep{sneden73} and 1D \citet{kurucz93} model atmospheres. The stellar parameters where determined by using excitation and ionisation equilibrium principles, and minimising the correlation between $\log \epsilon(\mathrm{Fe~\Rmnum{1}})$ and $\chi_l$ (excitation potential) as well as  $\log \epsilon(\mathrm{Fe~ \Rmnum{1}})$ and $\log(W_\lambda /  \lambda)$ making sure the mean abundance between the Fe I and Fe II lines were equivalent. 

The aforementioned stellar parameters were subsequently compared to interpolated isochrones by \cite{girardi00} using the method described in \cite{dasilva06}. The metallicities were computed using two methods, through the analysis of the spectra and from the CCF using the \cite{boisse10} calibration. The metallicities from the spectral analysis were obtained from a detailed analysis of carefully selected lines, which resulted in an typical accuracy of $\sim0.04$\,dex. The metallicities from the CCF were obtained from a global, empirical calibration and provide a typical accuracy of $\sim0.1$\,dex. We found that the two methods give consistent results but use the spectral analyses technique for all the targets as the obtained values are more accurate. The largest metallicity difference derived from the two methods was for HD\,77065 where there was a $2.6\,\sigma$ discrepancy. For this object we used the spectral analyses method for consitency with the other results. The stellar parameters, including metallicities, are presented in Table~\ref{tab:stellar_properties}.

Intrinsic phenomena caused by photospheric features associated with chromospheric activity such as starspots and convective inhomogeneities are known to influence RV measurements up to a few tens of m s$^{-1}$ \citep[e.g.][]{campbell91,saar97,saar98,santos00,boisse11}. The impact of stellar activity can affect the derived orbital parameters if the RV semi amplitudes are similar in size to that introduced by stellar noise or when the orbital period is comparable to the rotational period of the host star. Objects with periods longer than $\sim2$ years could also be affected by activity cycles or long term variation of the activity level of the star. The objects presented here all have RV amplitudes outside the domain where they might be mistaken for planets when they instead are caused by intrinsic phenomena (e.g. \citealt{saar97}). The individual stellar activity values ($\log R^\prime_{\mathrm{HK}}$), reported in Table~\ref{tab:stellar_properties}, were obtained following the method of \cite{boisse10}. The majority of stars in the sample show low-activity with only three stars showing an $\log R^\prime_{\mathrm{HK}}$ value larger than -4.7, namely HD\,6664, HD\,51813, and BD+24~4697.

\begin{figure*} 
\includegraphics[width=\hsize]{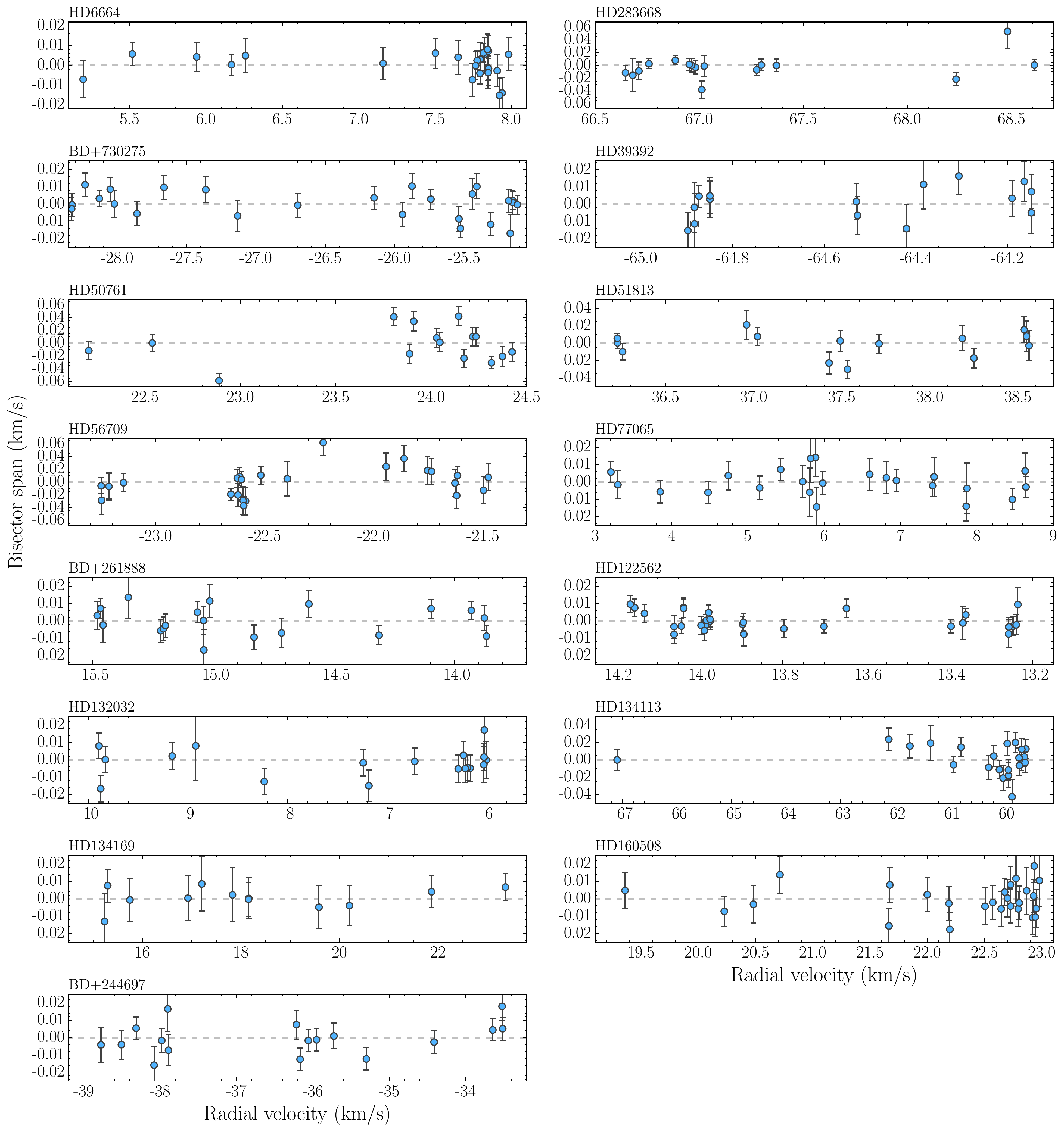}
\caption{Bisector span as a function of radial velocity for all the targets in the sample. The dashed line represents a zero bisector span and is shown as a reference.}
\label{fig:bisectors}
\end{figure*}

\subsection{Companion parameters \label{subsec:companion_params}}
The parameters of the companions were derived by fitting a single Keplerian orbit model to the radial velocity data. The free parameters were the orbital period P, the eccentricity $e$, the argument of periapsis $\omega$, the semi amplitude $K$, the periastron time T$_0$ and the RV shift between the pre ($\gamma_\textrm{SOPHIE}$) and post ($\gamma_\textrm{SOPHIE}^+$) instrument upgrade. As the radial velocity variations are large compared to the uncertainties on individual radial velocity points, no periodogram analysis was done. The best fit Keplerian orbit model was obtained by using a genetic algorithm to find the starting point for a Levenberg-Marquardt fit. The uncertainties of the individual parameters were calculated using a one million step Markov chain Monte Carlo (MCMC). The MCMC method obtains samples from the posterior distribution of the model parameters.

The 17$^{th}$ of June 2011 instrument upgrade introduced a systematic shift of $\Delta \mathrm{RV}=0.03\pm0.02$~km/s. This value was calculated by taking the weighted mean of the individual RV shifts ($\gamma$-values) available for six of the objects with observations done before and after the instrument upgrade (see $\gamma$ values in Table~\ref{tab:companion_properties}). The observed shifts measured are likely influenced by the airmass of the observations, the spectral type of the star, choice of cross correlation mask and the spectral orders used. The $\Delta \mathrm{RV}$ value has a negligible impact on the fits, as it is only marginally different from 0. HD\,51813 which also has observations before and after the upgrade was not included in the weighted mean as this object exhibited a shift of $\Delta \mathrm{RV}=-1.5\pm0.35$~km/s which was inconsistent compared to the weighted mean of the other six objects. This shift is thought to be due to the lack of pre-upgrade measurements which all occurred within 4 days, a short amount of time when compared to a much longer orbital period of $1437^{+11}_{-15}$~days. To account for such a large shift, we fix the pre-upgrade $\gamma$ value for HD\,51813 to a new value consistent with the six targets and recalculate the Keplerian parameters. To avoid underestimating the uncertainties for HD\,51813 by fixing the RV shift $\gamma$, we adopt the global $\gamma$ uncertainty of 0.02~km/s, calculated as the standard deviation of individual RV shifts and use this to inflate the uncertainties on the Keplerian parameters for this object. We find that despite having to adjust the pre-upgrade RV shift, a negligible shift in the derived Keplerian parameters is observed (consistent to within $1\,\sigma$), even for this object which has the largest shift.

The results of the Keplerian fits of the SOPHIE radial velocities are reported in Table~\ref{tab:companion_properties}. The data and their Keplerian fits are plotted in Fig.~\ref{fig:RV}.

\begin{landscape}
\begin{deluxetable}{lccccccccccc}
\tabletypesize{\tiny}
\tablecolumns{12} 
\tablewidth{0pt}
 \tablecaption{Low-mass binary stars and brown dwarf candidates analysed in this study.
 \label{tab:companion_properties}}
\centering
 \tablehead{
  \vspace{-0.3cm}	& \colhead{Period} 		&	&  \colhead{$\omega$} &   \colhead{$K$} & \colhead{T$_0$ } & \colhead{$\gamma_\textrm{SOPHIE}$} &  \colhead{$\gamma_\textrm{SOPHIE}^+$}& \colhead{$\textrm{O}-\textrm{C}$} & \colhead{$\textrm{O}-\textrm{C}^+$} & \colhead{$a$} &\colhead{$\textrm{m}_2 \sin i^\ddagger$}\\
Name	& 		&  \colhead{Eccentricity}	&   &    &  &  &  &  & & &\\
 & \colhead{[day]} &       & \colhead{[$^{\circ}$]}  & \colhead{[m/s]}    & \colhead{[$\textrm{BJD}-2400000 $]} & \colhead{[km/s]}     & \colhead{[km/s]}      & \colhead{[m/s]} & \colhead{[m/s]}    & \colhead{[AU]}   & \colhead{[$\textrm{M}_\textrm{Jup}$]}
 }

\startdata
  \\  
  \multicolumn{12}{c}{Low-mass binary stars}\\ 
  \hline
\multicolumn{12}{c}{\vspace{-0.2cm}}\\ 
HD\,6664B   & $1718\pm11$ & $0.289\pm0.002$ & $138.9 \pm 0.5 $ & $1417\pm4$ & $56227 \pm 2$  & $6.898 \pm 0.009$  & $6.932  \pm 0.007$  & $14.6$  & $7.0$   & $2.80\pm0.09$   & $80\pm5$\\  
BD+73~0275B & $1423.20\pm0.14 $ & $0.81386\pm0.00038 $ & $120.62\pm0.09 $ & $1624.9\pm1.3 $ & $57326.15\pm0.04 $  & $-26.0352\pm0.0035 $ & $-26.0297\pm 0.0014$  & $11.3$  & $4.3$   & $2.31\pm0.06$   & $44\pm3$  \\
HD\,50761B  & $2092\pm11$ & $0.377 \pm 0.007 $ & $-127.88^{+0.48}_{-0.49}$ & $2022^{+23}_{-21}$  & $57692^{+7}_{-7}$  &  & $23.89\pm0.01$  &  & $7.4$    & $3.5\pm0.2$   & $145^{+36}_{-29}$  \\ 
HD\,51813B & $1437^{+11}_{-15}$ & $0.735\pm0.003 $ & $-68.8^{+0.4}_{-0.7}$ & $1150^{+14}_{-8} $ & $55932.3^{+0.3}_{-0.6}$  & $37.07\pm0.02$ & $37.104^{+0.010}_{-0.009}$  & $11.2$ & $13.7$    & $2.6\pm0.1$   & $46\pm4$  \\
  
HD\,56709B  & $2499.0 \pm 5.6 $ & $0.106 \pm 0.006 $ & $153\pm4$ & $896 \pm 4 $ & $56958^{+30}_{-29}$  & $-22.296\pm 0.007$  & $-22.278  \pm 0.006$  & $11.4$ & $12.4$    & $4.03\pm0.11$   & $72\pm4$\\ 
HD\,134169B & $67.8578 \pm 0.0038 $ & $0.10300 \pm 0.00053 $ & $-5.02 \pm 0.48 $ & $4184.9 \pm 2.4 $ & $56243.940  \pm 0.090$  &  & $18.9395 \pm 0.0017$  &  & $4.0$   & $0.324\pm0.008$   & $83\pm4$  \\

  \\
  \multicolumn{12}{c}{Brown dwarf candidates to solar-type stars with projected masses between 10 and 80~$\textrm{M}_\textrm{Jup}$}\\ 
  \hline 

\multicolumn{12}{c}{\vspace{-0.2cm}}\\ 
HD\,283668b & $2558\pm8$ & $0.577\pm0.011$ & $-94.6\pm1.2$ & $1243^{+25}_{-24}$ & $57401^{+9}_{-8} $  & $67.80\pm0.011$  & $67.79\pm0.02$  & $10.8$ & $4.7$  & $3.30\pm0.11$   & $53\pm4$\\
HD\,39392b  & $394.3^{+1.4}_{-1.2}$ & $0.394 \pm 0.008 $ & $-31.9 \pm 1.3 $ & $374.2^{+2.4}_{-2.3}$ & $56322 \pm 1 $  &  &$-64.632\pm0.002$  & & $10.0$     & $1.08\pm0.03$   & $13.2\pm0.8$\\
HD\,77065b$^\dagger$  & $119.1135^{+0.0026}_{-0.0027}$ & $0.69397 \pm 0.00036 $ & $105.972\pm0.065$ & $2817.1 \pm 1.6 $ & $56881.239\pm0.006$  &  & $6.3822 \pm 0.0009$  & & $5.1$    & $0.438\pm0.013$   & $41\pm2$  \\
BD+26~1888b & $536.78\pm0.25$ & $0.2675\pm0.0016$ & $1.3\pm 0.7 $ & $805.1 \pm 1.3$ & $56548.3 \pm 0.9$  &  & $-14.8843 \pm 0.0012$  &  & $4.7$   & $1.19\pm0.05$   & $26\pm2$  \\
HD\,122562b & $2777^{+104}_{-81}$ & $0.71\pm0.01$ & $-53.9^{+1.2}_{-1.3}$ & $445^{+12}_{-10} $ & $55935\pm2$  & $-13.907^{+0.007}_{-0.008} $  & $-13.868^{+0.013}_{-0.011}$ & $8.0$ & $3.5$     & $4.1\pm0.2$   & $24\pm2$\\
HD\,132032b & $274.33 \pm 0.24 $ & $0.0844 \pm 0.0024 $ & $-0.9\pm0.9$ & $1950.7 \pm 2.2 $ & $56238.0\pm 0.7$  &  & $-8.118\pm0.005$  &  & $3.2$    & $0.87\pm0.02$   & $70\pm4$\\
HD\,134113b$^{\dagger}$ & $201.680 \pm 0.004$ & $0.891\pm0.002$ & $163.0 \pm 0.2 $ & $3965^{+55}_{-46}$ & $56808.54  \pm 0.05 $  &  & $-60.2027 \pm 0.0015$  &  & $9.7$   & $0.638\pm0.010$   & $47^{+2}_{-3}$  \\
HD\,160508b & $178.9049 \pm 0.0074$ & $0.5967 \pm 0.0009 $ & $-168.83 \pm 0.13 $ & $1825.3 \pm 2.7 $ & $55998.70 \pm 0.03$  & $22.2054 \pm 0.0035$ & $22.2440 \pm 0.0020$  & $11.0$ & $6.9$   & $0.68\pm0.02$   & $48\pm3$\\
BD+24~4697b& $ 145.081 \pm 0.016$ & $0.50048 \pm 0.00043 $ & $66.35 \pm 0.09 $ & $2728.4 \pm 1.6 $ & $55986.042 \pm 0.037$  &  & $-36.5913 \pm 0.0016$  &  & $3.5$    & $0.50\pm0.08$   & $53\pm3$\\

\enddata
\tablenotetext{\dagger}{These objects contain data from \cite{latham02} where HD\,77065 is published as G9-42 and HD\,134133 as G66-65.}
\tablenotetext{\ddagger}{The uncertainties in stellar mass were included in the uncertainty calculation of $\textrm{m}_2 \sin i$.}
\end{deluxetable}

\end{landscape}

\setlength{\tabcolsep}{4pt}
\begin{table}\caption{Parameters of the Hipparcos astrometric observations.}
\label{tab:HIPparams} 
\centering  
\begin{tabular}{r r r r r r r} 	
\hline\hline %
Name & HIP & $S_\mathrm{n}$& $N_\mathrm{orb}$ & $\sigma_{\Lambda}$  & $N_\mathrm{Hip}$ & $m_{2,\mathrm{max}}$\\  
       &     &  &           & (mas)                 &           &   ($M_\sun$)\\  
\hline 
 HD\,6664 & 005313 & 1 & 0.7 & 5.2 & 115 & $0.46$ \\ 
 HD\,283668 & 020834 & 5 & 0.4 & 4.5 & 44 & $\cdots$ \\ 
 BD+73~0275 & 024329 & 5 & 0.8 & 7.5 & 184 &$0.29$ \\ 
HD\,39392 & 027828 & 5 & 2.7 & 3.7 & 114 & 0.37 \\ 
 HD\,50761 & 033542 & 7 & 0.7 & 2.8 & 91 & $0.80$ \\ 
 HD\,51813 & 033608 & 5 & 0.8 & 3.4 & 125 & $0.55$ \\ 
 HD\,56709 & 035697 & 7 & 0.4 & 3.4 & 106 & $1.08$ \\ 
 HD\,77065 & 044259 & 5 & 8.1 & 3.6 & 49 & 0.68 \\ 
 BD+26~1888 & 044387 & 5 & 1.4 & 6.9 & 49 & 0.25 \\ 
 HD\,122562 & 068578 & 5 & 0.4 & 4.0 & 70 & $\cdots$ \\ 
 HD\,132032 & 073128 & 5 & 3.2 & 4.4 & 84 & 0.34 \\ 
 HD\,134113 & 074033 & 5 & 5.4 & 3.7 & 39 & 16.13 \\ 
 HD\,134169 & 074079 & 5 & 13.4 & 2.9 & 41 & 1.03 \\ 
 HD\,160508 & 086394 & 5 & 6.3 & 5.5 & 163 & 1.58 \\ 
 BD+24~4697 & 113698 & 5 & 7.7 & 6.3 & 196 & 0.51 \\ 
\hline
\end{tabular}
\tablefoot{The columns of this table are defined in \S~\ref{sect:hipp_analysis}.}
\end{table}

\subsection{Bisector analysis}
\label{sec:bisector_analysis}
Variations in the bisector velocity span can be caused by two or more stars blended together or stellar activity. Stellar activity, as discussed in \S\,~\ref{subsec:stellar}, is unlikely to be responsible for any large radial velocity variations since the semi-amplitudes measured are much larger than what variation caused by stellar activity. To look for possible blends we looked for trends in bisector velocity span as a function of radial velocity to see if any of the objects could be accompanied by an additional star blended in the CCF (for more details on this technique see \citealt{santerne15}). We calculated the Kendall rank correlation coefficient for all the objects and found no significant correlations with all $p$-values being greater than 0.10. However, the object which shows the largest amount of scatter, HD\,50761 (see Fig.~\ref{fig:bisectors}), also shows a tentative detection of being a low-mass binary. We looked for secondary peaks in the CCF using a M4 cross correlation mask \citep{bouchy15b} but found no clear detections. The lack of clear detections could in part be due to the low S/N ($\sim50$) of the individual spectra. Furthermore, HD\,56709 and HD\,50761 are the fastest rotating stars in the sample with $v\sin i > 5$\,km/s which may have broadened the CCF's to an extent where a clear detection is not possible.

\subsection{Analysis of the Hipparcos astrometry}
\label{sect:hipp_analysis}
All stars listed in Table \ref{tab:HIPparams} were observed by the Hipparcos satellite \citep{Perryman:1997kx}. We used the new Hipparcos reduction \citep{:2007kx} to search for signatures of orbital motion in the Intermediate Astrometric Data (IAD). The analysis was performed following \cite{sahlmann11}, where a detailed description of the method can be found. Using the parameters of the radial velocity orbit of a given star (Table \ref{tab:companion_properties}), the IAD was fitted with a seven-parameter model, where the free parameters are the inclination $i$, the longitude of the ascending node $\Omega$, the parallax $\varpi$, and offsets to the coordinates ($\Delta \alpha^{\star}$, $\Delta \delta$) and proper motions ($\Delta \mu_{\alpha^\star}$, $\Delta \mu_{\delta}$). A two-dimensional grid in $i$ and $\Omega$ was searched for its global $\chi^2$-minimum. The statistical significance of the derived astrometric orbit was determined with a permutation test employing 1000 pseudo orbits. Uncertainties in the solution parameters were derived by Monte Carlo simulations, which also propagate the uncertainties in the RV parameters. This method has proven to be reliable in detecting orbital signatures in the Hipparcos IAD \citep{sahlmann11, sahlmann11b, diaz12, sahlmann13}.

\begin{table*}\caption{Astrometric parameters determined for systems with a significant or tentative astrometric signal in the Hipparcos data.}
\label{tab:HIPparams2} 
\centering  
\begin{tabular}{r r r r r r rrrrr} 	
\hline\hline %
Source  & $\Delta \alpha^{\star}$ & $\Delta \delta$ & $\varpi$ &$\Delta \varpi$  & $\Delta \mu_{\alpha^\star}$ & $\Delta \mu_{\delta}$ & $i$ & $\Omega$  \\  
        &  (mas)                     & (mas)             &  (mas)    &  (mas)            &  (mas $\mathrm{yr}^{-1}$)       & (mas $\mathrm{yr}^{-1}$) & (deg)	      & (deg)	    \\  
\hline 
HD\,6664 & $-4.3^{+ 1.8}_{-1.8}$ & $-16.9^{+ 1.5}_{-1.5}$  & $27.29^{+ 0.90}_{-0.91}$ & $-4.50$ & $22.6^{+ 1.8}_{-1.8}$ & $-2.9^{+ 1.5}_{-1.5}$ & $ 165.1^{+ 1.0}_{-1.2}$ & $ 319.0^{+ 8.1}_{-8.1}$  \vspace{1mm} \\ 
HD\,56709$^{T}$ & $1.0^{+ 3.6}_{-3.6}$ & $15.6^{+ 2.6}_{-2.6}$  & $15.32^{+ 1.02}_{-1.02}$ & $0.34$ & $-13.7^{+ 2.0}_{-2.0}$ & $0.6^{+ 3.2}_{-3.2}$ & $ 170.6^{+ 1.5}_{-2.1}$ & $ 177.4^{+ 17.7}_{-11.0}$  \vspace{1mm} \\ 
BD+730275 & $7.3^{+ 5.7}_{-6.2}$ & $14.0^{+ 3.3}_{-3.1}$  & $26.18^{+ 1.06}_{-1.06}$ & $2.11$ & $0.4^{+ 1.4}_{-1.4}$ & $-7.8^{+ 1.8}_{-1.8}$ & $ 14.1^{+ 2.3}_{-1.7}$ & $ 259.3^{+ 20.9}_{-23.3}$  \vspace{1mm} \\ 
HD\,51813 & $8.9^{+ 2.1}_{-2.1}$ & $7.1^{+ 4.2}_{-4.0}$  & $15.13^{+ 1.03}_{-1.03}$ & $-1.00$ & $-4.8^{+ 1.3}_{-1.3}$ & $-1.9^{+ 1.0}_{-1.0}$ & $ 169.2^{+ 2.0}_{-3.2}$ & $ 5.3^{+ 22.6}_{-23.9}$  \vspace{1mm} \\  
HD\,50761$^{T}$ & $-5.5^{+ 3.3}_{-3.1}$ & $0.3^{+ 2.2}_{-2.0}$  & $12.06^{+ 0.77}_{-0.77}$ & $0.53$ & $2.3^{+ 1.8}_{-1.8}$ & $7.5^{+ 1.5}_{-1.5}$ & $ 13.2^{+ 4.8}_{-2.8}$ & $ 13.9^{+ 43.6}_{-43.3}$  \vspace{1mm} \\ 
 
\hline
\end{tabular}

{\footnotesize $^{T}$ These objects are regarded as systems with a tentative astrometric signal in the Hipparcos data.}\\
\end{table*}

\begin{table*}\caption{Solution parameters determined for systems with a significant or tentative astrometric signal in the Hipparcos data.}
\label{tab:HIPparams4} 
\centering  
\begin{tabular}{r r r r r c rccccrc} 	
\hline\hline %
Source &   $a \sin i$ & $a$ & $M_2$  & $M_2$ (3-$\sigma$)& $a_{\mathrm{rel}}$ &$O-C_5$ &$O-C_7$ & $\chi^2_{7,red}$ & p-value & Significance \\  
                              &(mas)     & (mas) & ($M_\sun$) & ($M_\sun$) &(mas)/(AU) & (mas) & (mas) & & (\%)  & (\%) \\  
\hline
HD\,6664 &   6.81 & $ 22.8^{+ 1.4}_{-1.4}$ & $ 0.35^{+ 0.03}_{-0.03}$ & $(0.26,0.46)$ & 83.0/3.0 & 7.00 & 4.03& 0.53& $1.7\times10^{-30}$ &$>$99.9 \vspace{1mm} \\ 
HD\,56709\tablefootmark{T}\tablefootmark{a}&   3.13 & $ 19.4^{+ 2.8}_{-2.8}$ & $ 0.54^{+ 0.12}_{-0.12}$ & $(0.25,1.08)$ & 68.4/4.5 & 4.97 & 4.27& 1.39& $1.9\times10^{-11}$ &86.8 \vspace{1mm} \\ 
BD+730275 &   2.90 & $ 12.9^{+ 1.7}_{-1.8}$ & $ 0.20^{+ 0.03}_{-0.03}$ & $(0.11,0.29)$ & 63.7/2.4 & 8.04 & 6.81& 0.75& $2.2\times10^{-11}$ &$>$99.9 \vspace{1mm} \\ 
HD\,51813 &   1.66 & $ 8.3^{+ 1.7}_{-1.7}$ & $ 0.27^{+ 0.07}_{-0.07}$ & $(0.10,0.55)$ & 41.5/2.7 & 4.28 & 3.67& 0.89& $1.1\times10^{-5}$ &$>$99.9 \vspace{1mm} \\ 
HD\,50761\tablefootmark{T} &   1.99 & $ 9.1^{+ 2.0}_{-2.0}$ & $ 0.38^{+ 0.11}_{-0.11}$ & $(0.16,0.80)$ & 38.5/3.2 & 3.74 & 3.00& 1.15& $2.7\times10^{-7}$ &99.4 \vspace{1mm} \\ 
\hline
\end{tabular}

{\footnotesize $^{T}$ These objects are regarded as systems with a tentative astrometric signal in the Hipparcos data.}\\
{\footnotesize $^{a}$ These parameters correspond to the tentative orbit detection discussed in the text.}\\
\end{table*}

Table~\ref{tab:HIPparams} lists the target names and the basic parameters of the Hipparcos observations relevant for the astrometric analysis. The solution type $S_\mathrm{n}$  indicates the astrometric model adopted by the new reduction. For the standard five-parameter solution it is '5', whereas it is '7' when the model included proper-motion derivatives of first order to obtain a reasonable fit, which is the case for HD\,56709 and HD\,50761. The parameter $N_\mathrm{orb}$ represents the number of orbital periods covered by the Hipparcos observation timespan and $N_\mathrm{Hip}$ is the number of astrometric measurements with a median precision of $\sigma_{\Lambda}$. The last column in Table~\ref{tab:HIPparams} shows the maximum companion mass ($m_{2,\mathrm{max}}$) that would be compatible with a non-detection in the Hipparcos astrometry. Outliers in the IAD have to be removed because they can substantially alter the outcome of the astrometric analysis. We eliminated one outlier for both HD\,160508 and HD\,77065.

Even if the astrometric signal is not detected in the astrometric data, i.e. the derived significance is low, we can use the Hipparcos observations to set an upper  limit to the companion mass by determining the minimum detectable astrometric signal $a_\mathrm{min}$ of the individual target. When the data cover at least one complete orbit, \cite{sahlmann11, sahlmann11b} have shown that an astrometric signal-to-noise of $\mathrm{S/N} \gtrsim 6-7$ is required to obtain a detection at the $3\,\sigma$ level, where $\mathrm{S/N}=a\, \sqrt{N_\mathrm{Hip}} / \sigma_{\Lambda}$ and $a$ is the semi-major axis of the detected orbit. Using a conservative S/N-limit of 8, we derive the upper companion mass limit $m_{2,\mathrm{max}}$ as the companion mass which introduces the astrometric signal 
\begin{equation}\label{ }
a_\mathrm{min} = 8 \frac{ \sigma_{\Lambda} } { \sqrt{N_\mathrm{Hip}}} \left( 1-e^2 \right),
\end{equation}
where the factor $1-e^2$ accounts for the most unfavourable case of $i=90\degr$ and $\Omega=90\degr$ in which the astrometric signal is given by the semi-minor axis of the orbit. The last column in Table~\ref{tab:HIPparams} lists this upper companion mass limit.

\section{Results}
\label{sect:results}

We characterise the orbital parameters of 15 companions of solar-type stars, 13 of which are newly discovered companions. Their minimum masses $\textrm{m}_2 \sin i$ range from 13.2~\MJ\ (HD\,39392b) to 145~\MJ\ (HD\,50761B). We further improve the parameters of HD\,134113 and HD\,77065 by combining our new observations with previous observations by \cite{latham02}. For three stars, we found significant ($>3\sigma$) orbit signatures in the Hipparcos IAD. Tentative detections ($1.5 \lesssim \sigma \leq 3$) were seen for an additional two objects. The largest residual dispersion is found for the most active stars in the sample such as HD\,6664 and HD\,51813 with the less active stars such as HD\,122562, HD\,132032 and HD\,134169 showing the smallest dispersion.

We divide the objects in this study into low-mass binary stars, low-mass binary stars candidates and brown dwarf candidates and present information concerning each object in the subsections below. The numerical results from astrometry are presented in Tables \ref{tab:HIPparams2}, \ref{tab:HIPparams4} with the Keplerian fit parameters are presented in Table~\ref{tab:companion_properties}. Objects we regard as likely to be stellar in nature are designated with the stellar name followed by the suffix 'B'. Objects with minimum masses below the hydrogen burning limit are designated as brown dwarf candidates with the suffix 'b' added to the name of the star. As the parameters improve through continued observations, especially for astrometric observations, these suffixes are likely to change for at least some of the companions which later might turn out to be low-mass stars as their inclination is determined.

\subsection{Low-mass binary stars}
The mass limit which divides brown dwarfs and stars is $\sim0.075$\,M$_{\sun}$ \citep{burrows97,chabrier00} (depending on metallicity). We detect the astrometric orbital solution for three low-mass stellar companions with a statistical significance greater than $3\sigma$ orbiting HD\,6664, BD+730275 and HD\,51813. Two additional tentative astrometric detections are made for HD\,56709 ($\simeq1.5\sigma$) and HD\,50761 ($2.7\sigma$). The astrometric analysis yields masses which range from 0.20 to 0.54\,M$_{\sun}$ (see Table~\ref{tab:HIPparams4}). Each companion is described in more detail in the subsections below, which includes the tentative detections. We categorise HD\,50761, HD\,56709 and HD\,134169 as low-mass binaries. In the case of HD\,50761 it is because we derive a minimum mass of $\textrm{m}_2 \sin i = 145\pm4$\,\MJ. For HD\,56709 and HD\,134169 it is because their large minimum masses of $\textrm{m}_2 \sin i = 72\pm4$\,\MJ\ and $\textrm{m}_2 \sin i = 83\pm4$\,\MJ\ respectively, make them much more likely to be low-mass stars rather than brown dwarfs.

\subsubsection{HD\,6664}
HD\,6664 is a G1V star with a mass of $0.92\pm0.07$\,M$_{\sun}$. RV measurements show the presence of a companion with an orbital period of $1718\pm11$ days and with a minimum mass of $80\pm5$\,\MJ, which by itself would indicate a companion which is stellar in nature. According to \cite{riddle15} HD\,6664 is a triple system. The astrometric analysis was done on the inner pair. The outer companion at $>$$26\arcsec$ has no influence on the detectability of the inner orbit with Hipparcos astrometry and is outside the SOPHIE aperture. We detected the astrometric signal in the Hipparcos data with high significance $>3.3\,\sigma$ (>99.9 \%) and measured a semimajor axis of $a=22.8 \pm 1.4$~mas. HD\,6664 is the only source in this study with solution type of `1', which already indicated non-standard astrometric motion. The system's effective inclination is small (15\,\degr\ out of the sky plane) with an orbital inclination of $i = 165.1^{+1.0}_{-1.2}$\,\degr\ and we measure the longitude of the ascending node of $\Omega=319.0\pm8.1$~$\degr$. Based on the orbit's inclination, we determine a companion mass of $M_2 = 0.35 \pm 0.03$ M$_{\sun}$, i.e.\ HD\,6664\,B is indeed a low-mass star.

\begin{figure}[ht]\begin{center} 
\includegraphics[width= 0.8\linewidth, trim= 0 0cm 0 0cm,clip=true]{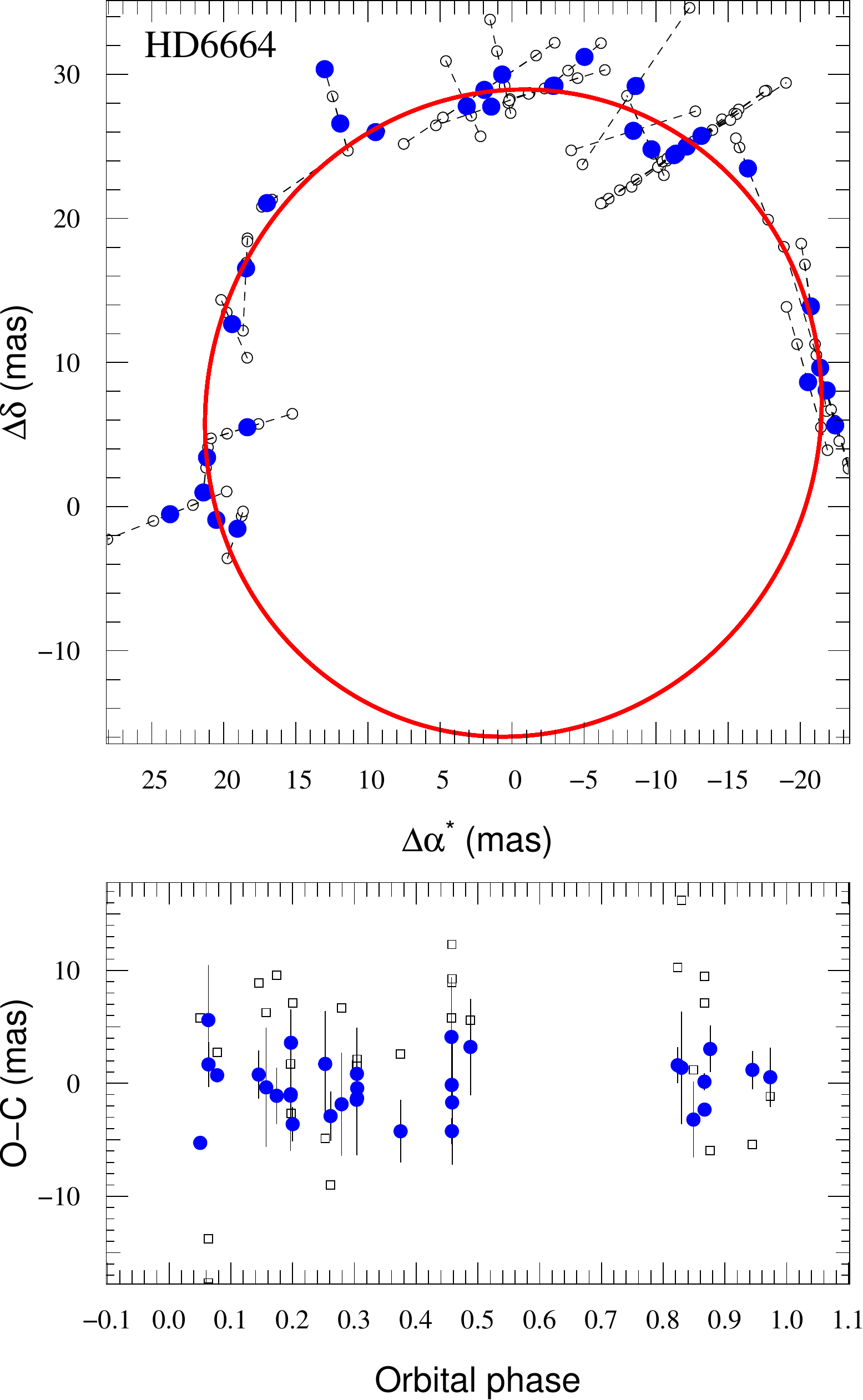}
 \caption{\emph{Top:} Astrometric orbit of HD\,6664 projected on the sky. North is up and east is left. The solid red line shows the orbital solution, which is orientated clockwise, and open circles mark the individual Hipparcos measurements. The dashed lines represents the scan angle which connects these one dimensional position measurements. The blue filled circles are the mean of the Hipparcos measurements obtained during one orbit. \emph{Bottom:} O--C residuals for the normal points of the orbital solution (filled blue circles) and of the five-parameter model without companion (open squares).} 
\label{fig:orbit}
 \end{center} \end{figure}

There is a significant offset in the updated proper motion in RA of $\Delta \mu_{\alpha^\star} = 22.6\pm 1.8$ mas $\mathrm{yr}^{-1}$. More importantly, the new parallax $\varpi = 27.29^{+0.90}_{-0.91}$ mas is significantly smaller (by $-4.5$ mas) than the catalogue value of $\varpi_\mathrm{HIP2} = 31.8\pm1.7$ mas \citep{:2007kx}, and the revised distance to HD\,6664 is $36.6\pm1.2$ pc. The sky-projected orbit of HD\,6664 and the Hipparcos measurements are shown in Fig.~\ref{fig:orbit}. 
 
\begin{figure}[ht]\begin{center} 
\includegraphics[width= 0.8\linewidth, trim= 0 0cm 0 0cm,clip=true]{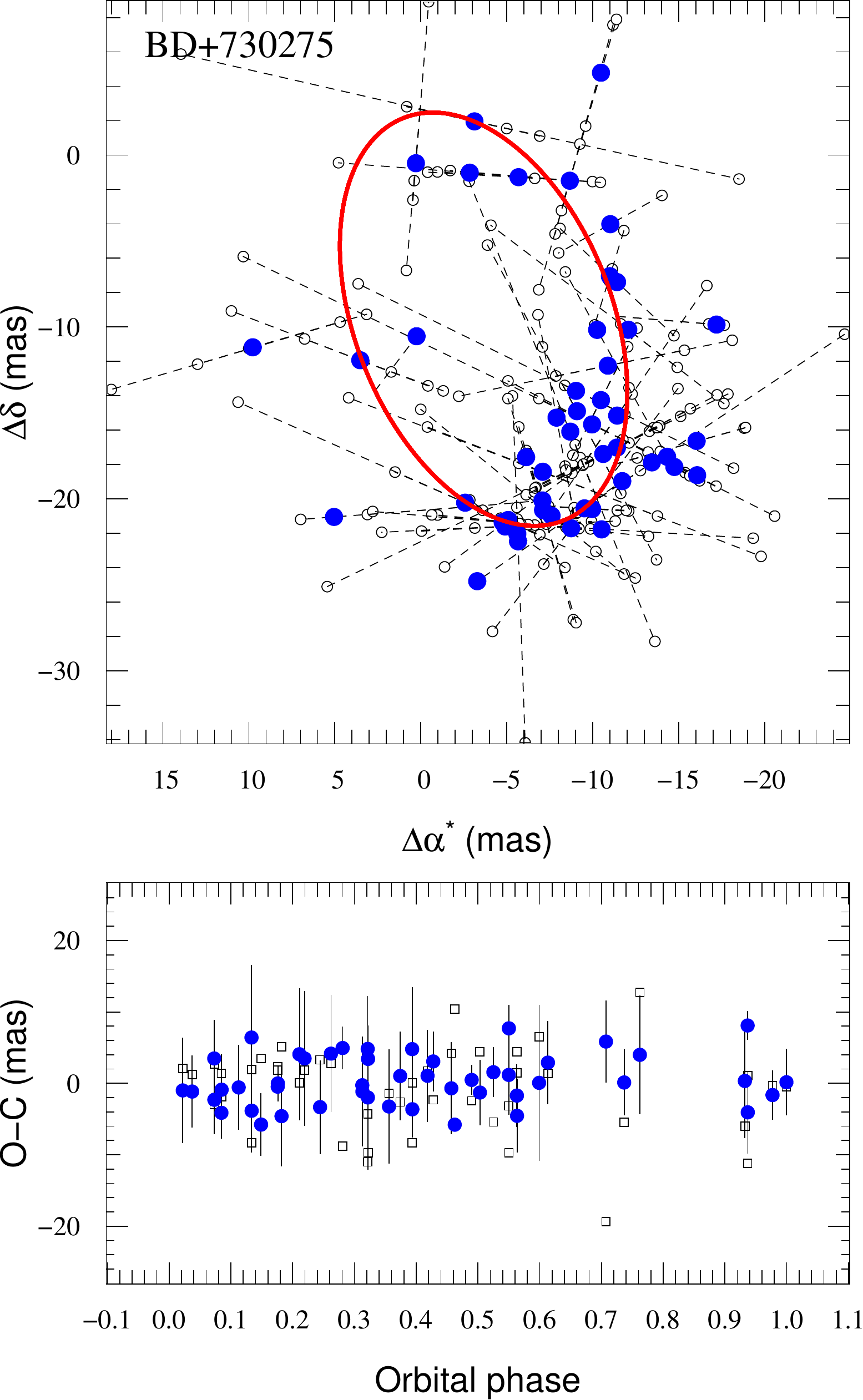}
 \caption{Astrometric orbit of BD$+$730275 projected on the sky, cf. Fig. \ref{fig:orbit}, corresponding to one set of RV parameters. The star's motion is counter-clockwise.} 
\label{fig:orbit_BD73}
 \end{center} \end{figure}

\subsubsection{BD$+$73 0275}
BD$+$73 0275 is a G5 star with a mass of $0.77\pm0.05$\,M$_{\sun}$. RV measurements show the presence of a $\textrm{m}_2 \sin i = 44\pm3~\textrm{M}_\textrm{Jup}$ companion on a $1423.20\pm0.14$~day orbit. We detected the Hipparcos astrometric orbit of BD$+$730275 with a significance $>3.3\,\sigma$ (>99.9 \%).  The orbit's size and orientation is determined by $i = 14.1^{+2.3}_{-1.7}$\,\degr, $\Omega=259.3^{+20.9}_{-23.3}$\,deg, and $a=12.9^{+1.7}_{-1.8}$\,mas. The sky-projected orbit of BD$+$730275 and the Hipparcos measurements are shown in Fig.~\ref{fig:orbit_BD73}. Consequently, we determined the secondary mass to be $M_2 = 0.20 \pm 0.03$ M$_{\sun}$, i.e.\ the companion of BD$+$730275 is a low-mass star. This low-mass star is the only one found within our sample (and that of \citealt{diaz12}) to orbit a star with an effective temperature ($5260\pm25$\,K) less than the Sun. The companion to BD$+$73 0275, similar to HD\,51813, has a minimum mass within the driest part of the brown dwarf desert before astrometry revealed the companion to be stellar in nature.

\subsubsection{HD\,50761}
HD\,50761 is a F8 star with a mass of $1.19\pm0.10$\,M$_{\sun}$, the second most massive target in the sample. RV measurements show a companion with an orbital period of $2092\pm11$~days and with a minimum mass of $\textrm{m}_2 \sin i = 145^{+36}_{-29}~\textrm{M}_\textrm{Jup}$. HD\,50761 is the only object which does not have RV observations spanning an entire orbital period of their companion. This makes the minimum mass estimate difficult to constrain. HD\,50761 is the object which shows the largest amount of scatter in the bisector span (see Fig.~\ref{fig:bisectors}). No clear secondary peaks in the CCF were detected using a M4 cross correlation mask. This may be caused by the fast rotation of the star ($8.2\pm1.0$\,km/s).

\begin{figure}[ht]\begin{center} 
\includegraphics[width= 0.9\linewidth, trim= 0 0cm 0 0cm,clip=true]{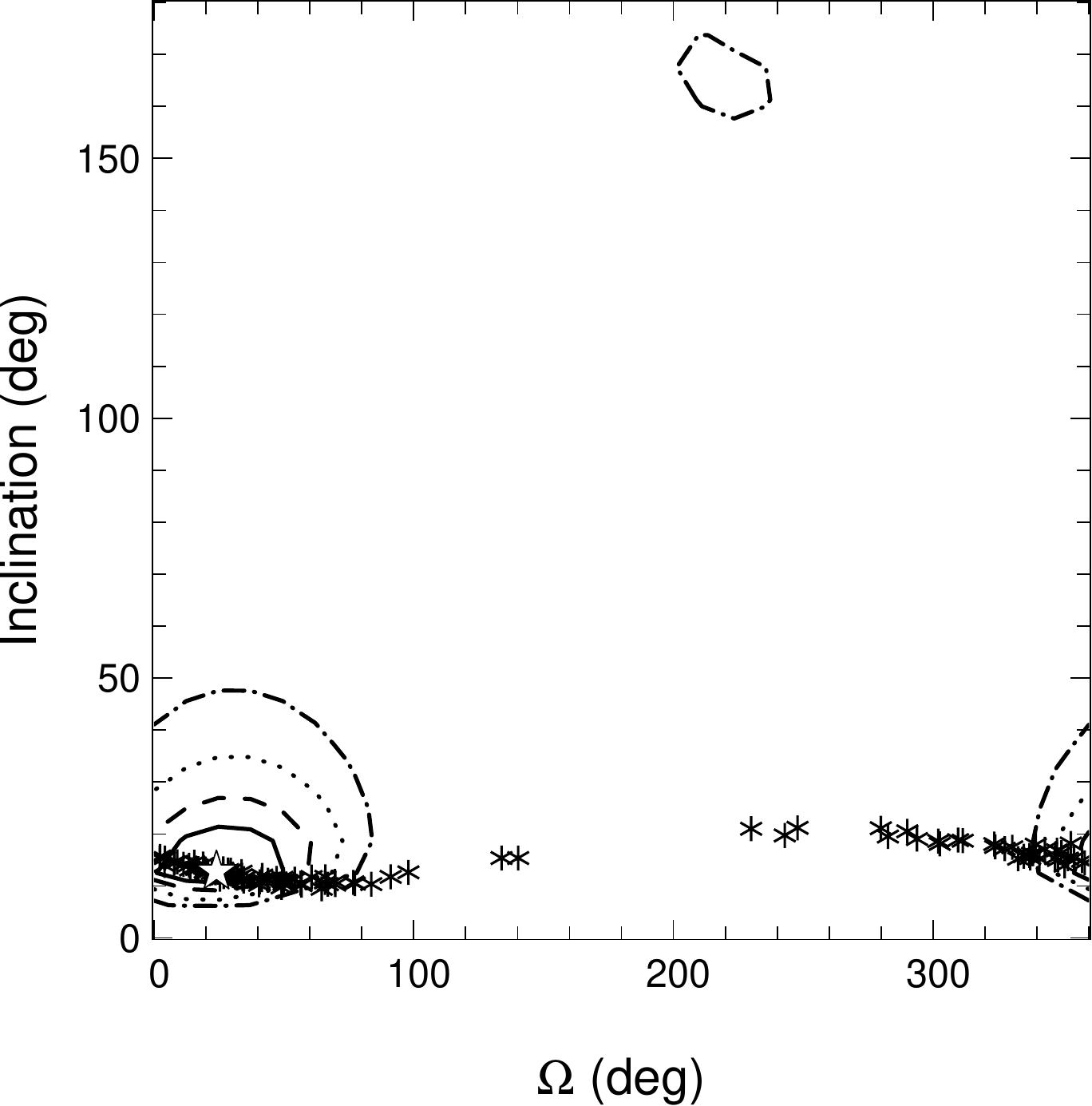}
 \caption{Joint confidence contours on the $i$-$\Omega$-grid for one set of spectroscopic parameters of HD\,50761. The contour lines correspond to confidences at 1-$\sigma$ (solid), 2-$\sigma$ (dashed), 3-$\sigma$ (dotted), and 4-$\sigma$ (dash-dotted) level. The crosses indicate the position of the best non-linear adjustment solution for each of the 100 Monte Carlo samples of spectroscopic parameters and the star corresponds to the orbit shown in Fig. \ref{fig:orbit_HD507}.} 
\label{fig:hipgrid}
 \end{center} \end{figure}

We detected the astrometric orbit with Hipparcos at $2.7\,\sigma$ significance (99.4 \%). However, the orbit orientation in the sky given by the angle $\Omega$ is poorly constrained by the astrometry. As a consequence, the best-fit astrometric parameters are strongly correlated with parameters derived from RV measurements, e.g.\ the period $P$ and the periastron time $T_0$. In our analysis, we vary those within the quoted uncertainties by randomly drawing 100 sets of RV parameters, thus we obtain 100 clearly different solutions in terms of inclination and $\Omega$. This is illustrated in Fig. \ref{fig:hipgrid}, where a large variation in the best-fit $\Omega$ can be seen. A large range of $\Omega$ values is acceptable for different sets of RV parameters, thus the final global distribution is particularly broad and its average value is ill-defined ($\Omega= 13.9^{+43.6}_{-43.3}$\,\degr)\footnote{The global distribution contains 100\,000 data points, because for each of the 100 RV parameter sets, we run the astrometric analysis 1000 times, see \cite{sahlmann11}.}. 

\begin{figure}[ht]\begin{center} 
\includegraphics[width= 0.8\linewidth, trim= 0 0cm 0 0cm,clip=true]{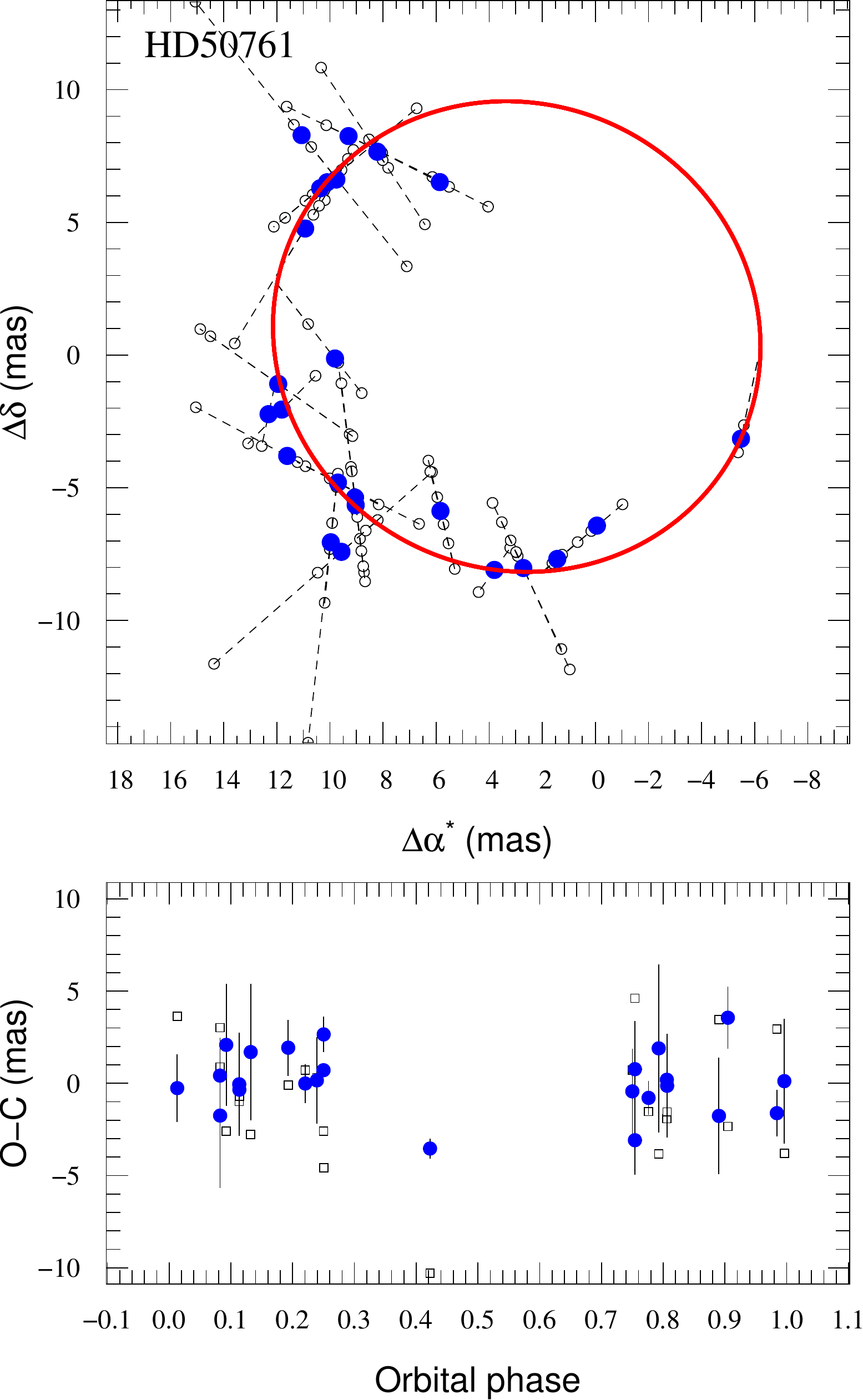}
 \caption{Astrometric orbit of HD\,50761 projected on the sky, cf. Fig. \ref{fig:orbit}, corresponding to one set of RV parameters. The star's motion is counter-clockwise.} 
\label{fig:orbit_HD507}
 \end{center} \end{figure}

This does not affect the significance of the astrometric orbit and the parameters listed in Tables \ref{tab:HIPparams2} and \ref{tab:HIPparams4} which naturally account for the correlations, i.e.\ the uncertainty of the orbital inclination $i = 13.2^{+4.8}_{-2.8}$\,\degr\ reflects the variation seen in Fig. \ref{fig:hipgrid}. However, it forces us to choose one single set of RV parameters to show one realisation of the astrometric orbit in Fig. \ref{fig:orbit_HD507}.

The measured semimajor axis of HD\,50761 is $a=9.1\pm2.0$~mas and we determine a companion mass of $M_2 = 0.38\pm0.11\,M_{\sun}$, i.e.\ HD\,50761\,B is a low-mass star. The large uncertainty in companion mass reflects the parameter correlations and the ill-defined orbit orientation $\Omega$ of the astrometric analysis. 

\subsubsection{HD\,51813}
\begin{figure}[ht]\begin{center} 
\includegraphics[width= 0.8\linewidth, trim= 0 0cm 0 0cm,clip=true]{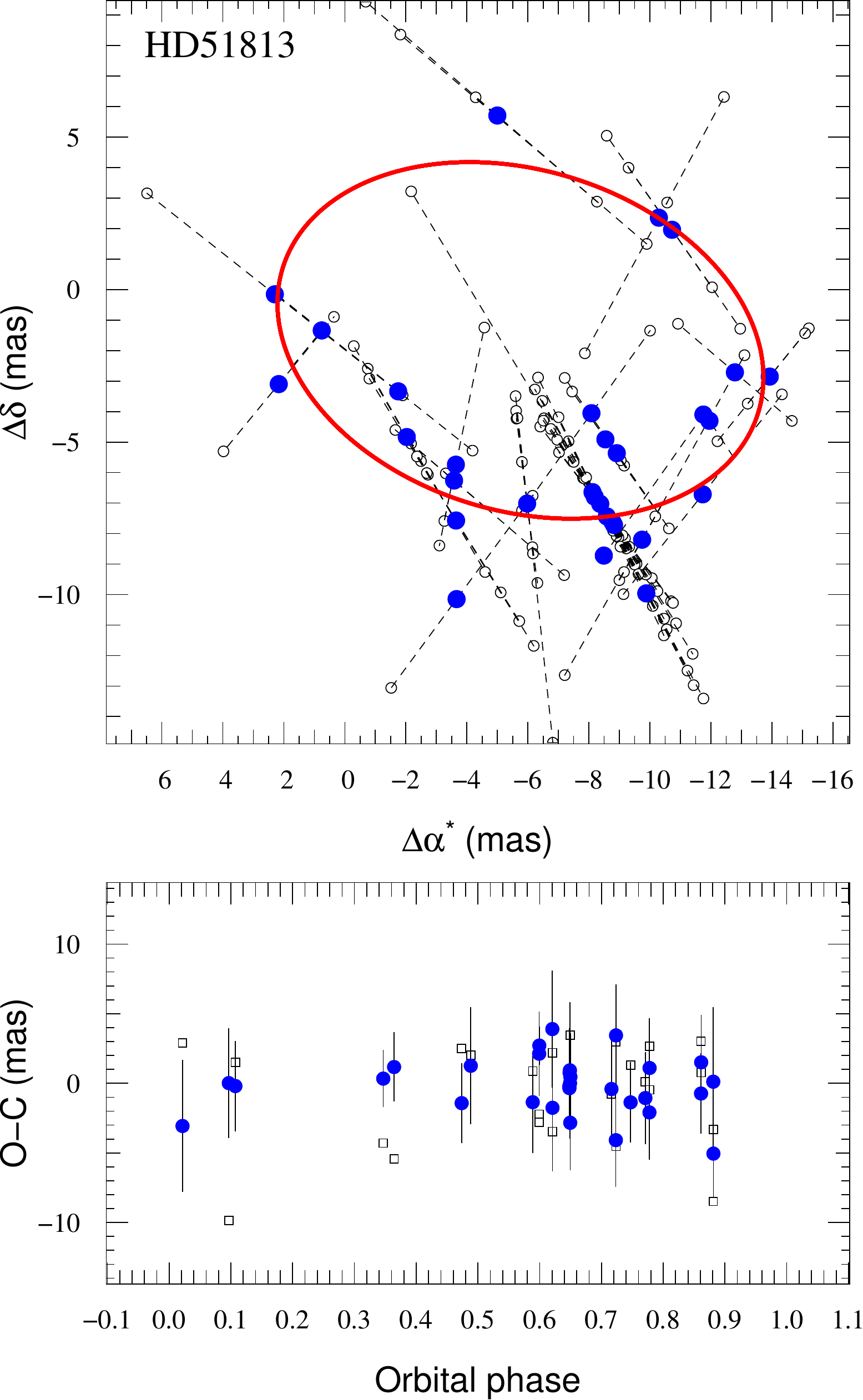}
 \caption{Astrometric orbit of HD\,51813 projected on the sky, cf. Fig. \ref{fig:orbit}. The star's motion is clockwise.} 
\label{fig:orbit3}
 \end{center} \end{figure}

\begin{figure*} 
\begin{center}

\includegraphics[width=0.45\textwidth]{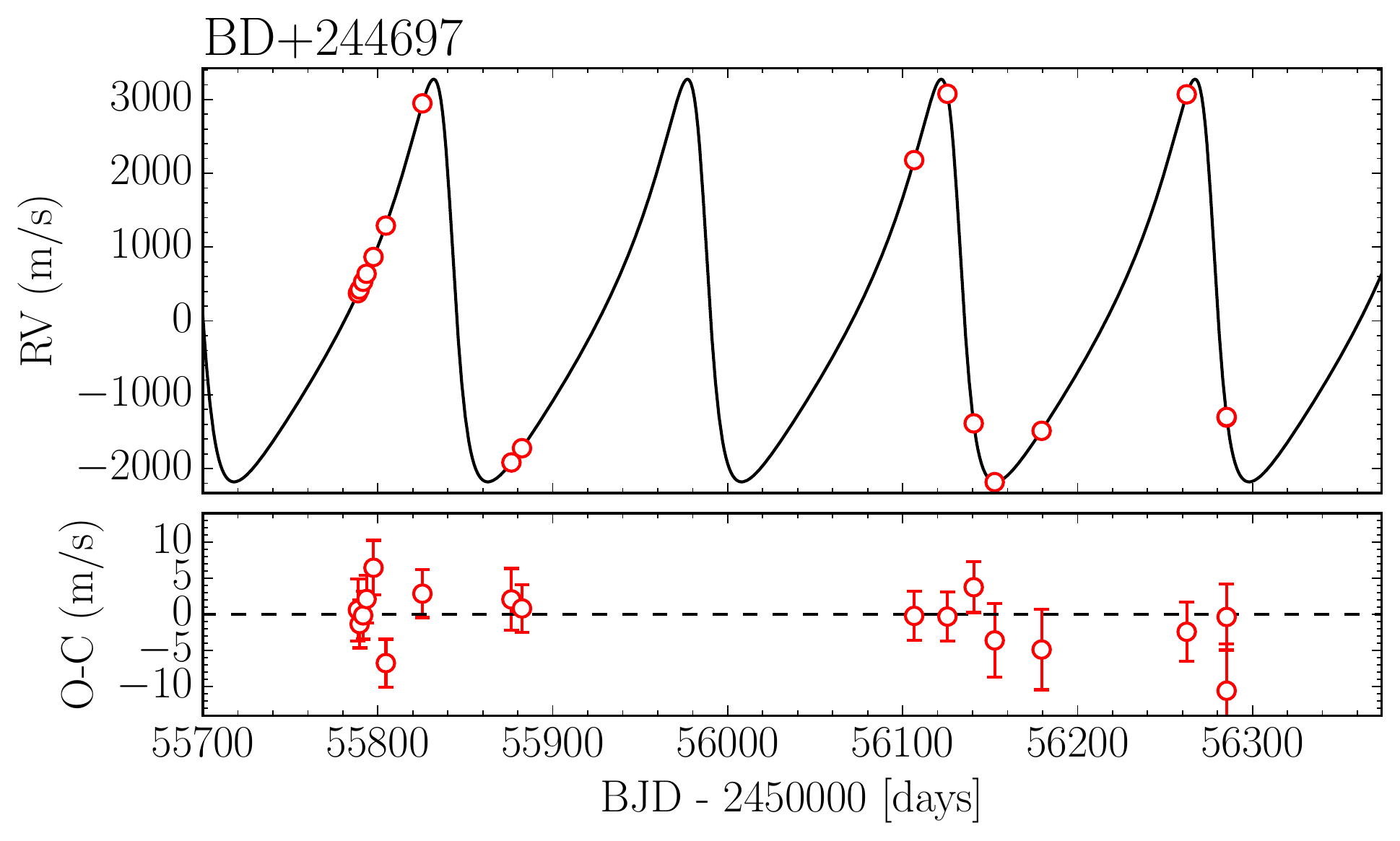}

\caption{Radial velocity measurements obtained with SOPHIE. The top panel shows the best fit to the data with the bottom panel showing the best fit residuals. Blue points indicate data taken prior to the 17$^{th}$ of June 2011 instrument upgrade. The grey points represent data from \cite{latham02}.}
\label{fig:RV}
\end{center}

\end{figure*}

HD\,51813 is an active ($\log R^\prime_{\mathrm{HK}}=-4.60\pm0.10$) G-type star with a mass of $1.06\pm0.07$\,M$_{\sun}$. The stellar activity is likely the cause of the large dispersion seen in the residuals (see Fig.~\ref{fig:RV}). This is evident when looking at the SOPHIE observations done after the instrument upgrade which typically show a residual dispersion 4-6 times smaller. RV measurements reveal a $\textrm{m}_2 \sin i = 46\pm4~\textrm{M}_\textrm{Jup}$ companion on a $1437^{+11}_{-15}$~day orbit. With only three pre instrument upgrade observations of HD\,51813 taken in close succession, the gamma value was adjusted as described in \S~\ref{subsec:companion_params}. We detected the astrometric orbit with Hipparcos with a significance $>3.3\,\sigma$ ($>$99.9 \%) and measured $\Omega= 5.3^{+22.6}_{-23.9}$\,\degr and $i = 169.2^{+2.0}_{-3.2}$\,\degr. The corresponding semimajor axis of HD\,51813 is $a=8.3 \pm 1.7$~mas and we determined a companion mass of $M_2 = 0.27 \pm 0.07$\,M$_{\sun}$, i.e.\ HD\,51813\,B is a low-mass star. The astrometric orbit is shown in Fig.~\ref{fig:orbit3}. HD\,51813, similar to BD$+$73 0275, has a minimum mass within the driest part of the brown dwarf desert before astrometry revealed the companion to be stellar in nature.
 
\subsubsection{HD\,56709}
HD\,56709 is a F5 star with a mass of $1.33\pm0.09$\,M$_{\sun}$, the most massive target in the sample. RV measurements show a companion with an orbital period of $2499.0\pm5.6$~days and with a minimum mass of $\textrm{m}_2 \sin i = 72\pm4~\textrm{M}_\textrm{Jup}$. 
No clear secondary peaks in the CCF were detected using a M4 cross correlation mask. Astrometry shows a tentative detection of a stellar companion with a permutation significance of only 86.8 \% ($\simeq\!1.5\,\sigma$), which is below our detection threshold. There are however, several indications that the orbit is well constrained by Hipparcos astrometry. First, significant acceleration was detected by Hipparcos alone, assigning this source a solution type of `7' \citep{:2007kx}. Second, our F-test asserts at high confidence that the orbital solution of HD\,56709 is a much better fit than the standard model (the p-value for the simpler model to be true given in Table \ref{tab:HIPparams4} is $1.9\times10^{-11}$\,\%), which is supported by a decrease in residual r.m.s. from $O-C_5=4.97$ mas to $O-C_7=4.27$ mas when including the orbital motion. On the other hand, only $\sim$40 \% of the orbit is covered by Hipparcos measurements.

\begin{figure}[ht]\begin{center} 
\includegraphics[width= 0.8\linewidth, trim= 0 0cm 0 0cm,clip=true]{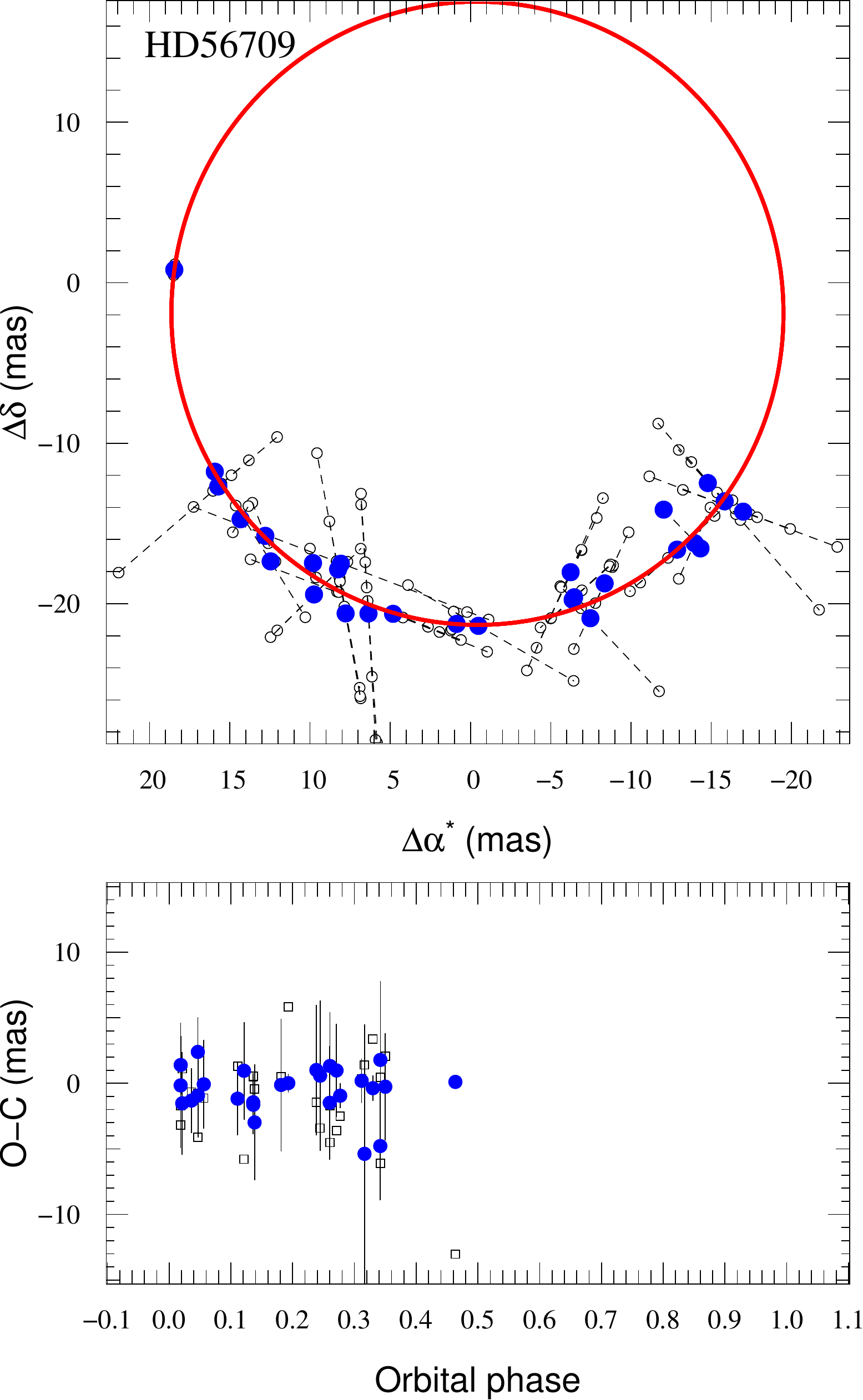}
 \caption{Astrometric orbit of HD\,56709 projected on the sky, cf. Fig.\,\ref{fig:orbit}. The star's motion is clockwise.} 
\label{fig:orbit2}
 \end{center} \end{figure}

We thus claim a tentative detection of the orbit and determine the system parameters, keeping in mind that these may have to be revised when better astrometry becomes available. The semimajor axis is $19.4 \pm 2.8$~mas. The orbital inclination of $i = 170.6^{+1.5}_{-2.1}$\degr and the  companion mass is $M_2 = 0.54 \pm 0.12$ M$_{\sun}$. The sky-projected orbit of HD\,56709 is  shown in Fig.~\ref{fig:orbit2}. Despite being fairly massive, the companion was not detected with speckle interferometry \citep{mason01}.

\subsubsection{HD\,134169}
HD\,134169 is a F8 star with a mass of $1.19\pm0.10$\,M$_{\sun}$. RV measurements show a companion with an orbital period of $67.8578\pm0.0038$~days which is the shortest orbital period amongst the objects in our sample. The detected companion has a minimum mass of $83\pm4~\textrm{M}_\textrm{Jup}$ shown by RV measurements and is therefore likely a stellar companion. Hipparcos measurements provide an upper limit on the companion at 1.03~M$_{\sun}$. This limit is likely too conservative as a companion with this mass would likely be detected by the bisector analysis described in \S\,\ref{sec:bisector_analysis}.

\subsection{Brown dwarf candidates}

\begin{figure} 
\includegraphics[width=\hsize]{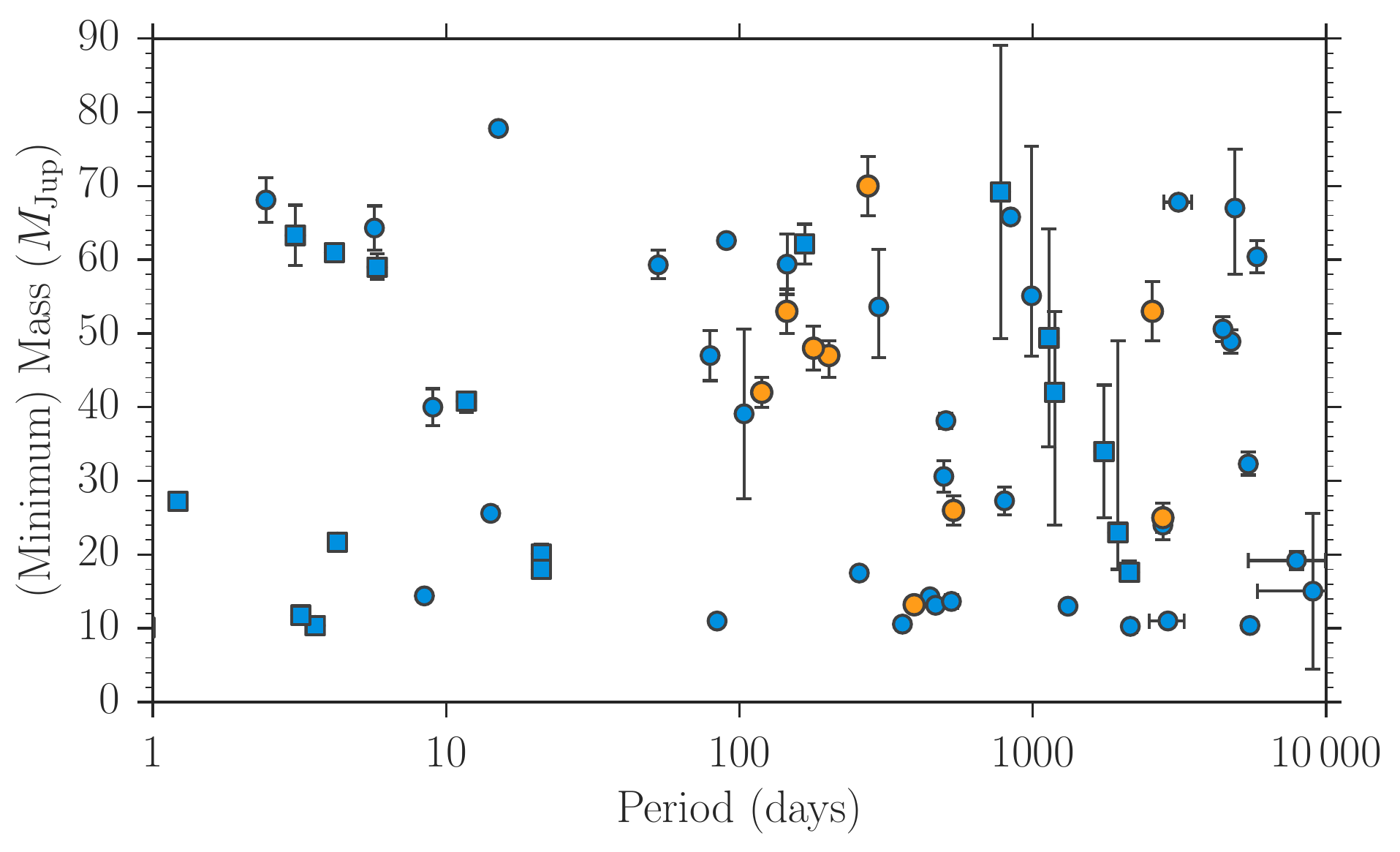}
\caption{The minimum mass as a function of orbital period. The data from this study are shown as orange points. The blue points represent data from the literature and are presented in Table~\ref{tab:appendix_table}. Objects with known masses are indicated using square symbols and objects with minimum mass measurements are shown as round points. All objects have 10~\MJ~$\leq m\sin i \leq 80$~\MJ\ and a Period $ < 10^5$~days with a host star mass 0.7~\msol~$\leq \rm{M} \leq1.5$~\msol.}
\label{fig:m_P}
\end{figure}

In the next subsections we describe the brown dwarf candidates in more detail. A list of brown dwarf candidates to solar-type stars with (projected) masses between 10 and 80~$\textrm{M}_\textrm{Jup}$ and periods $<10^5$ days currently found in litterature is presented in Table~\ref{tab:appendix_table}. The brown dwarf candidates from this study together with the objects from litterature are shown in Figure\,\ref{fig:m_P}.

\subsubsection{HD\,283668}
HD\,283668 is a K3V star which has both got the lowest mass ($0.68 \pm 0.06$~M$_\odot$) as well as the lowest metallicity ([Fe/H]$=-0.75 \pm 0.12$) amongst the star in our sample. The detected companion has a minimum mass of $53\pm4~\textrm{M}_\textrm{Jup}$ and a period of $2558\pm8$ days.

\subsubsection{HD\,39392}

HD\,39392 is a F8 star with a mass of $1.08 \pm 0.08$~M$_\odot$. The star is host to the lowest minimum mass companion at $\textrm{m}_2 \sin i = 13.2\pm0.8~\textrm{M}_\textrm{Jup}$ which orbits the star every $394.3^{+1.4}_{1.2}$ days. Hipparcos measurements provide an upper limit on the companion at 0.37~M$_{\sun}$. 

\subsubsection{HD\,77065}
HD\,77065 is a G star with a mass of $0.75 \pm 0.06$~M$_\odot$. The detected companion has a minimum mass of $\textrm{m}_2 \sin i = 41\pm2~\textrm{M}_\textrm{Jup}$ and an orbit lasting $119.1135^{+0.0026}_{0.0027}$ days. HD\,77065 has the most well constrained orbital parameters of all the objects in the sample due to the number of data points measured. HD\,77065 was first reported in \cite{latham02} as G9-42. Here we improve the reported parameters with additional data thereby reducing P by 1\%, K by 14\%, $e$ by 16\% and increasing the minimum mass by 2\%. Hipparcos measurements provide an upper limit on the companion at 0.68~M$_{\sun}$. This limit is likely too conservative as a companion with this mass would likely be detected by the bisector analysis described in \S\,\ref{sec:bisector_analysis}.

\subsubsection{BD$+$26 1888}
BD$+$26 1888 is K7 star with a mass of $0.76 \pm 0.08$~M$_\odot$ and is the coolest star in the sample with \teff$=4748\pm87$. The $\textrm{m}_2 \sin i = 26\pm2~\textrm{M}_\textrm{Jup}$ companion orbits the star with a period of $536.78\pm0.25$ days. Hipparcos measurements provide an upper limit on the companion at 0.25~M$_{\sun}$.

\subsubsection{HD\,122562}
HD\,122562 is a G5 star with a mass of $1.13 \pm 0.13$~M$_\odot$. The companions is found to have a mass of $\textrm{m}_2 \sin i = 24\pm2~\textrm{M}_\textrm{Jup}$ and has the longest period orbit with P$=2777^{+104}_{-81}$~days. HD\,122562 is also the least active star in the sample with $\log R^\prime_{\mathrm{HK}} = -5.27\pm0.16$~dex and shows the smallest dispersion in the residuals from the Keplerian fit. 

\subsubsection{HD\,132032}
HD\,132032 is a G5 star with a mass of $1.12 \pm 0.08$~M$_\odot$. The companions is found to have a mass of $\textrm{m}_2 \sin i = 70\pm4~\textrm{M}_\textrm{Jup}$ and to orbit the host with a period of $274.33\pm0.024$ days. The high companion mass results in a 50\% chance that the true companion mass is in the stellar regime. Hipparcos measurements provide an upper limit on the companion at 0.34~M$_{\sun}$.

\subsubsection{HD\,134113}
HD\,134113 is a F6/F7V type star with a mass of $0.81 \pm 0.01$~M$_\odot$. The companion is found to have a mass of $\textrm{m}_2 \sin i = 47^{+2}_{-3}~\textrm{M}_\textrm{Jup}$ and to orbit the host with a period of $201.680\pm0.004$ days. HD\,134113 was first reported in \cite{latham02} as G66-65. Here we improve the reported parameters by $\Delta \mathrm{P} = -0.035$~days, $\Delta  \mathrm{K} = -315$~m/s, $\Delta e = -0.009$. The minimum mass is found to be the same as that of \cite{latham02} at 45~\MJ. Hipparcos measurements provide an upper limit on the companion at 16.13~M$_{\sun}$. This upper mass limit can confidently be ruled out by the bisector analysis described in \S\,\ref{sec:bisector_analysis}.

\subsubsection{HD\,160508}
HD\,160508 is a F8V type star with a mass of $1.25 \pm 0.08$~M$_\odot$. The detected companion has a minimum mass of $\textrm{m}_2 \sin i = 48\pm3~\textrm{M}_\textrm{Jup}$ and orbits the star every $178.9049\pm0.0074$ days. Hipparcos measurements provide an upper limit on the companion at 1.58~M$_{\sun}$. This limit is likely too conservative as a companion with this mass would likely be detected by the bisector analysis described in \S\,\ref{sec:bisector_analysis}.

\subsubsection{BD$+$24\,4697}
HD\,160508 is a K2 type star with a mass of $0.75 \pm 0.05$~M$_\odot$. The detected companion has a minimum mass of $\textrm{m}_2 \sin i = 53\pm3~\textrm{M}_\textrm{Jup}$ and orbits the star every $145.081\pm0.016$ days. Hipparcos measurements provide an upper limit on the companion at 0.51~M$_{\sun}$.

\section{Conclusions \label{sect:conclusions}}

We characterise the orbital parameters of 15 companions to solar-type stars based on observations done with the SOPHIE spectrograph at the Haute-Provence Observatory. Using data from the Hipparcos satellite we constrain the true masses of all but two objects. Six companions are regarded as stellar in nature with masses ranging from a minimum mass of $76\pm4~\textrm{M}_\textrm{Jup}$ to a mass of $0.35\pm0.03~M_\sun$. The nine remaining companions that are not shown to be stellar are well distributed across the brown dwarf desert with minimum masses ranging from 13.2 to 70~\MJ. Seven of the nine objects are reported for the first time and mark a significant contribution to the number of objects in the brown dwarf desert. The increase in the number of brown dwarf desert objects is an important step required to eventually form a significant statistical sample which is necessary to be able to understand the formation and subsequent evolution of such objects.

Future characterisation of larger bias-corrected samples of high-mass planets and brown dwarf companions to solar type stars, confirmed through astrometric observations, are both welcomed and required to draw sound conclusions on the relationship between the orbital parameters in a quantifiable way. Not until the completion of the SOPHIE search for northern extrasolar planets sub-programme will we be able to derive the fraction of brown dwarfs within the SOPHIE sample. Continued observation will help constrain the different formation scenarios which could provide a way to discern exoplanets from brown dwarfs in the mass region where they overlap.

\begin{acknowledgements}
This work is based on observations made with the 1.93~m telescope at the Haute-Provence Observatory. P.A.W acknowledges the support of the French Agence Nationale de la Recherche (ANR), under program ANR-12-BS05-0012 "Exo-Atmos". This work was supported by Funda\c{c}\~ao para a Ci\^encia e a Tecnologia (FCT) through the research grant UID/FIS/04434/2013. NCS and AS acknowledge the support by the European Research Council/European Community under the FP7 through Starting Grant agreement number 239953 and support from Funda\c{c}\~ao para a Ci\^encia e a Tecnologia (FCT, Portugal) in the form of grants reference SFRH/BPD/70574/2010 and PTDC/FIS-AST/1526/2014. NCS also acknowledges the support from in the form of grant reference PTDC/CTE-AST/098528/2008 and through Investigador FCT contract of reference IF/00169/2012 as well as POPH/FSE (EC) by FEDER funding through the program ``Programa Operacional de Factores de Competitividade - COMPETE''. AS is supported by the European Union under a Marie Curie Intra-European Fellowship for Career Development with reference FP7-PEOPLE-2013-IEF, number 627202. J.S. is supported by an ESA Research Fellowship in Space Science.
\end{acknowledgements}

\bibliographystyle{aa}
\bibliography{bibliography}

\begin{thebibliography}{109}
\expandafter\ifx\csname natexlab\endcsname\relax\def\natexlab#1{#1}\fi

\bibitem[{{Anderson} {et~al.}(2011){Anderson}, {Collier Cameron}, {Hellier},
  {Lendl}, {Maxted}, {Pollacco}, {Queloz}, {Smalley}, {Smith}, {Todd},
  {Triaud}, {West}, {Barros}, {Enoch}, {Gillon}, {Lister}, {Pepe},
  {S{\'e}gransan}, {Street}, \& {Udry}}]{anderson11}
{Anderson}, D.~R., {Collier Cameron}, A., {Hellier}, C., {et~al.} 2011, \apjl,
  726, L19

\bibitem[{{Bakos} {et~al.}(2009){Bakos}, {Howard}, {Noyes}, {Hartman},
  {Torres}, {Kov{\'a}cs}, {Fischer}, {Latham}, {Johnson}, {Marcy}, {Sasselov},
  {Stefanik}, {Sip{\H o}cz}, {Kov{\'a}cs}, {Esquerdo}, {P{\'a}l},
  {L{\'a}z{\'a}r}, {Papp}, \& {S{\'a}ri}}]{bakos09}
{Bakos}, G.~{\'A}., {Howard}, A.~W., {Noyes}, R.~W., {et~al.} 2009, \apj, 707,
  446

\bibitem[{{Benedict} {et~al.}(2010){Benedict}, {McArthur}, {Bean}, {Barnes},
  {Harrison}, {Hatzes}, {Martioli}, \& {Nelan}}]{benedict10}
{Benedict}, G.~F., {McArthur}, B.~E., {Bean}, J.~L., {et~al.} 2010, \aj, 139,
  1844

\bibitem[{{Bodenheimer} {et~al.}(2003){Bodenheimer}, {Laughlin}, \&
  {Lin}}]{bodenheimer03}
{Bodenheimer}, P., {Laughlin}, G., \& {Lin}, D.~N.~C. 2003, \apj, 592, 555

\bibitem[{{Boisse} {et~al.}(2011){Boisse}, {Bouchy}, {H{\'e}brard}, {Bonfils},
  {Santos}, \& {Vauclair}}]{boisse11}
{Boisse}, I., {Bouchy}, F., {H{\'e}brard}, G., {et~al.} 2011, \aap, 528, A4

\bibitem[{{Boisse} {et~al.}(2010){Boisse}, {Eggenberger}, {Santos}, {Lovis},
  {Bouchy}, {H{\'e}brard}, {Arnold}, {Bonfils}, {Delfosse}, {Desort},
  {D{\'{\i}}az}, {Ehrenreich}, {Forveille}, {Gallenne}, {Lagrange}, {Moutou},
  {Udry}, {Pepe}, {Perrier}, {Perruchot}, {Pont}, {Queloz}, {Santerne},
  {S{\'e}gransan}, \& {Vidal-Madjar}}]{boisse10}
{Boisse}, I., {Eggenberger}, A., {Santos}, N.~C., {et~al.} 2010, \aap, 523, A88

\bibitem[{{Bonomo} {et~al.}(2010){Bonomo}, {Santerne}, {Alonso}, {Gazzano},
  {Havel}, {Aigrain}, {Auvergne}, {Baglin}, {Barbieri}, {Barge}, {Benz},
  {Bord{\'e}}, {Bouchy}, {Bruntt}, {Cabrera}, {Collier Cameron}, {Carone},
  {Carpano}, {Csizmadia}, {Deleuil}, {Deeg}, {Dvorak}, {Erikson},
  {Ferraz-Mello}, {Fridlund}, {Gandolfi}, {Gillon}, {Guenther}, {Guillot},
  {Hatzes}, {H{\'e}brard}, {Jorda}, {Lammer}, {Lanza}, {L{\'e}ger}, {Llebaria},
  {Mayor}, {Mazeh}, {Moutou}, {Ollivier}, {P{\"a}tzold}, {Pepe}, {Queloz},
  {Rauer}, {Rouan}, {Samuel}, {Schneider}, {Tingley}, {Udry}, \&
  {Wuchterl}}]{bonomo10}
{Bonomo}, A.~S., {Santerne}, A., {Alonso}, R., {et~al.} 2010, \aap, 520, A65

\bibitem[{{Bonomo} {et~al.}(2015){Bonomo}, {Sozzetti}, {Santerne}, {Deleuil},
  {Almenara}, {Bruno}, {D{\'{\i}}az}, {H{\'e}brard}, \& {Moutou}}]{bonomo15}
{Bonomo}, A.~S., {Sozzetti}, A., {Santerne}, A., {et~al.} 2015, \aap, 575, A85

\bibitem[{{Borucki} {et~al.}(2011){Borucki}, {Koch}, {Basri}, {Batalha},
  {Brown}, {Bryson}, {Caldwell}, {Christensen-Dalsgaard}, {Cochran}, {DeVore},
  {Dunham}, {Gautier}, {Geary}, {Gilliland}, {Gould}, {Howell}, {Jenkins},
  {Latham}, {Lissauer}, {Marcy}, {Rowe}, {Sasselov}, {Boss}, {Charbonneau},
  {Ciardi}, {Doyle}, {Dupree}, {Ford}, {Fortney}, {Holman}, {Seager},
  {Steffen}, {Tarter}, {Welsh}, {Allen}, {Buchhave}, {Christiansen}, {Clarke},
  {Das}, {D{\'e}sert}, {Endl}, {Fabrycky}, {Fressin}, {Haas}, {Horch},
  {Howard}, {Isaacson}, {Kjeldsen}, {Kolodziejczak}, {Kulesa}, {Li}, {Lucas},
  {Machalek}, {McCarthy}, {MacQueen}, {Meibom}, {Miquel}, {Prsa}, {Quinn},
  {Quintana}, {Ragozzine}, {Sherry}, {Shporer}, {Tenenbaum}, {Torres},
  {Twicken}, {Van Cleve}, {Walkowicz}, {Witteborn}, \& {Still}}]{borucki11}
{Borucki}, W.~J., {Koch}, D.~G., {Basri}, G., {et~al.} 2011, \apj, 736, 19

\bibitem[{{Bouchy} {et~al.}(2011{\natexlab{a}}){Bouchy}, {Bonomo}, {Santerne},
  {Moutou}, {Deleuil}, {D{\'{\i}}az}, {Eggenberger}, {Ehrenreich}, {Gry},
  {Guillot}, {Havel}, {H{\'e}brard}, \& {Udry}}]{bouchy11b}
{Bouchy}, F., {Bonomo}, A.~S., {Santerne}, A., {et~al.} 2011{\natexlab{a}},
  \aap, 533, A83

\bibitem[{{Bouchy} {et~al.}(2011{\natexlab{b}}){Bouchy}, {Deleuil}, {Guillot},
  {Aigrain}, {Carone}, {Cochran}, {Almenara}, {Alonso}, {Auvergne}, {Baglin},
  {Barge}, {Bonomo}, {Bord{\'e}}, {Csizmadia}, {de Bondt}, {Deeg},
  {D{\'{\i}}az}, {Dvorak}, {Endl}, {Erikson}, {Ferraz-Mello}, {Fridlund},
  {Gandolfi}, {Gazzano}, {Gibson}, {Gillon}, {Guenther}, {Hatzes}, {Havel},
  {H{\'e}brard}, {Jorda}, {L{\'e}ger}, {Lovis}, {Llebaria}, {Lammer},
  {MacQueen}, {Mazeh}, {Moutou}, {Ofir}, {Ollivier}, {Parviainen},
  {P{\"a}tzold}, {Queloz}, {Rauer}, {Rouan}, {Santerne}, {Schneider},
  {Tingley}, \& {Wuchterl}}]{bouchy11}
{Bouchy}, F., {Deleuil}, M., {Guillot}, T., {et~al.} 2011{\natexlab{b}}, \aap,
  525, A68

\bibitem[{{Bouchy} {et~al.}(2013){Bouchy}, {D{\'{\i}}az}, {H{\'e}brard},
  {Arnold}, {Boisse}, {Delfosse}, {Perruchot}, \& {Santerne}}]{bouchy13}
{Bouchy}, F., {D{\'{\i}}az}, R.~F., {H{\'e}brard}, G., {et~al.} 2013, \aap,
  549, A49

\bibitem[{{Bouchy} {et~al.}(2009){Bouchy}, {H{\'e}brard}, {Udry}, {Delfosse},
  {Boisse}, {Desort}, {Bonfils}, {Eggenberger}, {Ehrenreich}, {Forveille},
  {Lagrange}, {Le Coroller}, {Lovis}, {Moutou}, {Pepe}, {Perrier}, {Pont},
  {Queloz}, {Santos}, {S{\'e}gransan}, \& {Vidal-Madjar}}]{bouchy09}
{Bouchy}, F., {H{\'e}brard}, G., {Udry}, S., {et~al.} 2009, \aap, 505, 853

\bibitem[{{Bouchy} {et~al.}(2015){Bouchy}, {S{\'e}gransan}, {D{\'{\i}}az},
  {H{\'e}brard}, {Arnold}, {Boisse}, \& {Delfosse}}]{bouchy15b}
{Bouchy}, F., {S{\'e}gransan}, D., {D{\'{\i}}az}, R.~F., {et~al.} 2015,
  Submitted to \aap

\bibitem[{{Burrows} {et~al.}(2011){Burrows}, {Heng}, \&
  {Nampaisarn}}]{burrows11}
{Burrows}, A., {Heng}, K., \& {Nampaisarn}, T. 2011, \apj, 736, 47

\bibitem[{{Burrows} {et~al.}(2001){Burrows}, {Hubbard}, {Lunine}, \&
  {Liebert}}]{burrows01}
{Burrows}, A., {Hubbard}, W.~B., {Lunine}, J.~I., \& {Liebert}, J. 2001,
  Reviews of Modern Physics, 73, 719

\bibitem[{{Burrows} {et~al.}(2007){Burrows}, {Hubeny}, {Budaj}, {Knutson}, \&
  {Charbonneau}}]{burrows07}
{Burrows}, A., {Hubeny}, I., {Budaj}, J., {Knutson}, H.~A., \& {Charbonneau},
  D. 2007, \apjl, 668, L171

\bibitem[{{Burrows} {et~al.}(1997){Burrows}, {Marley}, {Hubbard}, {Lunine},
  {Guillot}, {Saumon}, {Freedman}, {Sudarsky}, \& {Sharp}}]{burrows97}
{Burrows}, A., {Marley}, M., {Hubbard}, W.~B., {et~al.} 1997, \apj, 491, 856

\bibitem[{{Butler} {et~al.}(2006){Butler}, {Wright}, {Marcy}, {Fischer},
  {Vogt}, {Tinney}, {Jones}, {Carter}, {Johnson}, {McCarthy}, \&
  {Penny}}]{butler06}
{Butler}, R.~P., {Wright}, J.~T., {Marcy}, G.~W., {et~al.} 2006, \apj, 646, 505

\bibitem[{{Campbell} {et~al.}(1991){Campbell}, {Yang}, {Irwin}, \&
  {Walker}}]{campbell91}
{Campbell}, B., {Yang}, S., {Irwin}, A.~W., \& {Walker}, G.~A.~H. 1991, in
  Lecture Notes in Physics, Berlin Springer Verlag, Vol. 390, Bioastronomy: The
  Search for Extraterrestial Life - The Exploration Broadens, ed. J.~{Heidmann}
  \& M.~J. {Klein}, 19--20

\bibitem[{{Chabrier} \& {Baraffe}(1997)}]{chabrier97}
{Chabrier}, G. \& {Baraffe}, I. 1997, \aap, 327, 1039

\bibitem[{{Chabrier} \& {Baraffe}(2000)}]{chabrier00}
{Chabrier}, G. \& {Baraffe}, I. 2000, \araa, 38, 337

\bibitem[{{Chabrier} {et~al.}(2014){Chabrier}, {Johansen}, {Janson}, \&
  {Rafikov}}]{chabrier14}
{Chabrier}, G., {Johansen}, A., {Janson}, M., \& {Rafikov}, R. 2014, Protostars
  and Planets VI, 619

\bibitem[{{Charbonneau} {et~al.}(2000){Charbonneau}, {Brown}, {Latham}, \&
  {Mayor}}]{charbonneau00}
{Charbonneau}, D., {Brown}, T.~M., {Latham}, D.~W., \& {Mayor}, M. 2000, \apjl,
  529, L45

\bibitem[{{Correia} {et~al.}(2005){Correia}, {Udry}, {Mayor}, {Laskar}, {Naef},
  {Pepe}, {Queloz}, \& {Santos}}]{correia05}
{Correia}, A.~C.~M., {Udry}, S., {Mayor}, M., {et~al.} 2005, \aap, 440, 751

\bibitem[{{Courcol} {et~al.}(2015){Courcol}, {Bouchy}, {Pepe}, {Santerne},
  {Delfosse}, {Arnold}, {Astudillo-Defru}, {Boisse}, {Bonfils}, {Borgniet},
  {Bourrier}, {Cabrera}, {Deleuil}, {Demangeon}, {D{\'{\i}}az}, {Ehrenreich},
  {Forveille}, {H{\'e}brard}, {Lagrange}, {Montagnier}, {Moutou}, {Rey},
  {Santos}, {S{\'e}gransan}, {Udry}, \& {Wilson}}]{courcol15}
{Courcol}, B., {Bouchy}, F., {Pepe}, F., {et~al.} 2015, \aap, 581, A38

\bibitem[{{Csizmadia} {et~al.}(2015){Csizmadia}, {Hatzes}, {Gandolfi},
  {Deleuil}, {Bouchy}, {Fridlund}, {Szabados}, {Parviainen}, {Cabrera},
  {Aigrain}, {Alonso}, {Almenara}, {Baglin}, {Bord{\'e}}, {Bonomo}, {Deeg},
  {D{\'{\i}}az}, {Erikson}, {Ferraz-Mello}, {Tadeu dos Santos}, {Guenther},
  {Guillot}, {Grziwa}, {H{\'e}brard}, {Klagyivik}, {Ollivier}, {P{\"a}tzold},
  {Rauer}, {Rouan}, {Santerne}, {Schneider}, {Mazeh}, {Wuchterl}, {Carpano}, \&
  {Ofir}}]{csizmadia15}
{Csizmadia}, S., {Hatzes}, A., {Gandolfi}, D., {et~al.} 2015, \aap, 584, A13

\bibitem[{{da Silva} {et~al.}(2006){da Silva}, {Udry}, {Bouchy}, {Mayor},
  {Moutou}, {Pont}, {Queloz}, {Santos}, {S{\'e}gransan}, \&
  {Zucker}}]{dasilva06}
{da Silva}, R., {Udry}, S., {Bouchy}, F., {et~al.} 2006, \aap, 446, 717

\bibitem[{{Deleuil} {et~al.}(2008){Deleuil}, {Deeg}, {Alonso}, {Bouchy},
  {Rouan}, {Auvergne}, {Baglin}, {Aigrain}, {Almenara}, {Barbieri}, {Barge},
  {Bruntt}, {Bord{\'e}}, {Collier Cameron}, {Csizmadia}, {de La Reza},
  {Dvorak}, {Erikson}, {Fridlund}, {Gandolfi}, {Gillon}, {Guenther}, {Guillot},
  {Hatzes}, {H{\'e}brard}, {Jorda}, {Lammer}, {L{\'e}ger}, {Llebaria},
  {Loeillet}, {Mayor}, {Mazeh}, {Moutou}, {Ollivier}, {P{\"a}tzold}, {Pont},
  {Queloz}, {Rauer}, {Schneider}, {Shporer}, {Wuchterl}, \&
  {Zucker}}]{deleuil08}
{Deleuil}, M., {Deeg}, H.~J., {Alonso}, R., {et~al.} 2008, \aap, 491, 889

\bibitem[{{D{\'{\i}}az} {et~al.}(2013){D{\'{\i}}az}, {Damiani}, {Deleuil},
  {Almenara}, {Moutou}, {Barros}, {Bonomo}, {Bouchy}, {Bruno}, {H{\'e}brard},
  {Montagnier}, \& {Santerne}}]{diaz13}
{D{\'{\i}}az}, R.~F., {Damiani}, C., {Deleuil}, M., {et~al.} 2013, \aap, 551,
  L9

\bibitem[{{D{\'{\i}}az} {et~al.}(2012){D{\'{\i}}az}, {Santerne}, {Sahlmann},
  {H{\'e}brard}, {Eggenberger}, {Santos}, {Moutou}, {Arnold}, {Boisse},
  {Bonfils}, {Bouchy}, {Delfosse}, {Desort}, {Ehrenreich}, {Forveille},
  {Lagrange}, {Lovis}, {Pepe}, {Perrier}, {Queloz}, {S{\'e}gransan}, {Udry}, \&
  {Vidal-Madjar}}]{diaz12}
{D{\'{\i}}az}, R.~F., {Santerne}, A., {Sahlmann}, J., {et~al.} 2012, \aap, 538,
  A113

\bibitem[{{Duquennoy} \& {Mayor}(1991)}]{duquennoy91}
{Duquennoy}, A. \& {Mayor}, M. 1991, \aap, 248, 485

\bibitem[{{Endl} {et~al.}(2004){Endl}, {Hatzes}, {Cochran}, {McArthur},
  {Allende Prieto}, {Paulson}, {Guenther}, \& {Bedalov}}]{endl04}
{Endl}, M., {Hatzes}, A.~P., {Cochran}, W.~D., {et~al.} 2004, \apj, 611, 1121

\bibitem[{{Feroz} {et~al.}(2011){Feroz}, {Balan}, \& {Hobson}}]{feroz11}
{Feroz}, F., {Balan}, S.~T., \& {Hobson}, M.~P. 2011, \mnras, 416, L104

\bibitem[{{Fischer} {et~al.}(2003){Fischer}, {Marcy}, {Butler}, {Vogt},
  {Henry}, {Pourbaix}, {Walp}, {Misch}, \& {Wright}}]{fischer03}
{Fischer}, D.~A., {Marcy}, G.~W., {Butler}, R.~P., {et~al.} 2003, \apj, 586,
  1394

\bibitem[{{Fischer} {et~al.}(2002){Fischer}, {Marcy}, {Butler}, {Vogt}, {Walp},
  \& {Apps}}]{fischer02}
{Fischer}, D.~A., {Marcy}, G.~W., {Butler}, R.~P., {et~al.} 2002, \pasp, 114,
  529

\bibitem[{{Fleming} {et~al.}(2012){Fleming}, {Ge}, {Barnes}, {Beatty}, {Crepp},
  {De Lee}, {Esposito}, {Femenia}, {Ferreira}, {Gary}, {Gaudi}, {Ghezzi},
  {Gonz{\'a}lez Hern{\'a}ndez}, {Hebb}, {Jiang}, {Lee}, {Nelson}, {Porto de
  Mello}, {Shappee}, {Stassun}, {Thompson}, {Tofflemire}, {Wisniewski},
  {Wood-Vasey}, {Agol}, {Allende Prieto}, {Bizyaev}, {Brewington}, {Cargile},
  {Coban}, {Costello}, {da Costa}, {Good}, {Hua}, {Kane}, {Lander}, {Liu},
  {Ma}, {Mahadevan}, {Maia}, {Malanushenko}, {Malanushenko}, {Muna}, {Nguyen},
  {Oravetz}, {Paegert}, {Pan}, {Pepper}, {Rebolo}, {Roebuck}, {Santiago},
  {Schneider}, {Shelden}, {Simmons}, {Sivarani}, {Snedden}, {Vincent}, {Wan},
  {Wang}, {Weaver}, {Weaver}, \& {Zhao}}]{fleming12}
{Fleming}, S.~W., {Ge}, J., {Barnes}, R., {et~al.} 2012, \aj, 144, 72

\bibitem[{{Fleming} {et~al.}(2010){Fleming}, {Ge}, {Mahadevan}, {Lee},
  {Eastman}, {Siverd}, {Gaudi}, {Niedzielski}, {Sivarani}, {Stassun},
  {Wolszczan}, {Barnes}, {Gary}, {Nguyen}, {Morehead}, {Wan}, {Zhao}, {Liu},
  {Guo}, {Kane}, {van Eyken}, {De Lee}, {Crepp}, {Shelden}, {Laws},
  {Wisniewski}, {Schneider}, {Pepper}, {Snedden}, {Pan}, {Bizyaev},
  {Brewington}, {Malanushenko}, {Malanushenko}, {Oravetz}, {Simmons}, \&
  {Watters}}]{fleming10}
{Fleming}, S.~W., {Ge}, J., {Mahadevan}, S., {et~al.} 2010, \apj, 718, 1186

\bibitem[{{Ford} {et~al.}(2011){Ford}, {Rowe}, {Fabrycky}, {Carter}, {Holman},
  {Lissauer}, {Ragozzine}, {Steffen}, {Batalha}, {Borucki}, {Bryson},
  {Caldwell}, {Dunham}, {Gautier}, {Jenkins}, {Koch}, {Li}, {Lucas}, {Marcy},
  {McCauliff}, {Mullally}, {Quintana}, {Still}, {Tenenbaum}, {Thompson}, \&
  {Twicken}}]{ford11}
{Ford}, E.~B., {Rowe}, J.~F., {Fabrycky}, D.~C., {et~al.} 2011, \apjs, 197, 2

\bibitem[{{Girardi} {et~al.}(2000){Girardi}, {Bressan}, {Bertelli}, \&
  {Chiosi}}]{girardi00}
{Girardi}, L., {Bressan}, A., {Bertelli}, G., \& {Chiosi}, C. 2000, \aaps, 141,
  371

\bibitem[{{Grether} \& {Lineweaver}(2006)}]{grether06}
{Grether}, D. \& {Lineweaver}, C.~H. 2006, \apj, 640, 1051

\bibitem[{{Halbwachs} {et~al.}(2000){Halbwachs}, {Arenou}, {Mayor}, {Udry}, \&
  {Queloz}}]{halbwachs00}
{Halbwachs}, J.~L., {Arenou}, F., {Mayor}, M., {Udry}, S., \& {Queloz}, D.
  2000, \aap, 355, 581

\bibitem[{{Haywood}(2008)}]{haywood08}
{Haywood}, M. 2008, \aap, 482, 673

\bibitem[{{H{\'e}brard} {et~al.}(2008){H{\'e}brard}, {Bouchy}, {Pont},
  {Loeillet}, {Rabus}, {Bonfils}, {Moutou}, {Boisse}, {Delfosse}, {Desort},
  {Eggenberger}, {Ehrenreich}, {Forveille}, {Lagrange}, {Lovis}, {Mayor},
  {Pepe}, {Perrier}, {Queloz}, {Santos}, {S{\'e}gransan}, {Udry}, \&
  {Vidal-Madjar}}]{hebrard08}
{H{\'e}brard}, G., {Bouchy}, F., {Pont}, F., {et~al.} 2008, \aap, 488, 763

\bibitem[{{Hellier} {et~al.}(2009){Hellier}, {Anderson}, {Collier Cameron},
  {Gillon}, {Hebb}, {Maxted}, {Queloz}, {Smalley}, {Triaud}, {West}, {Wilson},
  {Bentley}, {Enoch}, {Horne}, {Irwin}, {Lister}, {Mayor}, {Parley}, {Pepe},
  {Pollacco}, {Segransan}, {Udry}, \& {Wheatley}}]{hellier09}
{Hellier}, C., {Anderson}, D.~R., {Collier Cameron}, A., {et~al.} 2009, \nat,
  460, 1098

\bibitem[{{Howard} {et~al.}(2010){Howard}, {Marcy}, {Johnson}, {Fischer},
  {Wright}, {Isaacson}, {Valenti}, {Anderson}, {Lin}, \& {Ida}}]{howard10}
{Howard}, A.~W., {Marcy}, G.~W., {Johnson}, J.~A., {et~al.} 2010, Science, 330,
  653

\bibitem[{{Jenkins} {et~al.}(2009){Jenkins}, {Jones}, {Go{\'z}dziewski},
  {Migaszewski}, {Barnes}, {Jones}, {Rojo}, {Pinfield}, {Day-Jones}, \&
  {Hoyer}}]{jenkins09}
{Jenkins}, J.~S., {Jones}, H.~R.~A., {Go{\'z}dziewski}, K., {et~al.} 2009,
  \mnras, 398, 911

\bibitem[{{Jiang} {et~al.}(2013){Jiang}, {Ge}, {Cargile}, {Crepp}, {De Lee},
  {Porto de Mello}, {Esposito}, {Ferreira}, {Femenia}, {Fleming}, {Gaudi},
  {Ghezzi}, {Gonz{\'a}lez Hern{\'a}ndez}, {Hebb}, {Lee}, {Ma}, {Stassun},
  {Wang}, {Wisniewski}, {Agol}, {Bizyaev}, {Brewington}, {Chang}, {Nicolaci da
  Costa}, {Eastman}, {Ebelke}, {Gary}, {Kane}, {Li}, {Liu}, {Mahadevan},
  {Maia}, {Malanushenko}, {Malanushenko}, {Muna}, {Nguyen}, {Ogando},
  {Oravetz}, {Oravetz}, {Pan}, {Pepper}, {Paegert}, {Allende Prieto}, {Rebolo},
  {Santiago}, {Schneider}, {Shelden Bradley}, {Sivarani}, {Snedden}, {van
  Eyken}, {Wan}, {Weaver}, \& {Zhao}}]{jiang13}
{Jiang}, P., {Ge}, J., {Cargile}, P., {et~al.} 2013, \aj, 146, 65

\bibitem[{{Johns-Krull} {et~al.}(2008){Johns-Krull}, {McCullough}, {Burke},
  {Valenti}, {Janes}, {Heasley}, {Prato}, {Bissinger}, {Fleenor}, {Foote},
  {Garcia-Melendo}, {Gary}, {Howell}, {Mallia}, {Masi}, \&
  {Vanmunster}}]{johnskrull08}
{Johns-Krull}, C.~M., {McCullough}, P.~R., {Burke}, C.~J., {et~al.} 2008, \apj,
  677, 657

\bibitem[{{Johnson} {et~al.}(2011){Johnson}, {Clanton}, {Howard}, {Bowler},
  {Henry}, {Marcy}, {Crepp}, {Endl}, {Cochran}, {MacQueen}, {Wright}, \&
  {Isaacson}}]{johnson11}
{Johnson}, J.~A., {Clanton}, C., {Howard}, A.~W., {et~al.} 2011, \apjs, 197, 26

\bibitem[{{Kane} {et~al.}(2011{\natexlab{a}}){Kane}, {Henry}, {Dragomir},
  {Fischer}, {Howard}, {Wang}, \& {Wright}}]{kane11}
{Kane}, S.~R., {Henry}, G.~W., {Dragomir}, D., {et~al.} 2011{\natexlab{a}},
  \apjl, 735, L41

\bibitem[{{Kane} {et~al.}(2011{\natexlab{b}}){Kane}, {Howard}, {Pilyavsky},
  {Mahadevan}, {Henry}, {von Braun}, {Ciardi}, {Dragomir}, {Fischer}, {Jensen},
  {Laughlin}, {Ramirez}, \& {Wright}}]{kane2011b}
{Kane}, S.~R., {Howard}, A.~W., {Pilyavsky}, G., {et~al.} 2011{\natexlab{b}},
  \apj, 733, 28

\bibitem[{{Kane} {et~al.}(2009){Kane}, {Mahadevan}, {Cochran}, {Street},
  {Sivarani}, {Henry}, \& {Williamson}}]{kane2009}
{Kane}, S.~R., {Mahadevan}, S., {Cochran}, W.~D., {et~al.} 2009, \apj, 692, 290

\bibitem[{{Kurucz}(1993)}]{kurucz93}
{Kurucz}, R. 1993, ATLAS9 Stellar Atmosphere Programs and 2 km/s grid.~Kurucz
  CD-ROM No.~13.~ Cambridge, Mass.: Smithsonian Astrophysical Observatory,
  1993., 13

\bibitem[{{Latham} {et~al.}(1989){Latham}, {Stefanik}, {Mazeh}, {Mayor}, \&
  {Burki}}]{latham89}
{Latham}, D.~W., {Stefanik}, R.~P., {Mazeh}, T., {Mayor}, M., \& {Burki}, G.
  1989, \nat, 339, 38

\bibitem[{{Latham} {et~al.}(2002){Latham}, {Stefanik}, {Torres}, {Davis},
  {Mazeh}, {Carney}, {Laird}, \& {Morse}}]{latham02}
{Latham}, D.~W., {Stefanik}, R.~P., {Torres}, G., {et~al.} 2002, \aj, 124, 1144

\bibitem[{{Leconte} {et~al.}(2009){Leconte}, {Baraffe}, {Chabrier}, {Barman},
  \& {Levrard}}]{leconte09}
{Leconte}, J., {Baraffe}, I., {Chabrier}, G., {Barman}, T., \& {Levrard}, B.
  2009, \aap, 506, 385

\bibitem[{{Ma} {et~al.}(2013){Ma}, {Ge}, {Barnes}, {Crepp}, {De Lee},
  {Dutra-Ferreira}, {Esposito}, {Femenia}, {Fleming}, {Gaudi}, {Ghezzi},
  {Hebb}, {Gonzalez Hernandez}, {Lee}, {Porto de Mello}, {Stassun}, {Wang},
  {Wisniewski}, {Agol}, {Bizyaev}, {Cargile}, {Chang}, {Nicolaci da Costa},
  {Eastman}, {Gary}, {Jiang}, {Kane}, {Li}, {Liu}, {Mahadevan}, {Maia}, {Muna},
  {Nguyen}, {Ogando}, {Oravetz}, {Pepper}, {Paegert}, {Allende Prieto},
  {Rebolo}, {Santiago}, {Schneider}, {Shelden}, {Simmons}, {Sivarani}, {van
  Eyken}, {Wan}, {Weaver}, \& {Zhao}}]{ma13}
{Ma}, B., {Ge}, J., {Barnes}, R., {et~al.} 2013, \aj, 145, 20

\bibitem[{{Marcy} {et~al.}(2005){Marcy}, {Butler}, {Fischer}, {Vogt}, {Wright},
  {Tinney}, \& {Jones}}]{marcy05}
{Marcy}, G., {Butler}, R.~P., {Fischer}, D., {et~al.} 2005, Progress of
  Theoretical Physics Supplement, 158, 24

\bibitem[{{Marcy} \& {Butler}(2000)}]{marcy00}
{Marcy}, G.~W. \& {Butler}, R.~P. 2000, \pasp, 112, 137

\bibitem[{{Marcy} {et~al.}(2001){Marcy}, {Butler}, {Vogt}, {Liu}, {Laughlin},
  {Apps}, {Graham}, {Lloyd}, {Luhman}, \& {Jayawardhana}}]{marcy01}
{Marcy}, G.~W., {Butler}, R.~P., {Vogt}, S.~S., {et~al.} 2001, \apj, 555, 418

\bibitem[{{Marmier} {et~al.}(2013){Marmier}, {S{\'e}gransan}, {Udry}, {Mayor},
  {Pepe}, {Queloz}, {Lovis}, {Naef}, {Santos}, {Alonso}, {Alves}, {Berthet},
  {Chazelas}, {Demory}, {Dumusque}, {Eggenberger}, {Figueira}, {Gillon},
  {Hagelberg}, {Lendl}, {Mardling}, {M{\'e}gevand}, {Neveu}, {Sahlmann},
  {Sosnowska}, {Tewes}, \& {Triaud}}]{marmier13}
{Marmier}, M., {S{\'e}gransan}, D., {Udry}, S., {et~al.} 2013, \aap, 551, A90

\bibitem[{{Martioli} {et~al.}(2010){Martioli}, {McArthur}, {Benedict}, {Bean},
  {Harrison}, \& {Armstrong}}]{martioli10}
{Martioli}, E., {McArthur}, B.~E., {Benedict}, G.~F., {et~al.} 2010, \apj, 708,
  625

\bibitem[{{Mason} {et~al.}(2001){Mason}, {Hartkopf}, {Holdenried}, \&
  {Rafferty}}]{mason01}
{Mason}, B.~D., {Hartkopf}, W.~I., {Holdenried}, E.~R., \& {Rafferty}, T.~J.
  2001, \aj, 121, 3224

\bibitem[{{Mayor} {et~al.}(2011){Mayor}, {Marmier}, {Lovis}, {Udry},
  {S{\'e}gransan}, {Pepe}, {Benz}, {Bertaux}, {Bouchy}, {Dumusque}, {Lo Curto},
  {Mordasini}, {Queloz}, \& {Santos}}]{mayor11}
{Mayor}, M., {Marmier}, M., {Lovis}, C., {et~al.} 2011, ArXiv e-prints

\bibitem[{{Moutou} {et~al.}(2013){Moutou}, {Bonomo}, {Bruno}, {Montagnier},
  {Bouchy}, {Almenara}, {Barros}, {Deleuil}, {D{\'{\i}}az}, {H{\'e}brard}, \&
  {Santerne}}]{moutou13}
{Moutou}, C., {Bonomo}, A.~S., {Bruno}, G., {et~al.} 2013, \aap, 558, L6

\bibitem[{{Moutou} {et~al.}(2011){Moutou}, {Mayor}, {Lo Curto},
  {S{\'e}gransan}, {Udry}, {Bouchy}, {Benz}, {Lovis}, {Naef}, {Pepe}, {Queloz},
  {Santos}, \& {Sousa}}]{moutou11}
{Moutou}, C., {Mayor}, M., {Lo Curto}, G., {et~al.} 2011, \aap, 527, A63

\bibitem[{{Moutou} {et~al.}(2009){Moutou}, {Mayor}, {Lo Curto}, {Udry},
  {Bouchy}, {Benz}, {Lovis}, {Naef}, {Pepe}, {Queloz}, \& {Santos}}]{moutou09}
{Moutou}, C., {Mayor}, M., {Lo Curto}, G., {et~al.} 2009, \aap, 496, 513

\bibitem[{{Nidever} {et~al.}(2002){Nidever}, {Marcy}, {Butler}, {Fischer}, \&
  {Vogt}}]{nidever02}
{Nidever}, D.~L., {Marcy}, G.~W., {Butler}, R.~P., {Fischer}, D.~A., \& {Vogt},
  S.~S. 2002, \apjs, 141, 503

\bibitem[{{Parviainen} {et~al.}(2014){Parviainen}, {Gandolfi}, {Deleuil},
  {Moutou}, {Deeg}, {Ferraz-Mello}, {Samuel}, {Csizmadia}, {Pasternacki},
  {Wuchterl}, {Havel}, {Fridlund}, {Angus}, {Tingley}, {Grziwa}, {Korth},
  {Aigrain}, {Almenara}, {Alonso}, {Baglin}, {Barros}, {Bord{\'e}}, {Bouchy},
  {Cabrera}, {D{\'{\i}}az}, {Dvorak}, {Erikson}, {Guillot}, {Hatzes},
  {H{\'e}brard}, {Mazeh}, {Montagnier}, {Ofir}, {Ollivier}, {P{\"a}tzold},
  {Rauer}, {Rouan}, {Santerne}, \& {Schneider}}]{parviainen14}
{Parviainen}, H., {Gandolfi}, D., {Deleuil}, M., {et~al.} 2014, \aap, 562, A140

\bibitem[{{Patel} {et~al.}(2007){Patel}, {Vogt}, {Marcy}, {Johnson}, {Fischer},
  {Wright}, \& {Butler}}]{patel07}
{Patel}, S.~G., {Vogt}, S.~S., {Marcy}, G.~W., {et~al.} 2007, \apj, 665, 744

\bibitem[{{Perrier} {et~al.}(2003){Perrier}, {Sivan}, {Naef}, {Beuzit},
  {Mayor}, {Queloz}, \& {Udry}}]{perrier03}
{Perrier}, C., {Sivan}, J.-P., {Naef}, D., {et~al.} 2003, \aap, 410, 1039

\bibitem[{{Perruchot} {et~al.}(2011){Perruchot}, {Bouchy}, {Chazelas},
  {D{\'{\i}}az}, {H{\'e}brard}, {Arnaud}, {Arnold}, {Avila}, {Delfosse},
  {Boisse}, {Moreaux}, {Pepe}, {Richaud}, {Santerne}, {Sottile}, \&
  {T{\'e}zier}}]{perruchot11}
{Perruchot}, S., {Bouchy}, F., {Chazelas}, B., {et~al.} 2011, in Society of
  Photo-Optical Instrumentation Engineers (SPIE) Conference Series, Vol. 8151,
  Society of Photo-Optical Instrumentation Engineers (SPIE) Conference Series,
  15

\bibitem[{{Perruchot} {et~al.}(2008){Perruchot}, {Kohler}, {Bouchy}, {Richaud},
  {Richaud}, {Moreaux}, {Merzougui}, {Sottile}, {Hill}, {Knispel}, {Regal},
  {Meunier}, {Ilovaisky}, {Le Coroller}, {Gillet}, {Schmitt}, {Pepe}, {Fleury},
  {Sosnowska}, {Vors}, {M{\'e}gevand}, {Blanc}, {Carol}, {Point}, {Laloge}, \&
  {Brunel}}]{perruchot08}
{Perruchot}, S., {Kohler}, D., {Bouchy}, F., {et~al.} 2008, in Society of
  Photo-Optical Instrumentation Engineers (SPIE) Conference Series, Vol. 7014,
  Society of Photo-Optical Instrumentation Engineers (SPIE) Conference Series,
  0

\bibitem[{{Perryman} {et~al.}(1997){Perryman}, {Lindegren}, {Kovalevsky},
  {Hoeg}, {Bastian}, {Bernacca}, {Cr{\'e}z{\'e}}, {Donati}, {Grenon},
  {Grewing}, {van Leeuwen}, {van der Marel}, {Mignard}, {Murray}, {Le Poole},
  {Schrijver}, {Turon}, {Arenou}, {Froeschl{\'e}}, \&
  {Petersen}}]{Perryman:1997kx}
{Perryman}, M.~A.~C., {Lindegren}, L., {Kovalevsky}, J., {et~al.} 1997, \aap,
  323, L49

\bibitem[{{Pilyavsky} {et~al.}(2011){Pilyavsky}, {Mahadevan}, {Kane}, {Howard},
  {Ciardi}, {de Pree}, {Dragomir}, {Fischer}, {Henry}, {Jensen}, {Laughlin},
  {Marlowe}, {Rabus}, {von Braun}, {Wright}, \& {Wang}}]{pilyavsky11}
{Pilyavsky}, G., {Mahadevan}, S., {Kane}, S.~R., {et~al.} 2011, \apj, 743, 162

\bibitem[{{Queloz} {et~al.}(2001){Queloz}, {Henry}, {Sivan}, {Baliunas},
  {Beuzit}, {Donahue}, {Mayor}, {Naef}, {Perrier}, \& {Udry}}]{queloz01}
{Queloz}, D., {Henry}, G.~W., {Sivan}, J.~P., {et~al.} 2001, \aap, 379, 279

\bibitem[{{Reffert} \& {Quirrenbach}(2006)}]{reffert06}
{Reffert}, S. \& {Quirrenbach}, A. 2006, \aap, 449, 699

\bibitem[{{Riddle} {et~al.}(2015){Riddle}, {Tokovinin}, {Mason}, {Hartkopf},
  {Roberts}, {Baranec}, {Law}, {Bui}, {Burse}, {Das}, {Dekany}, {Kulkarni},
  {Punnadi}, {Ramaprakash}, \& {Tendulkar}}]{riddle15}
{Riddle}, R.~L., {Tokovinin}, A., {Mason}, B.~D., {et~al.} 2015, \apj, 799, 4

\bibitem[{{Saar} {et~al.}(1998){Saar}, {Butler}, \& {Marcy}}]{saar98}
{Saar}, S.~H., {Butler}, R.~P., \& {Marcy}, G.~W. 1998, \apjl, 498, L153

\bibitem[{{Saar} \& {Donahue}(1997)}]{saar97}
{Saar}, S.~H. \& {Donahue}, R.~A. 1997, \apj, 485, 319

\bibitem[{{Sahlmann} \& {Fekel}(2013)}]{sahlmann13}
{Sahlmann}, J. \& {Fekel}, F.~C. 2013, \aap, 556, A145

\bibitem[{{Sahlmann} {et~al.}(2011{\natexlab{a}}){Sahlmann}, {Lovis}, {Queloz},
  \& {S{\'e}gransan}}]{sahlmann11b}
{Sahlmann}, J., {Lovis}, C., {Queloz}, D., \& {S{\'e}gransan}, D.
  2011{\natexlab{a}}, \aap, 528, L8

\bibitem[{{Sahlmann} {et~al.}(2011{\natexlab{b}}){Sahlmann}, {S{\'e}gransan},
  {Queloz}, {Udry}, {Santos}, {Marmier}, {Mayor}, {Naef}, {Pepe}, \&
  {Zucker}}]{sahlmann11}
{Sahlmann}, J., {S{\'e}gransan}, D., {Queloz}, D., {et~al.} 2011{\natexlab{b}},
  \aap, 525, A95

\bibitem[{{Santerne} {et~al.}(2015){Santerne}, {D{\'{\i}}az}, {Almenara},
  {Bouchy}, {Deleuil}, {Figueira}, {H{\'e}brard}, {Moutou}, {Rodionov}, \&
  {Santos}}]{santerne15}
{Santerne}, A., {D{\'{\i}}az}, R.~F., {Almenara}, J.-M., {et~al.} 2015, \mnras,
  451, 2337

\bibitem[{{Santos} {et~al.}(2005){Santos}, {Israelian}, {Mayor}, {Bento},
  {Almeida}, {Sousa}, \& {Ecuvillon}}]{santos05}
{Santos}, N.~C., {Israelian}, G., {Mayor}, M., {et~al.} 2005, \aap, 437, 1127

\bibitem[{{Santos} {et~al.}(2010){Santos}, {Mayor}, {Benz}, {Bouchy},
  {Figueira}, {Lo Curto}, {Lovis}, {Melo}, {Moutou}, {Naef}, {Pepe}, {Queloz},
  {Sousa}, \& {Udry}}]{santos10}
{Santos}, N.~C., {Mayor}, M., {Benz}, W., {et~al.} 2010, \aap, 512, A47

\bibitem[{{Santos} {et~al.}(2000){Santos}, {Mayor}, {Naef}, {Pepe}, {Queloz},
  {Udry}, \& {Blecha}}]{santos00}
{Santos}, N.~C., {Mayor}, M., {Naef}, D., {et~al.} 2000, \aap, 361, 265

\bibitem[{{Santos} {et~al.}(2013){Santos}, {Sousa}, {Mortier}, {Neves},
  {Adibekyan}, {Tsantaki}, {Delgado Mena}, {Bonfils}, {Israelian}, {Mayor}, \&
  {Udry}}]{santos13}
{Santos}, N.~C., {Sousa}, S.~G., {Mortier}, A., {et~al.} 2013, \aap, 556, A150

\bibitem[{{Sato} {et~al.}(2009){Sato}, {Fischer}, {Ida}, {Harakawa}, {Omiya},
  {Johnson}, {Marcy}, {Toyota}, {Hori}, {Isaacson}, {Howard}, \&
  {Peek}}]{sato09}
{Sato}, B., {Fischer}, D.~A., {Ida}, S., {et~al.} 2009, \apj, 703, 671

\bibitem[{{Sivan} {et~al.}(2004){Sivan}, {Mayor}, {Naef}, {Queloz}, {Udry},
  {Perrier-Bellet}, \& {Beuzit}}]{sivan04}
{Sivan}, J.-P., {Mayor}, M., {Naef}, D., {et~al.} 2004, in IAU Symposium, Vol.
  202, Planetary Systems in the Universe, ed. A.~{Penny}, 124

\bibitem[{{Siverd} {et~al.}(2012){Siverd}, {Beatty}, {Pepper}, {Eastman},
  {Collins}, {Bieryla}, {Latham}, {Buchhave}, {Jensen}, {Crepp}, {Street},
  {Stassun}, {Gaudi}, {Berlind}, {Calkins}, {DePoy}, {Esquerdo}, {Fulton},
  {F{\H u}r{\'e}sz}, {Geary}, {Gould}, {Hebb}, {Kielkopf}, {Marshall}, {Pogge},
  {Stanek}, {Stefanik}, {Szentgyorgyi}, {Trueblood}, {Trueblood}, {Stutz}, \&
  {van Saders}}]{siverd12}
{Siverd}, R.~J., {Beatty}, T.~G., {Pepper}, J., {et~al.} 2012, \apj, 761, 123

\bibitem[{{Sneden}(1973)}]{sneden73}
{Sneden}, C.~A. 1973, PhD thesis, THE UNIVERSITY OF TEXAS AT AUSTIN.

\bibitem[{{Sousa} {et~al.}(2011){Sousa}, {Santos}, {Israelian}, {Lovis},
  {Mayor}, {Silva}, \& {Udry}}]{sausa11}
{Sousa}, S.~G., {Santos}, N.~C., {Israelian}, G., {et~al.} 2011, \aap, 526, A99

\bibitem[{{Southworth} {et~al.}(2012){Southworth}, {Bruni}, {Mancini}, \&
  {Gregorio}}]{southworth12}
{Southworth}, J., {Bruni}, I., {Mancini}, L., \& {Gregorio}, J. 2012, \mnras,
  420, 2580

\bibitem[{{Sozzetti} \& {Desidera}(2010)}]{sozzetti10}
{Sozzetti}, A. \& {Desidera}, S. 2010, \aap, 509, A103

\bibitem[{{Spiegel} {et~al.}(2011){Spiegel}, {Burrows}, \&
  {Milsom}}]{spiegel11}
{Spiegel}, D.~S., {Burrows}, A., \& {Milsom}, J.~A. 2011, \apj, 727, 57

\bibitem[{{Tamuz} {et~al.}(2008){Tamuz}, {S{\'e}gransan}, {Udry}, {Mayor},
  {Eggenberger}, {Naef}, {Pepe}, {Queloz}, {Santos}, {Demory}, {Figuera},
  {Marmier}, \& {Montagnier}}]{tamuz08}
{Tamuz}, O., {S{\'e}gransan}, D., {Udry}, S., {et~al.} 2008, \aap, 480, L33

\bibitem[{{Triaud} {et~al.}(2010){Triaud}, {Collier Cameron}, {Queloz},
  {Anderson}, {Gillon}, {Hebb}, {Hellier}, {Loeillet}, {Maxted}, {Mayor},
  {Pepe}, {Pollacco}, {S{\'e}gransan}, {Smalley}, {Udry}, {West}, \&
  {Wheatley}}]{triaud10}
{Triaud}, A.~H.~M.~J., {Collier Cameron}, A., {Queloz}, D., {et~al.} 2010,
  \aap, 524, A25

\bibitem[{{Udry} {et~al.}(2002){Udry}, {Mayor}, {Naef}, {Pepe}, {Queloz},
  {Santos}, \& {Burnet}}]{udry02}
{Udry}, S., {Mayor}, M., {Naef}, D., {et~al.} 2002, \aap, 390, 267

\bibitem[{{Valenti} \& {Fischer}(2005)}]{valenti05}
{Valenti}, J.~A. \& {Fischer}, D.~A. 2005, \apjs, 159, 141

\bibitem[{{van Leeuwen}(2007)}]{:2007kx}
{van Leeuwen}, F., ed. 2007, Astrophysics and Space Science Library, Vol. 350,
  {Hipparcos, the New Reduction of the Raw Data}

\bibitem[{{Winn} {et~al.}(2008){Winn}, {Holman}, {Torres}, {McCullough},
  {Johns-Krull}, {Latham}, {Shporer}, {Mazeh}, {Garcia-Melendo}, {Foote},
  {Esquerdo}, \& {Everett}}]{winn08}
{Winn}, J.~N., {Holman}, M.~J., {Torres}, G., {et~al.} 2008, \apj, 683, 1076

\bibitem[{{Winn} {et~al.}(2009){Winn}, {Johnson}, {Fabrycky}, {Howard},
  {Marcy}, {Narita}, {Crossfield}, {Suto}, {Turner}, {Esquerdo}, \&
  {Holman}}]{winn09}
{Winn}, J.~N., {Johnson}, J.~A., {Fabrycky}, D., {et~al.} 2009, \apj, 700, 302

\bibitem[{{Winn} {et~al.}(2010){Winn}, {Johnson}, {Howard}, {Marcy}, {Bakos},
  {Hartman}, {Torres}, {Albrecht}, \& {Narita}}]{winn10}
{Winn}, J.~N., {Johnson}, J.~A., {Howard}, A.~W., {et~al.} 2010, \apj, 718, 575

\bibitem[{{Wittenmyer} {et~al.}(2009{\natexlab{a}}){Wittenmyer}, {Endl},
  {Cochran}, {Levison}, \& {Henry}}]{wittenmyer09b}
{Wittenmyer}, R.~A., {Endl}, M., {Cochran}, W.~D., {Levison}, H.~F., \&
  {Henry}, G.~W. 2009{\natexlab{a}}, \apjs, 182, 97

\bibitem[{{Wittenmyer} {et~al.}(2009{\natexlab{b}}){Wittenmyer}, {Endl},
  {Cochran}, {Ram{\'{\i}}rez}, {Reffert}, {MacQueen}, \&
  {Shetrone}}]{wittenmyer09a}
{Wittenmyer}, R.~A., {Endl}, M., {Cochran}, W.~D., {et~al.} 2009{\natexlab{b}},
  \aj, 137, 3529

\bibitem[{{Wright} {et~al.}(2009){Wright}, {Upadhyay}, {Marcy}, {Fischer},
  {Ford}, \& {Johnson}}]{wright09}
{Wright}, J.~T., {Upadhyay}, S., {Marcy}, G.~W., {et~al.} 2009, \apj, 693, 1084

\bibitem[{{Zucker} \& {Mazeh}(2001)}]{zucker01}
{Zucker}, S. \& {Mazeh}, T. 2001, \apj, 562, 549

\end{thebibliography}

\Online

\begin{figure*} 
\begin{center}
\includegraphics[width=0.45\textwidth]{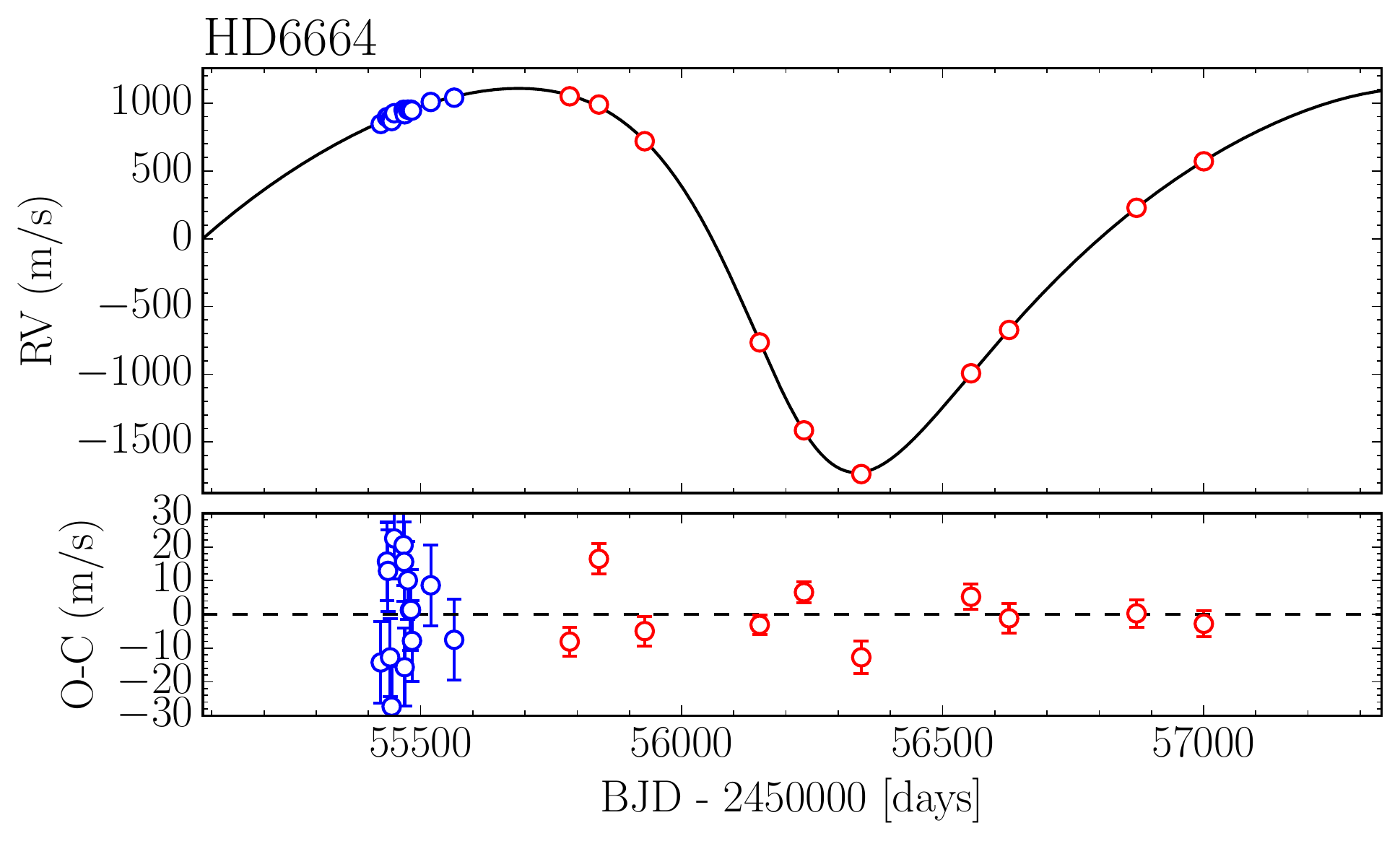}
\hspace*{.25in}
\includegraphics[width=0.45\textwidth]{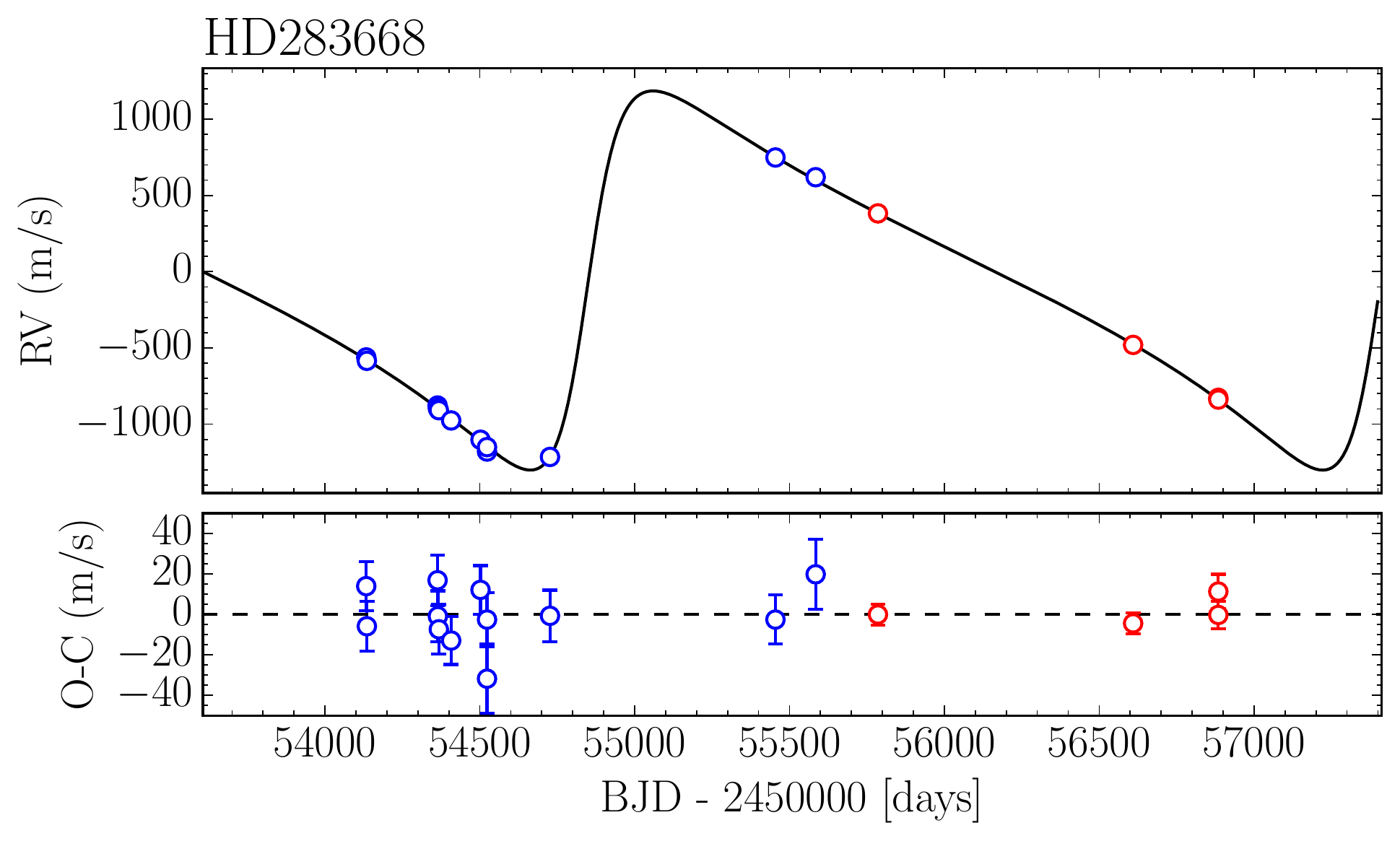}

\vspace*{.25in}

\includegraphics[width=0.45\textwidth]{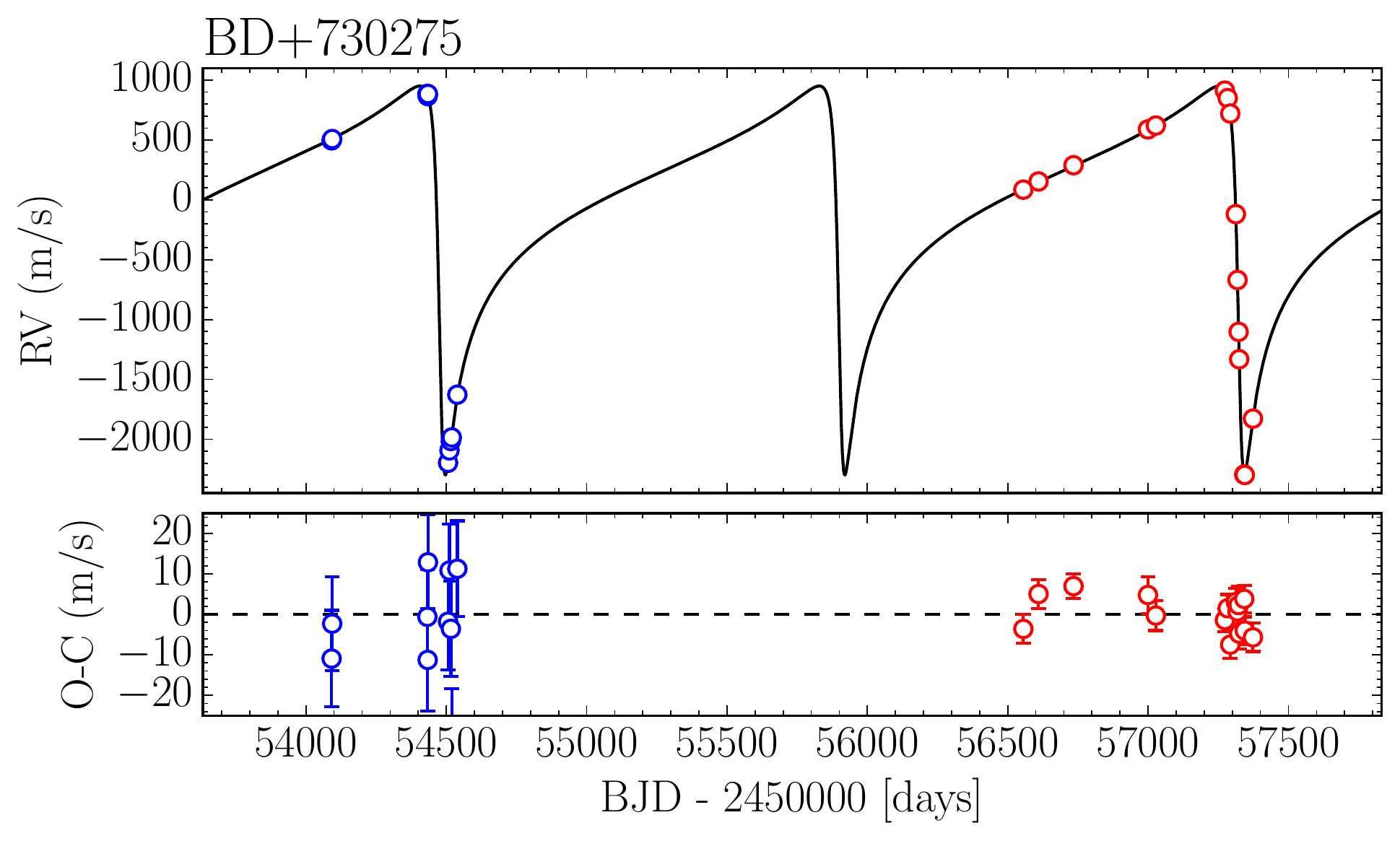}
\hspace*{.25in}
\includegraphics[width=0.45\textwidth]{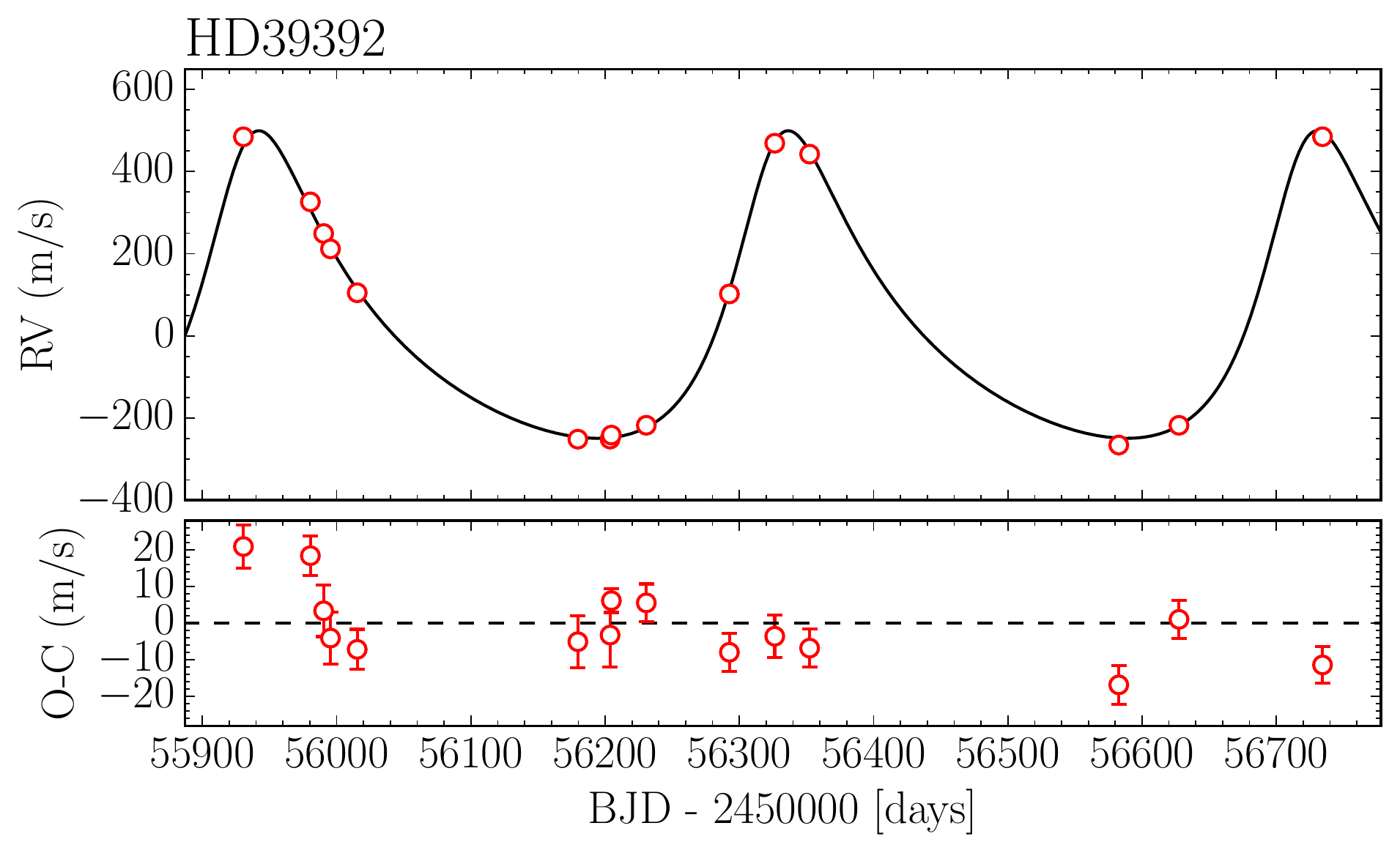}

\vspace*{.25in}

\includegraphics[width=0.45\textwidth]{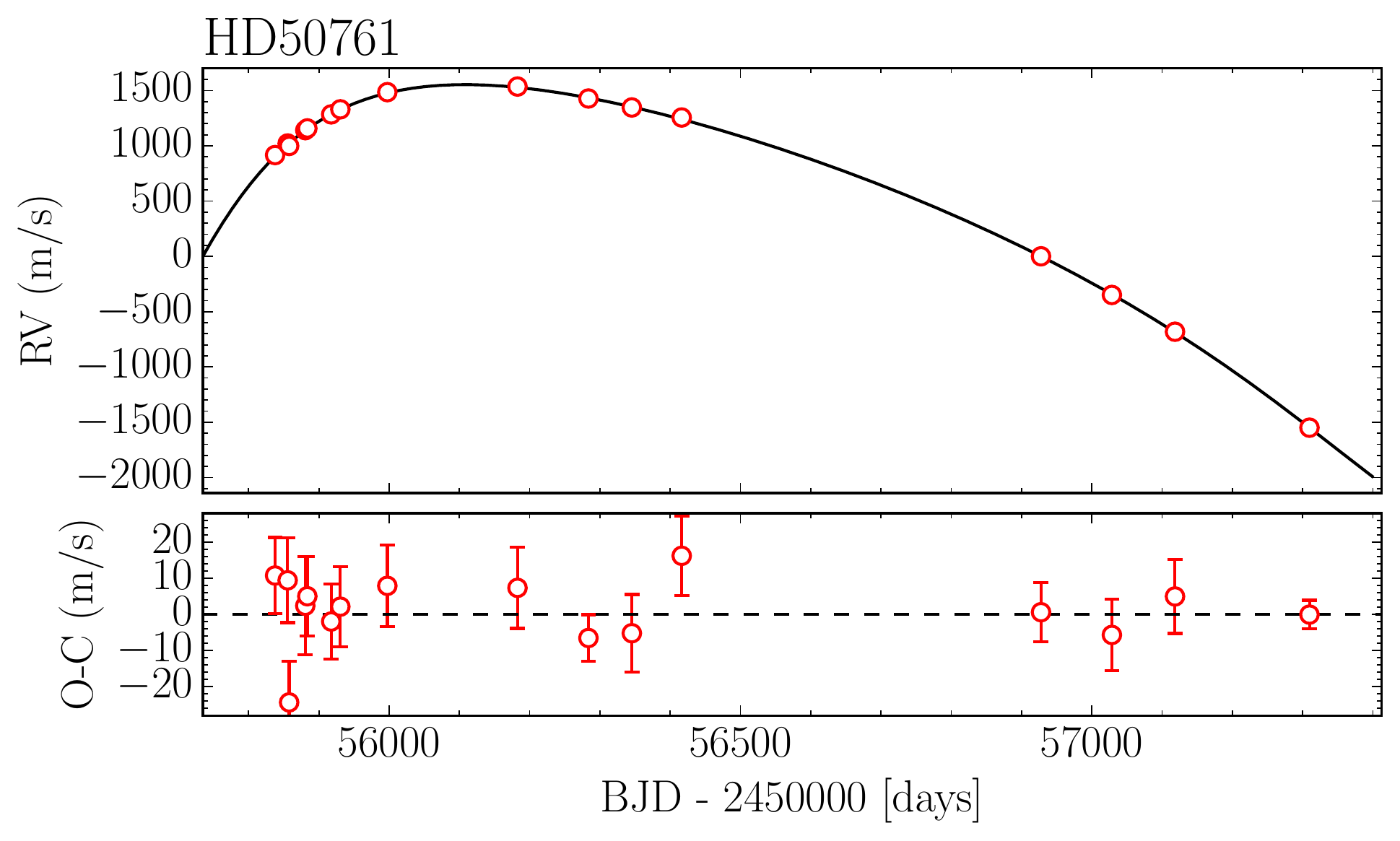}
\hspace*{.25in}
\includegraphics[width=0.45\textwidth]{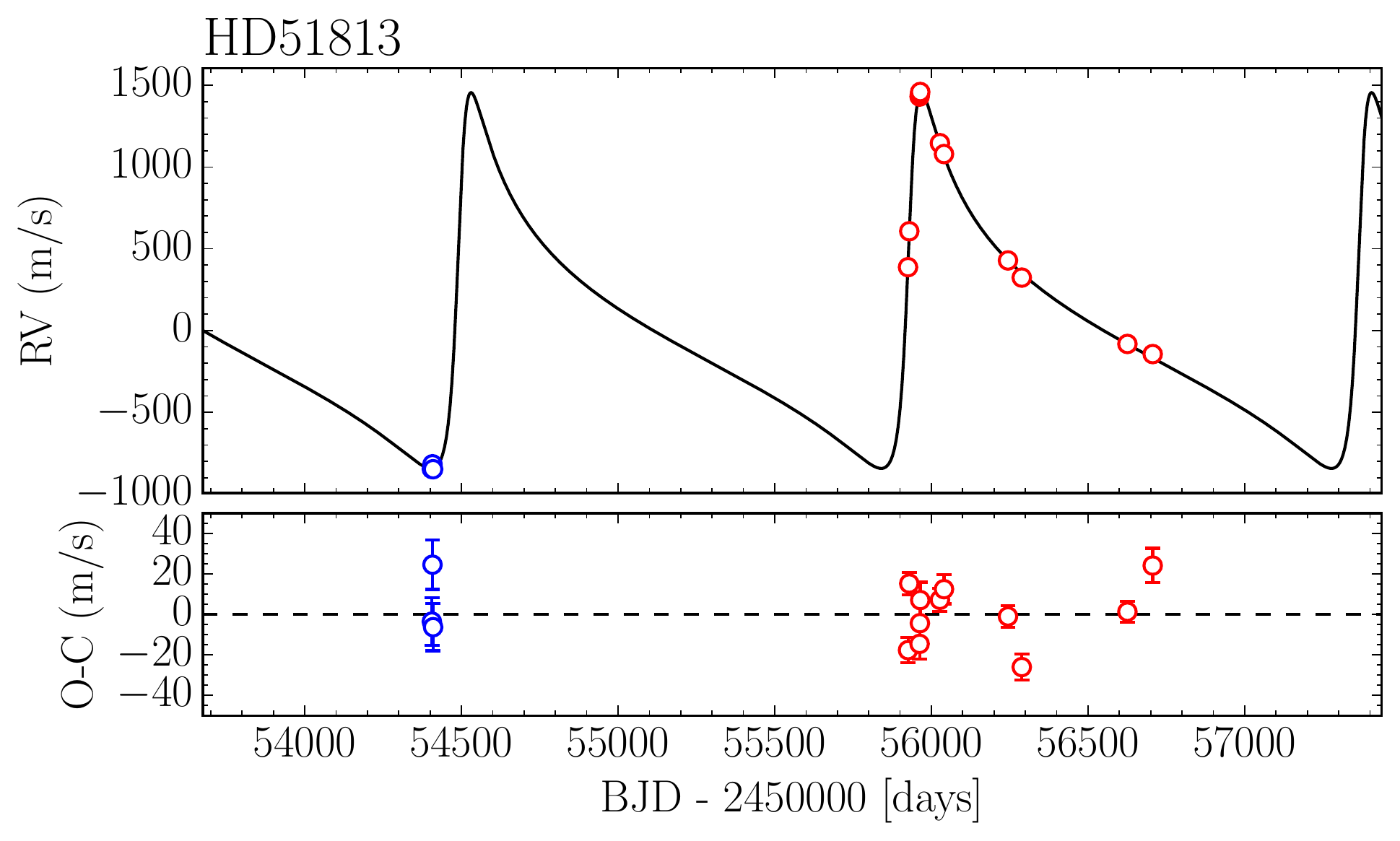}

\vspace*{.25in}

\includegraphics[width=0.45\textwidth]{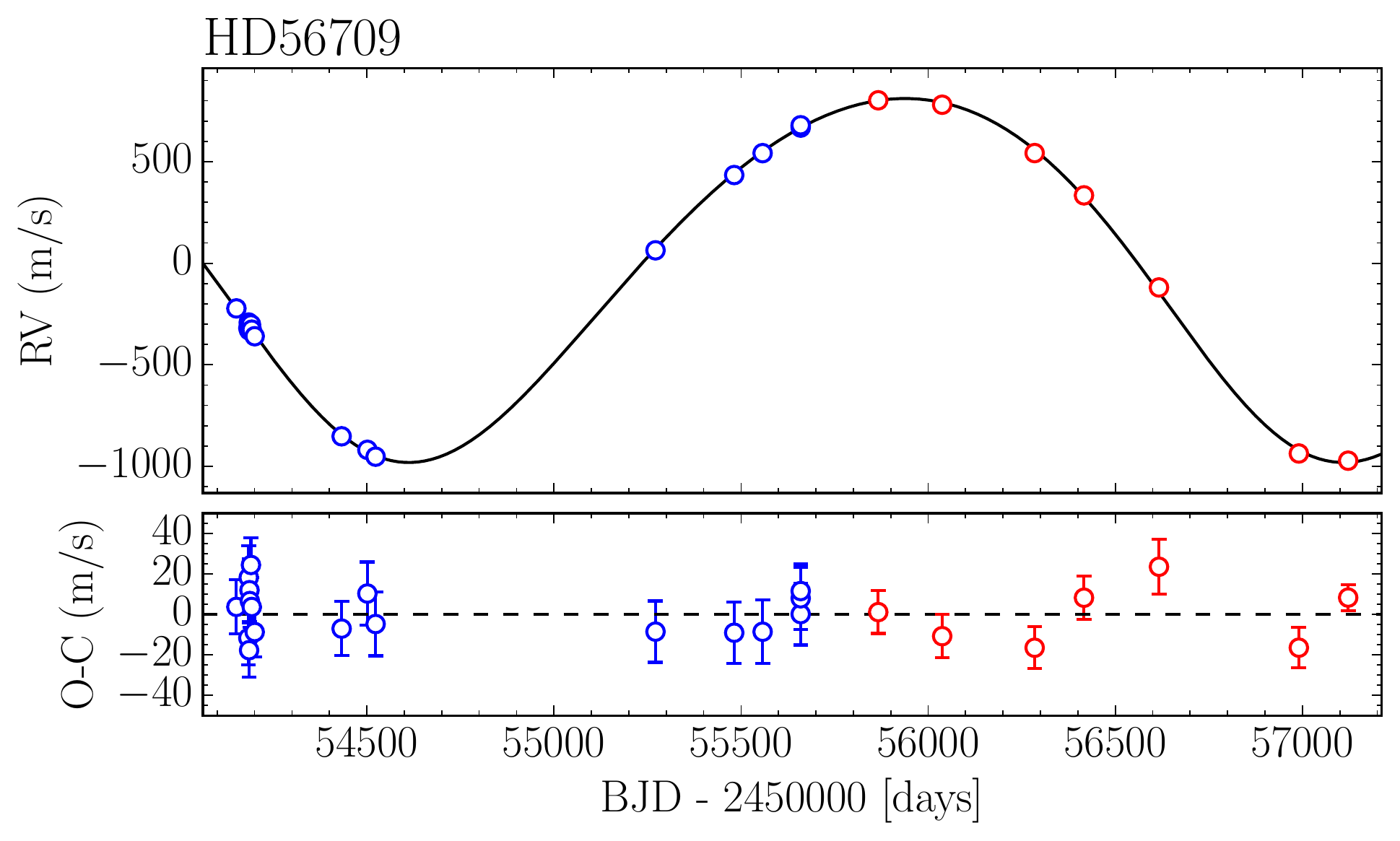}
\hspace*{.25in}
\includegraphics[width=0.45\textwidth]{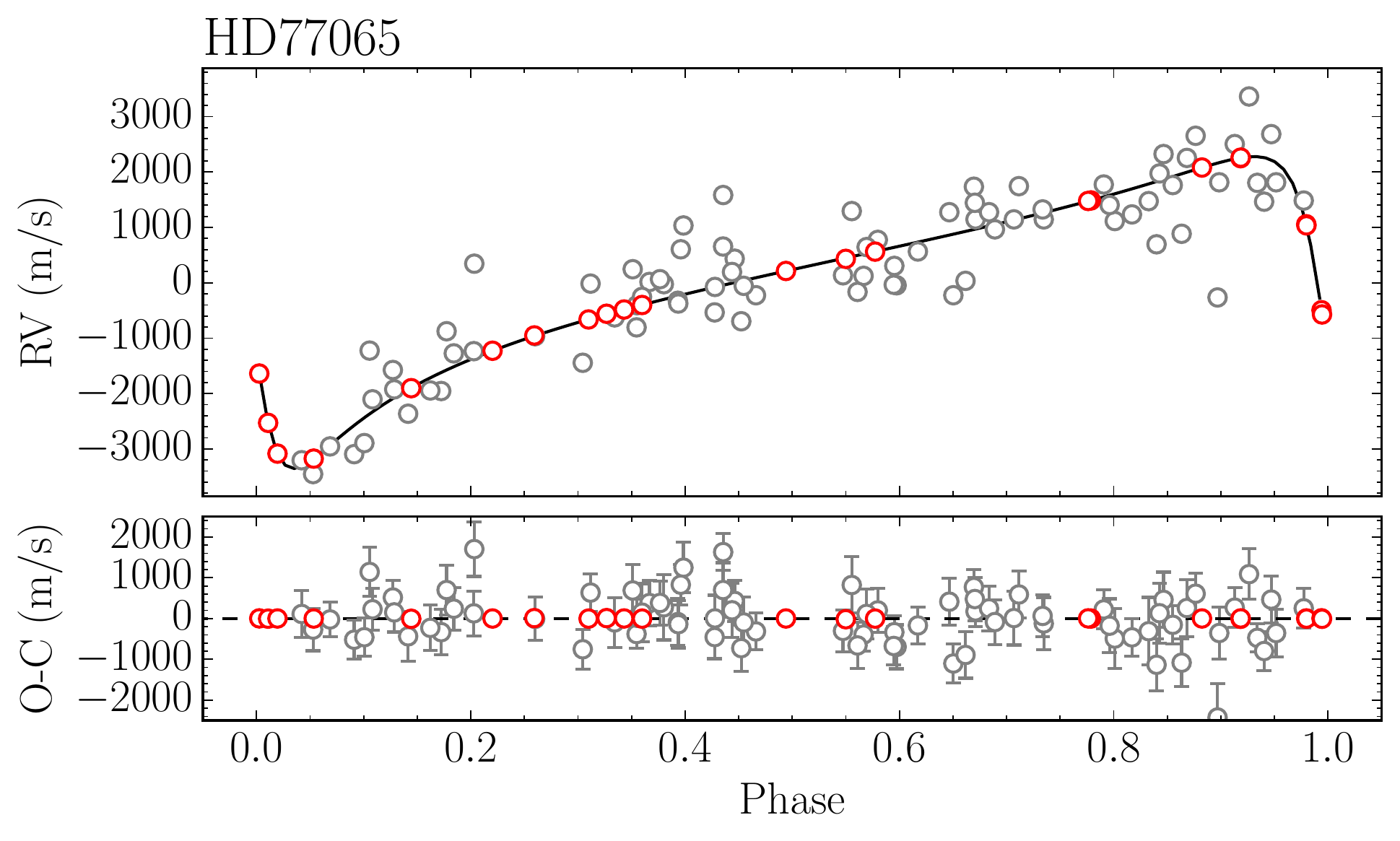}

\end{center}

\end{figure*}

\newpage
\begin{figure*} 
\begin{center}

\includegraphics[width=0.45\textwidth]{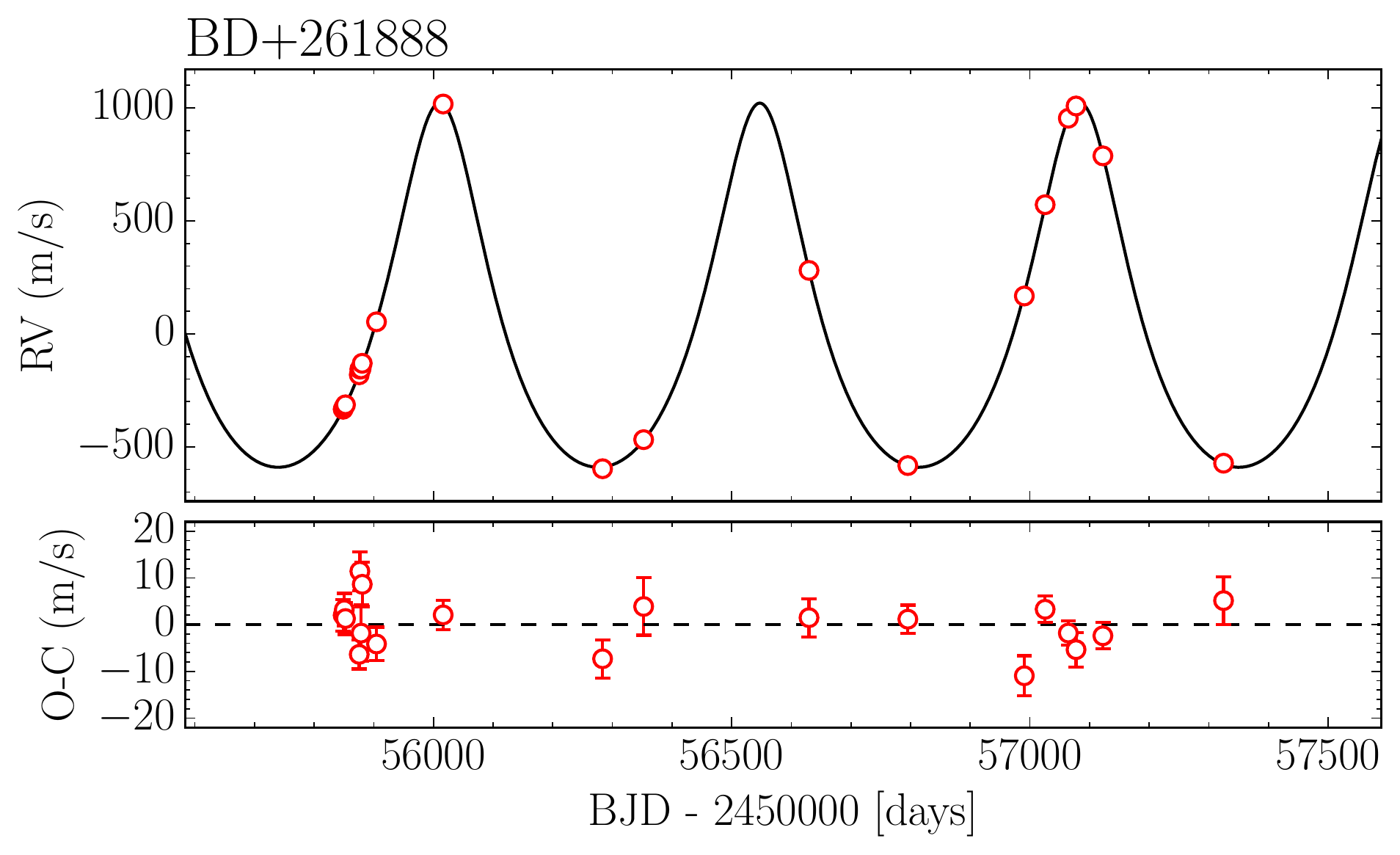}
\hspace*{.25in}
\includegraphics[width=0.45\textwidth]{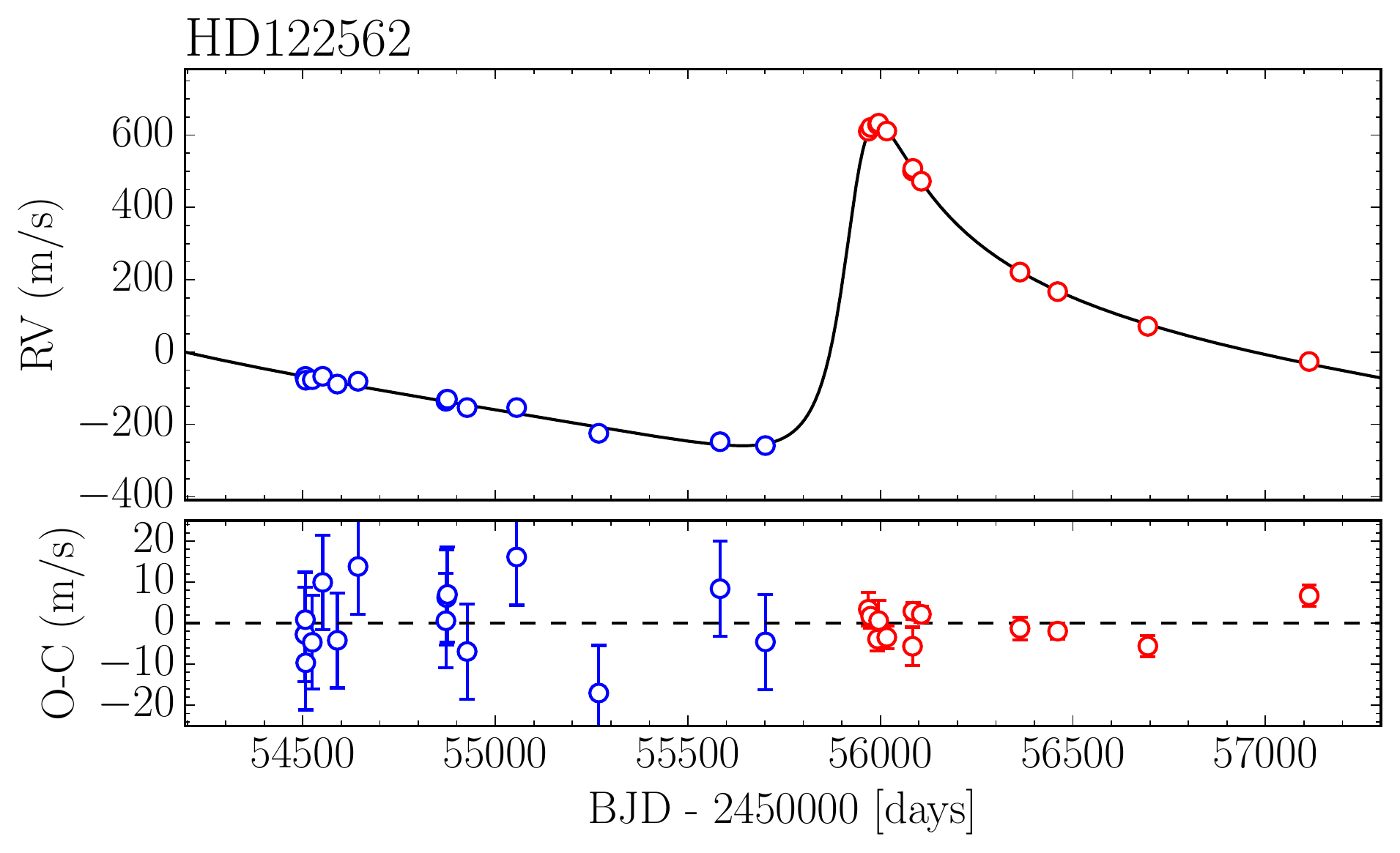}

\vspace*{.25in}

\includegraphics[width=0.45\textwidth]{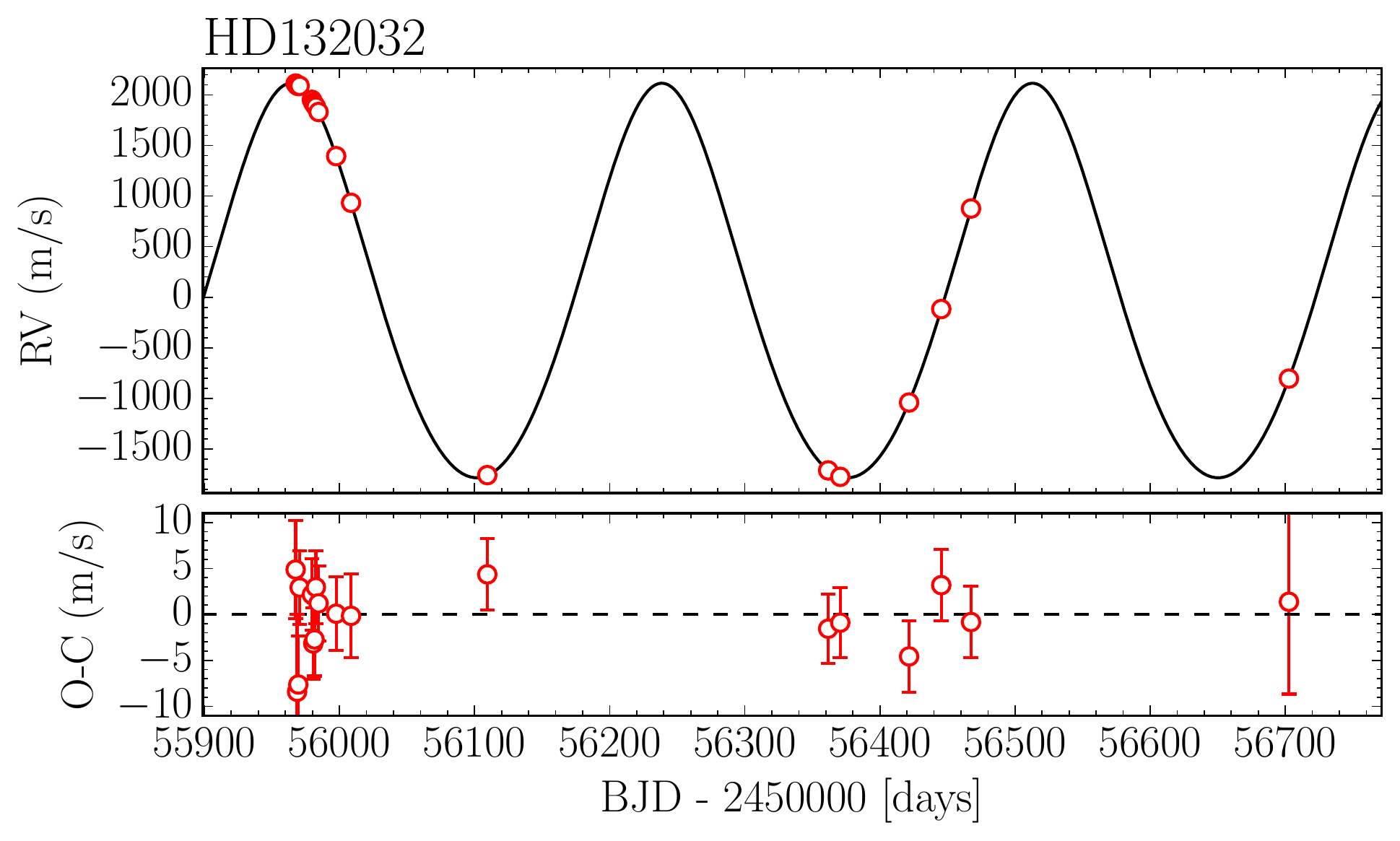}
\hspace*{.25in}
\includegraphics[width=0.45\textwidth]{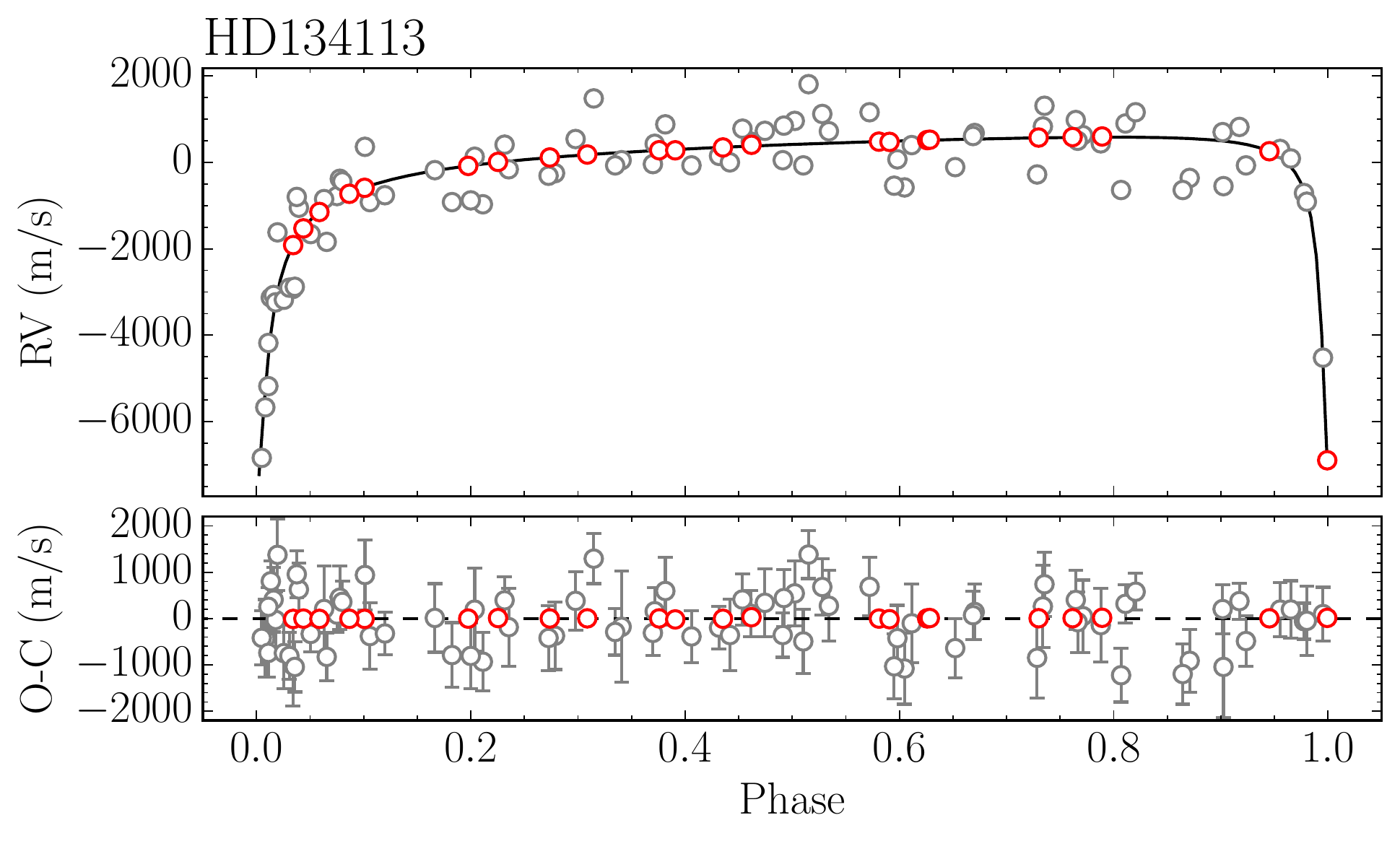}

\vspace*{.25in}

\includegraphics[width=0.45\textwidth]{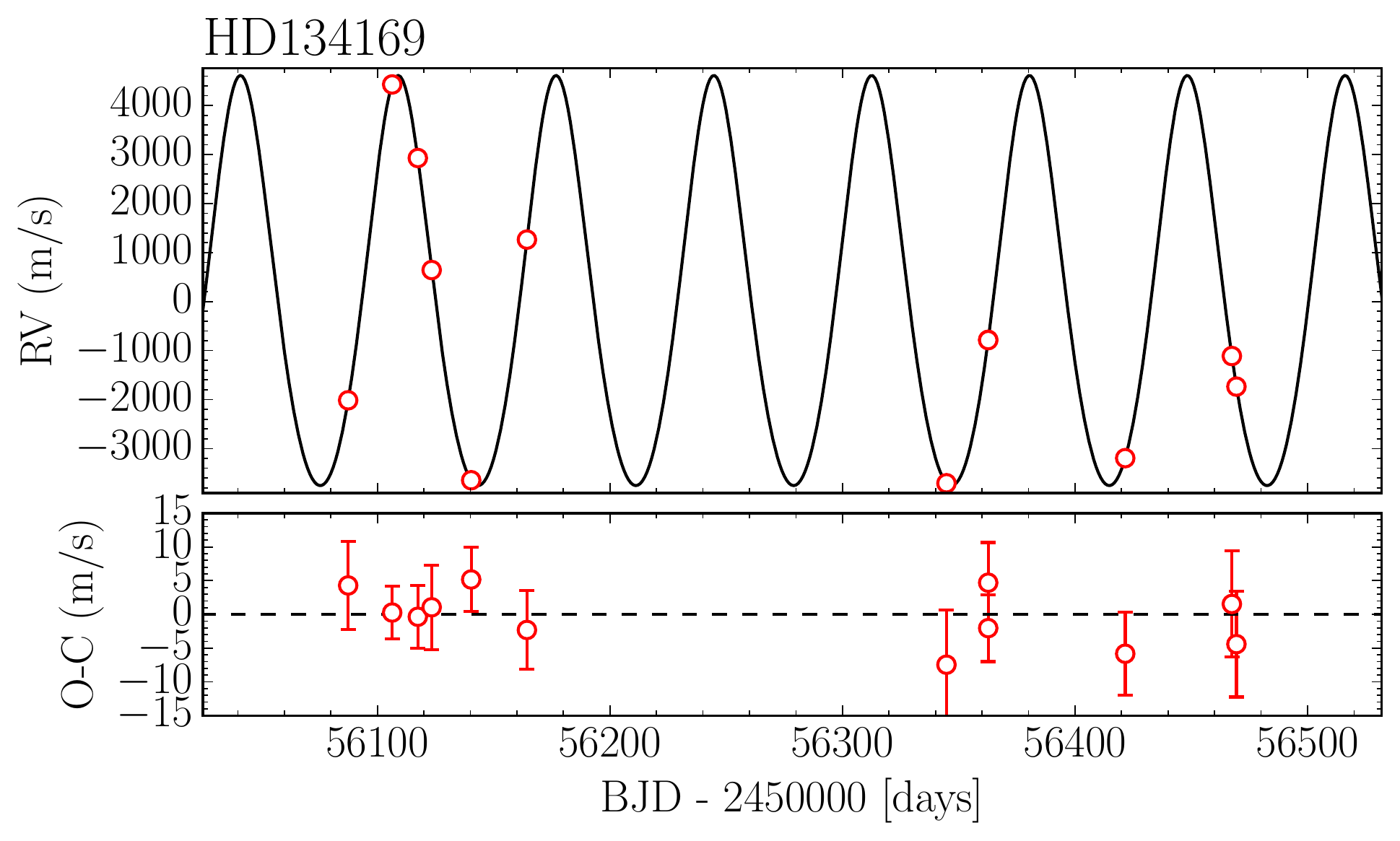}
\hspace*{.25in}
\includegraphics[width=0.45\textwidth]{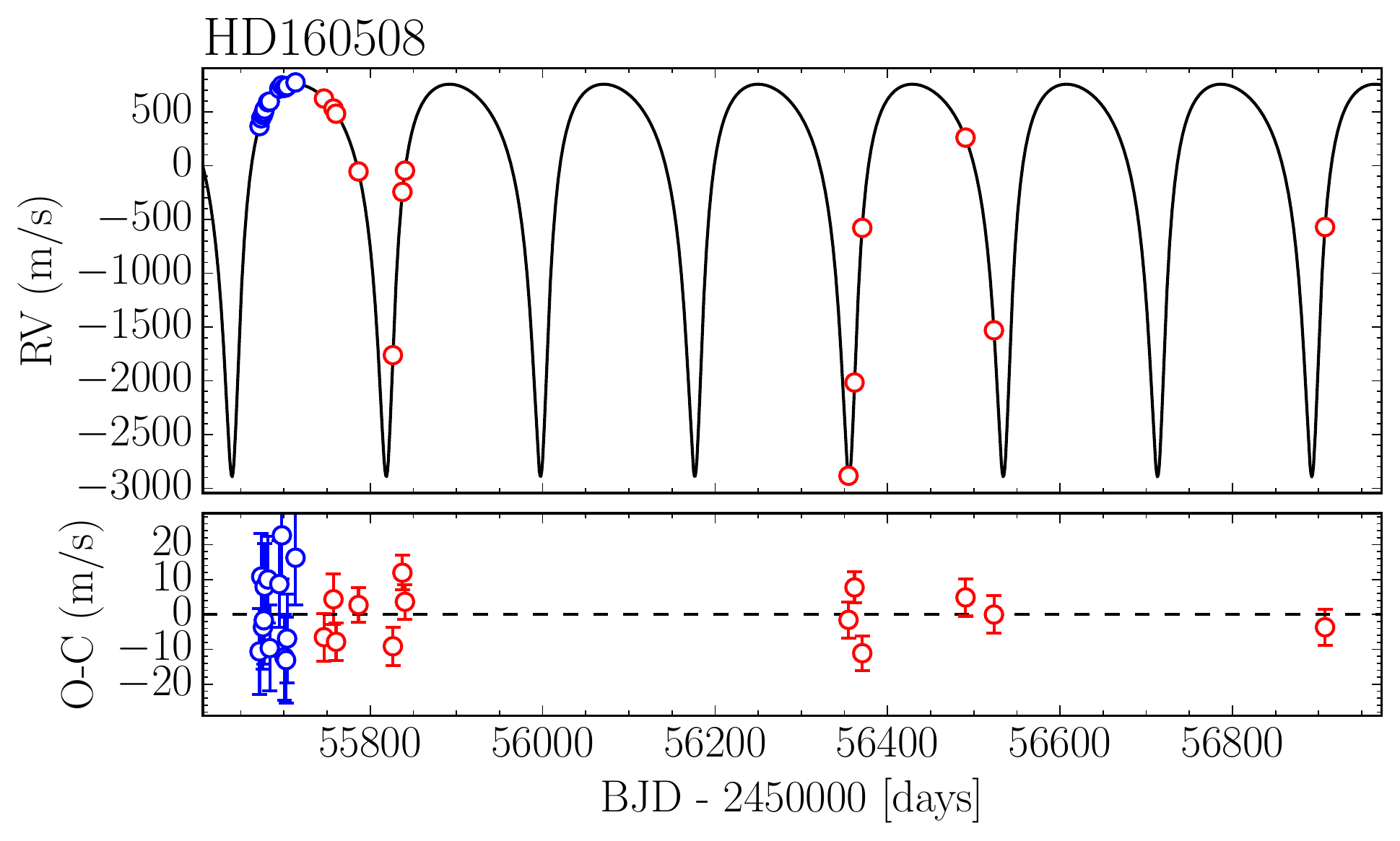}

\end{center}
\end{figure*}

\begin{landscape}
\begin{deluxetable}{lccccccccccc}

\tabletypesize{\tiny}
\tablecolumns{10} 
\tablewidth{0pt}
 \tablecaption{Brown dwarf candidates to solar-type stars with (projected) masses between 10 and 80~$\textrm{M}_\textrm{Jup}$ and periods $<10^5$ days.
 \label{tab:appendix_table}}

 \tablehead{

  \vspace{-0.1cm}	& \colhead{$m_2$} 			& $\textrm{m}_2 \sin i$	& Period & \vspace{-0.2cm} & M$_\star$ &\colhead{\teff} & \colhead{$\log \mathrm{g}$} & [Fe/H] &  &  &\\
   Name			& 					& 			&  & Eccentricity  & &  & & &$R_p/R_\star$ & Sp.T. & Reference \\
			& \colhead{[$\textrm{M}_\textrm{Jup}$]} & \colhead{[$\textrm{M}_\textrm{Jup}$]} & \colhead{[days]} & \colhead{}& \colhead{[M$_\sun$]} & \colhead{[K]} & \colhead{[cgs]} & \colhead{[dex]} & \colhead{}} 

\startdata 
BD+24 4697b	&	$							$	&	$	53	\pm	3				$	&	$	145.081	\pm	0.016				$	&	$		0.50048	\pm	0.00043				$	&	$	0.75	\pm	0.05				$	&	$	5077	\pm	32	$	&	$	4.67	\pm	0.07				$	&	$	-0.11	\pm	0.02	$	&	$							$	&	K2		&	1	\\
BD+26 1888b	&	$							$	&	$	26	\pm	2				$	&	$	536.78	\pm	0.25				$	&	$		0.2675	\pm	0.0016				$	&	$	0.76	\pm	0.08				$	&	$	4748	\pm	87	$	&	$	4.3	\pm	0.2				$	&	$	0.02	\pm	0.04	$	&	$							$	&	K7		&	1	\\
BD+48 2155b	&	$							$	&	$	62.6	\pm	0.6				$	&	$	90.2695	^{+	0.0188	}_{-	0.0187	}	$	&	$		0.4375	\pm	0.0040				$	&	$	1.11	\pm	0.08				$	&	$	6004	\pm	34	$	&	$	4.53	^{+	0.15	}_{-	0.13	}	$	&	$	0.04	\pm	0.06	$	&	$							$	&	G0		&	2	\\
CoRoT-15b	&	$	63.3	\pm	4.1				$	&	$							$	&	$	3.06036	\pm	0.00003				$	&	$		0		~\mathrm{(fixed)}				$	&	$	1.32	\pm	0.12				$	&	$	6350	\pm	200	$	&	$	4.3	\pm	0.2				$	&	$	0.1	\pm	0.2	$	&	$	1.12	^{+	0.30	}_{-	0.15	}	$	&	F7V		&	3	\\
CoRoT-27b	&	$	10.39	\pm	0.55				$	&	$							$	&	$	3.57532	\pm	0.00006				$	&	$	<	0.065						$	&	$	1.05	\pm	0.11				$	&	$	5900	\pm	120	$	&	$	4.4	\pm	0.1				$	&	$	-0.1	\pm	0.1	$	&	$	0.0990	\pm	0.0020				$	&	G2		&	4	\\
CoRoT-33b	&	$	59.0	^{+	1.8	}_{-	1.7	}	$	&	$							$	&	$	5.819143	\pm	0.000018				$	&	$		0.0700		\pm	0.0016				$	&	$	0.86	\pm	0.04				$	&	$	5225	\pm	80	$	&	$	4.4	\pm	0.1				$	&	$	0.44	\pm	0.10	$	&	$	0.12	\pm0.04	$	&	G9V		&	5	\\
CoRoT-3b	&	$	21.7	\pm	1.0				$	&	$							$	&	$	4.25680	\pm	0.00001				$	&	$		0		~\mathrm{(fixed)}				$	&	$	1.37	\pm	0.09				$	&	$	6740	\pm	140	$	&	$	4.22	\pm	0.07				$	&	$	-0.02	\pm	0.06	$	&	$	1.01	\pm	0.07				$	&	F3V		&	6	\\
G 268-114b / HIP 5158b	&	$							$	&	$	15.04	\pm	10.55				$	&	$	9017.76	\pm	3180.74				$	&	$		0.14	\pm	0.10				$	&	$	0.780	\pm	0.021				$	&	$	4962	\pm	89	$	&	$	4.37	\pm	0.20				$	&	$	0.10	\pm	0.07	$	&	$							$	&	K5V		&	7	\\
HAT-P-13c	&	$							$	&	$	14.28	\pm	0.28				$	&	$	446.27	\pm	0.22				$	&	$		0.6616	\pm	0.0054				$	&	$	1.32	\pm	0.20				$	&	$	5640	\pm	90	$	&	$	4.13	\pm	0.04				$	&	$	0.41	\pm	0.08	$	&	$							$	&	G4		&	8, 9, 10	\\
HD\,106270b	&	$							$	&	$	11.0	\pm	0.8				$	&	$	2890	\pm	390				$	&	$		0.402	\pm	0.054				$	&	$	1.320	\pm	0.092				$	&	$	5638	\pm	44	$	&	$	3.90	\pm	0.06				$	&	$	0.08	\pm	0.03	$	&	$							$	&	G5		&	11	\\
HD\,114762b	&	$							$	&	$	10.99	\pm	0.09				$	&	$	83.9152	\pm	0.0028				$	&	$		0.3325	\pm	0.0048				$	&	$	0.83	\pm	0.01				$	&	$	5673	\pm	44	$	&	$	4.135	\pm	0.060				$	&	$	-0.774	\pm	0.030	$	&	$							$	&	F9V		&	12, 13, 14, 15, 16	\\
HD\,122562b	&	$							$	&	$	24	\pm	2				$	&	$	2777	^{+	104	}_{-	81	}	$	&	$		0.71	\pm	0.01				$	&	$	1.13	\pm	0.13				$	&	$	4958	\pm	67	$	&	$	3.74	\pm	0.14				$	&	$	0.31	\pm	0.04	$	&	$							$	&	G5		&	1	\\
HD\,131664b	&	$	23	^{+	26	}_{-	5	}	$	&	$							$	&	$	1951	\pm	41				$	&	$		0.638	\pm	0.020				$	&	$	1.10	\pm	0.03				$	&	$	5886	\pm	21	$	&	$	4.44	\pm	0.10				$	&	$	0.32	\pm	0.02	$	&	$							$	&	G3V		&	17, 18	\\
HD\,132032b	&	$							$	&	$	70	\pm	4				$	&	$	274.33	\pm	0.24				$	&	$		0.0844	\pm	0.0024				$	&	$	1.12	\pm	0.08				$	&	$	6035	\pm	33	$	&	$	4.45	\pm	0.05				$	&	$	0.22	\pm	0.03	$	&	$							$	&	G5		&	1, 19	\\
HD\,13507b	&	$							$	&	$	67	^{+	8	}_{-	9	}	$	&	$	4890	^{+	209	}_{-	185	}	$	&	$		0.20	\pm	0.04				$	&	$	1.050	\pm	0.150				$	&	$	5755	\pm	44	$	&	$	4.56	\pm	0.06				$	&	$	0.00	\pm	0.03	$	&	$							$	&	G0		&	1, 14, 20	\\
HD\,136118b	&	$	42	^{+	11	}_{-	18	}	$	&	$	12.00	\pm	0.47				$	&	$	1188	\pm	2				$	&	$		0.34	\pm	0.01				$	&	$	1.24	\pm	0.07				$	&	$	6097	\pm	44	$	&	$	4.05	\pm	0.06				$	&	$	-0.050	\pm	0.030	$	&	$							$	&	F9V		&	13, 14, 21, 22, 23	\\
HD\,137510b	&	$	27.3~\mathrm{to}~59.5						$	&	$	27.3	\pm	1.9				$	&	$	801.30	\pm	0.45				$	&	$		0.3985	\pm	0.0073				$	&	$	1.36	\pm	0.04				$	&	$	6131	\pm	50	$	&	$	4.02	\pm	0.04				$	&	$	0.38	\pm	0.13	$	&	$							$	&	G0 IV		&	13, 14, 24, 25	\\
HD\,14348b	&	$							$	&	$	48.9	\pm	1.6				$	&	$	4740	^{+	6	}_{-	5	}	$	&	$		0.455	\pm	0.004				$	&	$	1.10	\pm	0.08				$	&	$	6237	\pm	47	$	&	$	4.51	\pm	0.07				$	&	$	0.28	\pm	0.03	$	&	$							$	&	F5V		&	26	\\
HD\,14651b	&	$							$	&	$	47.0	\pm	3.4				$	&	$	79.4179	\pm	0.0021				$	&	$		0.4751	\pm	0.0010				$	&	$	0.96	\pm	0.03				$	&	$	5491	\pm	26	$	&	$	4.45	\pm	0.03				$	&	$	-0.04	\pm	0.06	$	&	$							$	&	G6		&	25	\\
HD\,156846b	&	$							$	&	$	10.57	\pm	0.29				$	&	$	359.5546	\pm	0.0071				$	&	$		0.84785	\pm	0.00050				$	&	$	1.35	\pm	0.05				$	&	$	5969	\pm	44	$	&	$	3.92	\pm	0.08				$	&	$	0.17	\pm	0.04	$	&	$							$	&	G0V		&	27, 28	\\
HD\,160508b	&	$							$	&	$	48	\pm	3				$	&	$	178.9049	\pm	0.0074				$	&	$		0.5967	\pm	0.0009				$	&	$	1.25	\pm	0.08				$	&	$	6212	\pm	30	$	&	$	4.16	\pm	0.03				$	&	$	0.01	\pm	0.02	$	&	$							$	&	F8V		&	1	\\
HD\,162020b	&	$							$	&	$	14.4						$	&	$	8.428198	\pm	0.000056				$	&	$		0.277	\pm	0.002				$	&	$	1.450	\pm	0.240				$	&	$	4845	\pm	44	$	&	$	4.90	\pm	0.06				$	&	$	0.00	\pm	0.03	$	&	$							$	&	K3V		&	13, 29	\\
HD\,16760b	&	$							$	&	$	13.13	\pm	0.56				$	&	$	466.47	\pm	0.35				$	&	$		0.084	\pm	0.003				$	&	$	0.78	\pm	0.05				$	&	$	5629	\pm	44	$	&	$	4.47	\pm	0.06				$	&	$	0.067	\pm	0.050	$	&	$							$	&	G5V		&	30, 31	\\
HD 167665b	&	$							$	&	$	50.6	\pm	1.7				$	&	$	4451.8	^{+	27.6	}_{-	27.3	}	$	&	$		0.340	\pm	0.005				$	&	$	1.14	\pm	0.03				$	&	$	6224	\pm	50	$	&	$	4.44	\pm	0.10				$	&	$	-0.05	\pm	0.06	$	&	$							$	&	F9		&	13, 32, 33	\\
HD\,168443c	&	$	34	\pm	9				$	&	$	17.19	\pm	0.21				$	&	$	1749.83	\pm	0.57				$	&	$		0.2113	\pm	0.0017				$	&	$	0.995	^{+	0.019	}_{-	0.050	}	$	&	$	5491	\pm	44	$	&	$	4.07	\pm	0.06				$	&	$	0.04	\pm	0.03	$	&	$							$	&	G6IV		&	29, 35, 36, 37	\\
HD\,174457b	&	$							$	&	$	65.8						$	&	$	840.80	\pm	0.05				$	&	$		0.23	\pm	0.01				$	&	$	1.030	\pm	0.170				$	&	$	5852	\pm	44	$	&	$	4.08	\pm	0.06				$	&	$	-0.15	\pm	0.03	$	&	$							$	&	F8		&	14, 38	\\
HD\,189310b	&	$							$	&	$	25.6	^{+	0.9	}_{-	0.8	}	$	&	$	14.18643	\pm	0.00002				$	&	$		0.359	\pm	0.001				$	&	$	0.83	\pm	0.02				$	&	$	5188	\pm	50	$	&	$	4.49	\pm	0.1				$	&	$	-0.01	\pm	0.06	$	&	$							$	&	K2V		&	33	\\
HD\,190228b	&	$	49.4	\pm	14.8				$	&	$							$	&	$	1136.1	\pm	9.9				$	&	$		0.531	\pm	0.028				$	&	$	1.821	\pm	0.050				$	&	$	5348	\pm	44	$	&	$	3.98	\pm	0.06				$	&	$	-0.25	\pm	0.03	$	&	$							$	&	G5IV		&	13, 20, 22, 35, 39	\\
HD\,191760b	&	$							$	&	$	38.17	\pm	1.02				$	&	$	505.65	\pm	0.42				$	&	$		0.63	\pm	0.01				$	&	$	1.28	^{+	0.02	}_{-	0.10	}	$	&	$	5821	\pm	82	$	&	$	4.13	^{+	0.05	}_{-	0.04	}	$	&	$	0.29	\pm	0.07	$	&	$							$	&	G3IV		&	33, 40	\\
HD\,202206b	&	$							$	&	$	17.5						$	&	$	256.20	\pm	0.03				$	&	$		0.433	\pm	0.001				$	&	$	1.180	\pm	0.200				$	&	$	5788	\pm	44	$	&	$	4.49	\pm	0.06				$	&	$	0.31	\pm	0.03	$	&	$							$	&	G6V		&	13, 29, 41	\\
HD\,209262b	&	$							$	&	$	32.3	^{+	1.6	}_{-	1.5	}	$	&	$	5431	^{+	135	}_{-	95	}	$	&	$		0.347	^{+	0.010	}_{-	0.008	}	$	&	$	1.02	\pm	0.07				$	&	$	5815	\pm	28	$	&	$	4.41	\pm	0.04				$	&	$	0.10	\pm	0.02	$	&	$							$	&	G5V		&	26	\\
HD\,211847b	&	$							$	&	$	19.2	\pm	1.2				$	&	$	7929.4	^{+	1999.1	}_{-	2500.2	}	$	&	$		0.685	^{+	0.068	}_{-	0.067	}	$	&	$	0.94	\pm	0.04				$	&	$	5715	\pm	50	$	&	$	4.49	\pm	0.10				$	&	$	-0.08	\pm	0.06	$	&	$							$	&	G5V		&	33	\\
HD\,217786b	&	$							$	&	$	13.0	\pm	0.8				$	&	$	1319	\pm	4				$	&	$		0.40	\pm	0.05				$	&	$	1.02	\pm	0.03				$	&	$	5966	\pm	65	$	&	$	4.35	\pm	0.11				$	&	$	-0.135	\pm	0.043	$	&	$							$	&	F8V		&	42	\\
HD\,219077b	&	$							$	&	$	10.39	\pm	0.09				$	&	$	5501	^{+	130	}_{-	119	}	$	&	$		0.770	\pm	0.003				$	&	$	1.05	\pm	0.02				$	&	$	5362	\pm	18	$	&	$	4.00	\pm	0.03				$	&	$	-0.13	\pm	0.01	$	&	$							$	&	G8V+		&	43	\\
HD\,22781b	&	$							$	&	$	13.65	\pm	0.97				$	&	$	528.07	\pm	0.14				$	&	$		0.8191	\pm	0.0023				$	&	$	0.75	\pm	0.03				$	&	$	5027	\pm	50	$	&	$	4.60	\pm	0.02				$	&	$	-0.37	\pm	0.12	$	&	$							$	&	K0		&	25	\\
HD\,283668b	&	$							$	&	$	53	\pm	4				$	&	$	2558	\pm	8				$	&	$		0.577	\pm	0.011				$	&	$	0.68	\pm	0.06				$	&	$	4845	\pm	66	$	&	$	4.35	\pm	0.12				$	&	$	-0.75	\pm	0.12	$	&	$							$	&	K3V		&	1	\\
HD\,30246b	&	$							$	&	$	55.1	^{+	20.3	}_{-	8.2	}	$	&	$	990.7	\pm	5.6				$	&	$		0.838	\pm	0.081				$	&	$	1.05	\pm	0.04				$	&	$	5833	\pm	44	$	&	$	4.39	\pm	0.04				$	&	$	0.17	\pm	0.10	$	&	$							$	&	G1		&	25	\\
HD\,30339b	&	$							$	&	$	77.8						$	&	$	15.0778	\pm	0.0003				$	&	$		0.25	\pm	0.01				$	&	$	1.580	\pm	0.330				$	&	$	6074	\pm	44	$	&	$	4.37	\pm	0.06				$	&	$	0.21	\pm	0.03	$	&	$							$	&	F8V		&	13, 38	\\
HD\,38529c	&	$	17.6	^{+	1.5	}_{-	1.2	}	$	&	$	13.99	\pm	0.59				$	&	$	2136.14	\pm	0.29				$	&	$		0.362	\pm	0.002				$	&	$	1.34	^{+	0.02	}_{-	0.05	}	$	&	$	5674	\pm	40	$	&	$	3.94	\pm	0.12				$	&	$	0.40	\pm	0.05	$	&	$							$	&	G4V		&	13, 22, 36, 44, 45, 46	\\
HD\,39091b / * pi. Men b	&	$							$	&	$	10.27	\pm	0.84				$	&	$	2151	\pm	85				$	&	$		0.6405	\pm	0.0072				$	&	$	1.118	\pm	0.088				$	&	$	5950	\pm	44	$	&	$	4.36	\pm	0.06				$	&	$	0.04	\pm	0.03	$	&	$							$	&	G0V		&	13, 15	\\
HD\,39392b	&	$							$	&	$	13.2	\pm	0.8				$	&	$	394.3	^{+	1.4	}_{-	1.2	}	$	&	$		0.394	\pm	0.008				$	&	$	1.08	\pm	0.08				$	&	$	5951	\pm	42	$	&	$	4.08	\pm	0.03				$	&	$	-0.32	\pm	0.03	$	&	$							$	&	F8		&	1	\\
HD\,4747b	&	$							$	&	$	39.6 - 58.1						$	&	$	7900 - 29000						$	&	$		0.676 - 0.831						$	&	$	0.81	\pm	0.02				$	&	$	5316	\pm	50	$	&	$	4.48	\pm	0.1				$	&	$	-0.21	\pm	0.05	$	&	$							$	&	G8		&	13, 33, 38, 47	\\
HD\,52756b	&	$							$	&	$	59.3	^{+	2.0	}_{-	1.9	}	$	&	$	52.8657	\pm	0.0001				$	&	$		0.6780	\pm	0.0003				$	&	$	0.83	\pm	0.01				$	&	$	5216	\pm	65	$	&	$	4.47	\pm	0.11				$	&	$	0.13	\pm	0.05	$	&	$							$	&	K1V		&	33	\\
HD\,5388b	&	$	69.2	\pm	19.9				$	&	$	1.96						$	&	$	777.0	\pm	4.0				$	&	$		0.40	\pm	0.02				$	&	$	1.21	\pm	0.12				$	&	$	6297	\pm	32	$	&	$	4.28	\pm	0.06				$	&	$	-0.27	\pm	0.02	$	&	$							$	&	F6V		&	50, 51	\\
HD\,72946b	&	$							$	&	$	60.4	\pm	2.2				$	&	$	5814.45	^{+	55	}_{-	47	}	$	&	$		0.495	\pm	0.006				$	&	$	0.96	\pm	0.07				$	&	$	5656	\pm	40	$	&	$	4.5	\pm	0.06				$	&	$	0.11	\pm	0.03	$	&	$							$	&	G5V		&	26	\\
HD\,77065b	&	$							$	&	$	41	\pm	2				$	&	$	119.1135	^{+	0.0026	}_{-	0.0027	}	$	&	$		0.69397	\pm	0.00036				$	&	$	0.75	\pm	0.06				$	&	$	4990	\pm	100	$	&	$	4.28	\pm	0.20				$	&	$	-0.51	\pm	0.10	$	&	$							$	&	G		&	1, 19	\\
HD\,89707b	&	$							$	&	$	53.6	^{+	7.8	}_{-	6.9	}	$	&	$	298.5	\pm	0.1				$	&	$		0.900	^{+	0.039	}_{-	0.035	}	$	&	$	0.96	\pm	0.04				$	&	$	6047	\pm	50	$	&	$	4.52	\pm	0.10				$	&	$	-0.33	\pm	0.06	$	&	$							$	&	G1V		&	33, 52, 53	\\
HD\,91669b	&	$							$	&	$	30.6	\pm	2.1				$	&	$	497.5	\pm	0.6				$	&	$		0.448	\pm	0.002				$	&	$	0.914	^{+	0.018	}_{-	0.087	}	$	&	$	5185	\pm	87	$	&	$	4.48	\pm	0.20				$	&	$	0.31	\pm	0.08	$	&	$							$	&	K0/K1		&	54	\\
HD\,92320b	&	$							$	&	$	59.4	\pm	4.1				$	&	$	145.402	\pm	0.013				$	&	$		0.3226	\pm	0.0014				$	&	$	0.92	\pm	0.04				$	&	$	5664	\pm	24	$	&	$	4.48	\pm	0.03				$	&	$	-0.10	\pm	0.06	$	&	$							$	&	G1		&	25	\\
KELT-1b	&	$	27.23	^{+	0.50	}_{-	0.48	}	$	&	$							$	&	$	1.217513	\pm	0.000015				$	&	$		0.0100	^{+	0.0100	}_{-	0.0070	}	$	&	$	1.335	\pm	0.063				$	&	$	6516	\pm	49	$	&	$	4.736	^{+	0.017	}_{-	0.025	}	$	&	$	0.052	\pm	0.079	$	&	$	0.07806	^{+	0.00061	}_{-	0.00058	}	$	&	Mid-F		&	55	\\
Kepler-39b	&	$	20.1	^{+	1.3	}_{-	1.2	}	$	&	$							$	&	$	21.087210	\pm	0.000037				$	&	$		0.112	\pm	0.057				$	&	$	1.29	^{+	0.06	}_{-	0.07	}	$	&	$	6350	\pm	100	$	&	$	4.25	\pm	0.06				$	&	$	0.10	\pm	0.14	$	&	$	0.0910	^{+	0.0006	}_{-	0.0008	}	$	&	F7V		&	56, 57	\\
KOI-205b	&	$	40.8	^{+	1.1	}_{-	1.5	}	$	&	$							$	&	$	11.720126	\pm	0.000011				$	&	$	<	0.015						$	&	$	0.96	^{+	0.03	}_{-	0.04	}	$	&	$	5400	\pm	75	$	&	$	4.558	\pm	0.014				$	&	$	0.18	\pm	0.12	$	&	$	0.09906	\pm	0.00094				$	&	K0		&	58, 59, 60, 61	\\
KOI-415b	&	$	62.14	\pm	2.69				$	&	$							$	&	$	166.78805	\pm	0.00022				$	&	$		0.698	\pm	0.002				$	&	$	0.94	\pm	0.06				$	&	$	5810	\pm	80	$	&	$	4.5	\pm	0.2				$	&	$	-0.24	\pm	0.11	$	&	$	0.0649	^{+	0.0017	}_{-	0.0013	}	$	&	G0IV		&	62	\\
KOI-423b	&	$	18.00	^{+	0.93	}_{-	0.91	}	$	&	$							$	&	$	21.0874	\pm	0.0002				$	&	$		0.121	^{+	0.022	}_{-	0.023	}	$	&	$	1.10	^{+	0.07	}_{-	0.06	}	$	&	$	6260	\pm	140	$	&	$	4.1	\pm	0.2				$	&	$	-0.29	\pm	0.10	$	&	$	0.0896	^{+	0.0011	}_{-	0.0012	}	$	&	F7IV		&	63	\\
TYC 2087-255-1b	&	$							$	&	$	40.0	\pm	2.5				$	&	$	9.0090	\pm	0.0004				$	&	$		0.226	\pm	0.011				$	&	$	1.16	\pm	0.08				$	&	$	5903	\pm	42	$	&	$	4.07	\pm	0.16				$	&	$	-0.23	\pm	0.04	$	&	$							$	&	G0IV		&	64	\\
TYC 2534-698-1b	&	$							$	&	$	39.1	\pm	11.5				$	&	$	103.698	\pm	0.111				$	&	$		0.385	\pm	0.011				$	&	$	0.998	\pm	0.040				$	&	$	5700	\pm	80	$	&	$	4.5	\pm	0.1				$	&	$	-0.25	\pm	0.06	$	&	$							$	&	G2V		&	65	\\
TYC 2930-872-1b	&	$							$	&	$	68.1	\pm	3.0				$	&	$	2.430420	\pm	0.000006				$	&	$		0.0066	\pm	0.0010				$	&	$	1.21	\pm	0.08				$	&	$	6427	\pm	33	$	&	$	4.52	\pm	0.14				$	&	$	-0.04	\pm	0.05	$	&	$							$	&	F8		&	66	\\
TYC 2949-557-1b	&	$							$	&	$	64.3	\pm	3.0				$	&	$	5.69459	\pm	0.00029				$	&	$		0.0017	^{+	0.0019	}_{-	0.0017	}	$	&	$	1.25	\pm	0.09				$	&	$	6135	\pm	40	$	&	$	4.4	\pm	0.1				$	&	$	0.32	\pm	0.01	$	&	$							$	&	F		&	67	\\
WASP-18b	&	$	10.11	^{+	0.24	}_{-	0.21	}	$	&	$							$	&	$	0.94145290	^{+	0.00000078	}_{-	0.00000086	}	$	&	$		0.00848	^{+	0.00085	}_{-	0.00095	}	$	&	$	1.24	\pm	0.04				$	&	$	6400	\pm	100	$	&	$	4.4	\pm	0.2				$	&	$	0.00	\pm	0.09	$	&	$	0.09576	^{+	0.00105	}_{-	0.00063	}	$	&	F6		&	68, 69	\\
WASP-30b	&	$	60.96	\pm	0.89				$	&	$							$	&	$	4.156736	\pm	0.000013				$	&	$		0		~\mathrm{(fixed)}				$	&	$	1.166	\pm	0.026				$	&	$	6201	\pm	97	$	&	$	4.280	\pm	0.010				$	&	$	-0.03	\pm	0.10	$	&	$	0.07057	\pm	0.01304				$	&	F8V		&	70	\\
XO-3b	&	$	11.79	\pm	0.59				$	&	$							$	&	$	3.1915239	\pm	0.0000068				$	&	$		0.260	\pm	0.017				$	&	$	1.213	\pm	0.066				$	&	$	6429	\pm	100	$	&	$	4.244	\pm	0.041				$	&	$	-0.177	\pm	0.080	$	&	$	0.92	\pm	0.04				$	&	F5V		&	71, 72, 73, 74	\\
 \enddata
 \vspace{-0.5cm}
\tablerefs{
(\,1\,)~This work	;
(\,2\,)~\cite{jiang13};
(\,3\,)~\cite{bouchy11};
(\,4\,)~\cite{parviainen14};
(\,5\,)~\cite{csizmadia15};
(\,6\,)~\cite{deleuil08};
(\,7\,)~\cite{feroz11};
(\,8\,)~\cite{winn10};
(\,9\,)~\cite{bakos09};
(\,10\,)~\cite{southworth12};
(\,11\,)~\cite{johnson11};
(\,12\,)~\cite{latham89};
(\,13\,)~\cite{valenti05};
(\,14\,)~\cite{butler06};
(\,15\,)~\cite{haywood08};
(\,16\,)~\cite{kane11};
(\,17\,)~\cite{sozzetti10};
(\,18\,)~\cite{moutou09};
(\,19\,)~\cite{latham02};
(\,20\,)~\cite{perrier03};
(\,21\,)~\cite{fischer02};
(\,22\,)~\cite{wittenmyer09b};
(\,23\,)~\cite{martioli10};
(\,24\,)~\cite{endl04};
(\,25\,)~\cite{diaz12};
(\,26\,)~\cite{bouchy15b};
(\,27\,)~\cite{tamuz08};
(\,28\,)~\cite{kane2011b};
(\,29\,)~\cite{udry02};
(\,30\,)~\cite{sato09};
(\,31\,)~\cite{bouchy09};
(\,32\,)~\cite{patel07};
(\,33\,)~\cite{sahlmann11};
(\,34\,)~\cite{marcy01};
(\,35\,)~\cite{zucker01};
(\,36\,)~\cite{reffert06};
(\,37\,)~\cite{pilyavsky11};
(\,38\,)~\cite{nidever02};
(\,39\,)~\cite{sivan04};
(\,40\,)~\cite{jenkins09};
(\,41\,)~\cite{correia05};
(\,42\,)~\cite{moutou11};
(\,43\,)~\cite{marmier13};
(\,44\,)~\cite{fischer03};
(\,45\,)~\cite{wright09};
(\,46\,)~\cite{benedict10};
(\,47\,)~\cite{santos05};
(\,50\,)~\cite{sahlmann11b};
(\,51\,)~\cite{santos10};
(\,52\,)~\cite{duquennoy91};
(\,53\,)~\cite{halbwachs00};
(\,54\,)~\cite{wittenmyer09a};
(\,55\,)~\cite{siverd12};
(\,56\,)~\cite{bouchy11b};
(\,57\,)~\cite{bonomo15};
(\,58\,)~\cite{siverd12};
(\,59\,)~\cite{borucki11};
(\,60\,)~\cite{ford11};
(\,61\,)~\cite{diaz13};
(\,62\,)~\cite{moutou13};
(\,63\,)~\cite{bouchy11b};
(\,64\,)~\cite{ma13};
(\,65\,)~\cite{kane2009};
(\,66\,)~\cite{fleming12};
(\,67\,)~\cite{fleming10};
(\,68\,)~\cite{hellier09};
(\,69\,)~\cite{triaud10};
(\,70\,)~\cite{anderson11};
(\,71\,)~\cite{johnskrull08};
(\,72\,)~\cite{winn08};
(\,73\,)~\cite{hebrard08};
(\,74\,)~\cite{winn09};
}
\end{deluxetable}

\end{landscape}

\end{document}